\documentclass[aps,prr,amsmath,amssymb,floatfix,twocolumn]{revtex4-1}

\usepackage{multirow}
\usepackage{bbold}
\usepackage{graphicx}
\usepackage{subfigure}
\usepackage{color}
\usepackage{hyperref}
\usepackage[justification=raggedright]{caption}
\usepackage{filecontents}

\begin{document}
\preprint{APS/123-QED}

\title{Machine learning guided discovery of stable, spin-resolved topological insulators}
\author{Alexander C. Tyner$^{1,2}$}

\affiliation{$^{1}$ Nordita, KTH Royal Institute of Technology and Stockholm University 106 91 Stockholm, Sweden}
\affiliation{$^2$ Department of Physics, University of Connecticut, Storrs, Connecticut 06269, USA}

\date{\today}

\begin{abstract} 
Identification of a non-trivial $\mathbb{Z}_{2}$ index in a spinful two dimensional insulator indicates the presence of an odd, quantized (pseudo)spin-resolved Chern number, $\mathcal{C}_{s}=(\mathcal{C}_{\uparrow}-\mathcal{C}_{\downarrow})/2$. However, the statement is not biconditional. An odd spin-Chern number can survive when the familiar $\mathbb{Z}_{2}$ index vanishes. Identification of solid-state systems hosting an odd, quantized $\mathcal{C}_{s}$ and trivial $\mathbb{Z}_{2}$ index is a pressing issue due to the potential for such insulators to admit band gaps optimal for experiments and quantum devices. Nevertheless, they have proven elusive due to the computational expense associated with their discovery. In this work, a neural network capable of identifying the spin-Chern number is developed and used to identify the first solid-state systems hosting a trivial $\mathbb{Z}_{2}$ index and odd $\mathcal{C}_{s}$. We demonstrate the potential of one such system, Ti$_{2}$CO$_{2}$, to support Majorana corner modes via the superconducting proximity effect.   
\end{abstract}

\maketitle
\par 
\section{Introduction} 
The comprehensive cataloguing of materials supporting non-trivial band-topology, as diagnosed via elementary band representations (EBRs) and other efficient protocols relying on analysis of symmetry eigenvalues represents a remarkable milestone\cite{tang2019efficient,zhang2019catalogue,vergniory2019complete,tang2019comprehensive,xu2020high}. However, it is now clear that these works can not represent a comprehensive analysis of bulk topology. This is due to the existence of symmetry non-indicative phases (SNIPs)\cite{Prodan2009,MooreHopf,PhysRevB.105.125115,PhysRevLett.126.216404,PhysRevResearch.3.033045,lange2023,Bai2022Doubled,bansilspin,tyner2020topology,tynerbismuthene}. Importantly, it has been demonstrated that there exist a number of two-dimensional higher-order insulators which fall in this category for which the ground-state bulk invariant is a non-zero spin-Chern number, as defined by Prodan\cite{Prodan2009}. 
\par 
The spin-Chern number, $\mathcal{C}_{s}=(\mathcal{C}_{\uparrow}-\mathcal{C}_{\downarrow})/2$, has been shown to be robust both in the presence of impurity effects and the absence of spin-rotation symmetry. It is protected by both the energetic bulk gap as well as the spin-gap. The spin-gap is identified by constructing the projected spin operator, $PSO=\mathbf{P}\hat{s}\mathbf{P}$, where $\mathbf{P}$ is the projector over occupied states and $\hat{s}$ is the preferred spin direction. In the presence of spin-rotation symmetry, the eigenvalues of the PSO are fixed to be $\pm 1$. When spin-rotation symmetry is broken, the eigenvalues adiabatically deviate, but as long as the spectra of the PSO remains gapped the spin-Chern number is robust. It is further shown by Lin et. al\cite{Lin2022Spin}, that for an odd, ground-state spin-Chern number, only the bulk-gap need be maintained to protect the band topology. Under this formulation, the $\mathbb{Z}_{2}$ index and mirror Chern number both represent special cases of the spin-Chern number, however this statement is not biconditional. A finite spin-Chern number need not imply a finite $\mathbb{Z}_{2}$ index or mirror Chern number.
\par 
Identification of a non-trivial spin-Chern number in systems lacking enhanced symmetry has proven computationally demanding, particularly in the context of \emph{ab initio} simulations of many-band systems\cite{Lin2022Spin,tynerbismuthene,tyner2023solitons}. While density-functional theory software exists for efficient symmetry analysis of the wavefunctions and computation of Wannier center spectra\cite{Taherinejad2014,Pizzi2020,Z2pack,WU2017}, these systems often fall under the category of higher-order topological insulators (HOTIs) and can thus be ``invisible" to the $\mathbb{Z}_{2}$ index. In correspondence with a trivialized $\mathbb{Z}_{2}$ index, the surface spectra does not support gapless modes which can be used to diagnose topology. In certain cases, corner localized states can be used to diagnose topology\cite{Benalcazar61,BenalacazarCn,schindler2018higher,Schindlereaat0346,sodequist2022abundance}, however corner modes are notoriously difficult to identify as their presence depends on the geometry of the sample and in the absence of chiral symmetry they are not pinned to zero energy, often becoming hidden among bulk states.
\par 

\begin{figure*}
\includegraphics[scale=0.63]{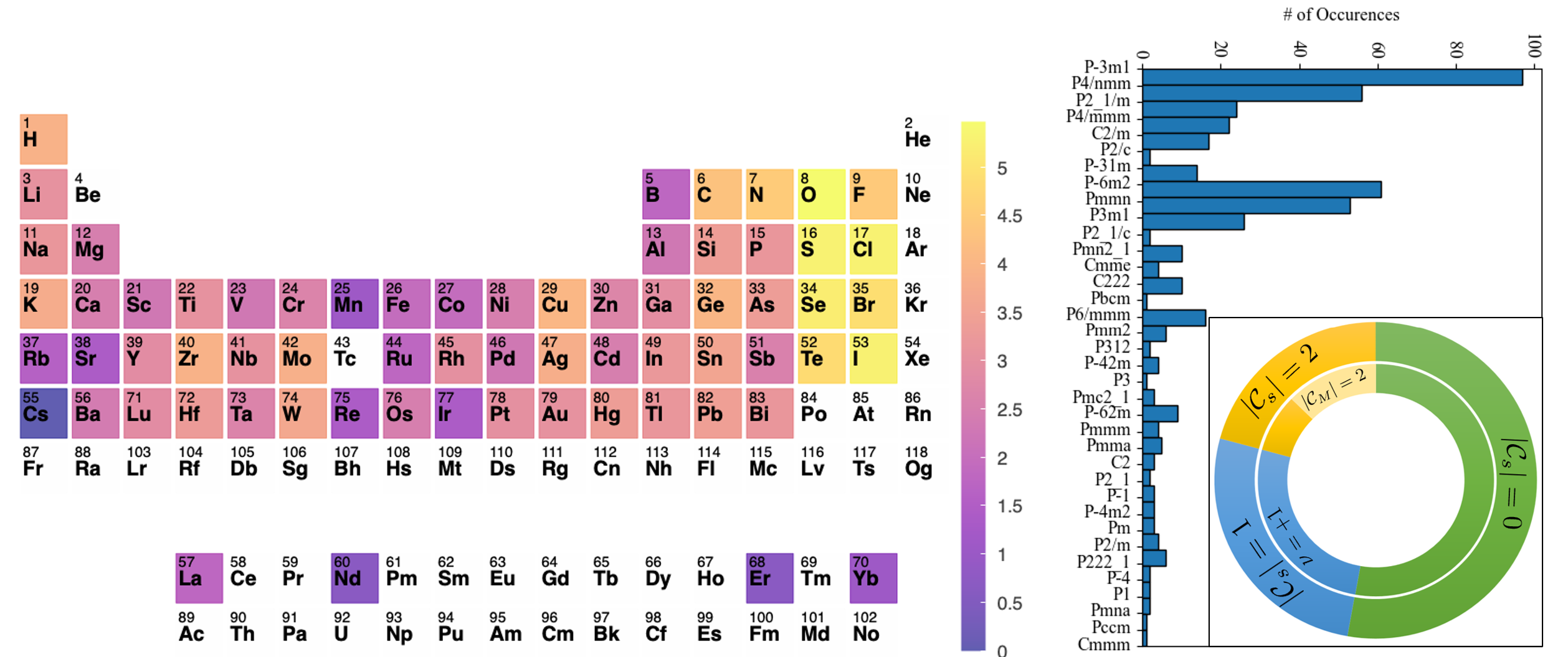}
\caption{\textbf{Statistics of training dataset:} Statistics of compounds contained within the training dataset. The periodic table is color coded according to the number of times each element occurs in the dataset on a log scale. Elements in white do not appear in the dataset. The circular plot details the breakdown of how many compounds support $|\mathcal{C}_{s}|=2,1,0$, and how many of the non-trivial materials are further identified by a $\mathbb{Z}_{2}$ index or mirror Chern number. The breakdown of spacegroups present and their occurrences is further provided. }
\label{fig:DataStats}
\end{figure*}

Alternatively, in a recent work by Tyner and Goswami\cite{tyner2023solitons}, the proposal of Qi and Zhang\cite{QiSpinCharge} and Ran et. al\cite{SpinChargeVishwanath} to utilize magnetic flux tubes ($\pi$-flux vortices) as real-space probes of both Chern and spin-Chern number was expanded to \emph{ab initio} simulations. It was shown to be robust in both symmetry indicative and non-indicative phases. An automated workflow was developed to scan 141 two-dimensional, spinful ($\mathcal{T}^2=-1$), insulators. The result was identification of 21, novel quantum spin-Hall insulators, falling under the category of SNIPs, but supporting a non-trivial, even spin-Chern number. While automated and basis agnostic, this work remained computationally demanding. Furthermore, a two-dimensional insulator supporting \emph{odd spin-Chern number and trivial $\mathbb{Z}_{2}$ index} remained elusive.
\par 

This situation motivates a novel search strategy relying on development of a neural network capable of diagnosing the presence of bulk topological order in $\mathcal{T}$ preserving systems with $\mathcal{T}^2=-1$. {\color{black} While a machine learning based approach does not eliminate the need for density functional theory computations as they are necessary to validate the results; this approach allows for creation of an optimal list of candidate materials that can be validated within a reasonable timescale. Furthermore, prior intuition or knowledge of the electronic properties of a given material is not necessary to determine whether it is an optimal candidate for non-trivial topology, allowing for a more diverse material landscape to be considered.} This neural network should posses the capability to identify topological order of multiple types, including: (a) first-order insulators supporting a non-trivial Fu-Kane index, (b) topological crystalline insulators admitting finite mirror Chern numbers, and (c) higher-order or generalized quantum spin-Hall insulators (QSHIs). Such broad capabilities are crucial as any system supporting a non-trivial Fu-Kane index or mirror Chern number also supports a non-trivial (pseudo)spin-resolved Chern number. Therefore, in the above list, systems belonging to set (a) and (b) are subsets of (c).
\par
In this work we construct two convolutional neural networks (CNNs), both relying on voxel encoding of the crystal structure\cite{noh2019inverse,long2021constrained}. The first produces a binary classification, dictating whether the spin-Chern number is zero or finite. The second produces a multi-class classification, dictating whether the spin-Chern number is zero, odd, or non-zero and even. These CNNs are then applied to a set of experimentally synthesized, two-dimensional insulators in the computational two-dimensional materials database (C2DB)\cite{rasmussen2015computational,haastrup2018computational}. Materials of interest predicted to support non-trivial, odd ground-state spin-Chern number are subsequently selected and analyzed in depth, leading to identification of the first two-dimensional insulator supporting $|\mathcal{C}_{s,G}|=1$ and a trivial $\mathbb{Z}_{2}$ index. This work serves as a proof that the apparent computational expense present in identification of spin-resolved band topology can be overcome through the use of machine learning techniques. 
\par 
There is a particular need for such efficient tools to diagnose (pseudo)spin topology given the ongoing experimental interest in two-dimensional hetero-structures and twisted materials. In these systems additional degrees of freedom such as valley can give rise to pseudospin-resolved topology\cite{mak2014valley,tao2024giant}. 
\par
\section{Construction of training data-set}
The training dataset consists of spinful ($\mathcal{T}^2=-1$), two-dimensional insulators. The corresponding ground-state spin-Chern number is labeled as zero, odd, or finite and even. This dataset is made distinct from existing databases by enhanced criteria for labelling a system as topologically trivial. Following Ref. \cite{tyner2023solitons}, a compound is labelled topologically trivial only if it is demonstrated that an inserted magnetic flux tube (vortex) admits no mid-gap bound modes.

\begin{figure*}[t]
\centering

\includegraphics[width=15cm]{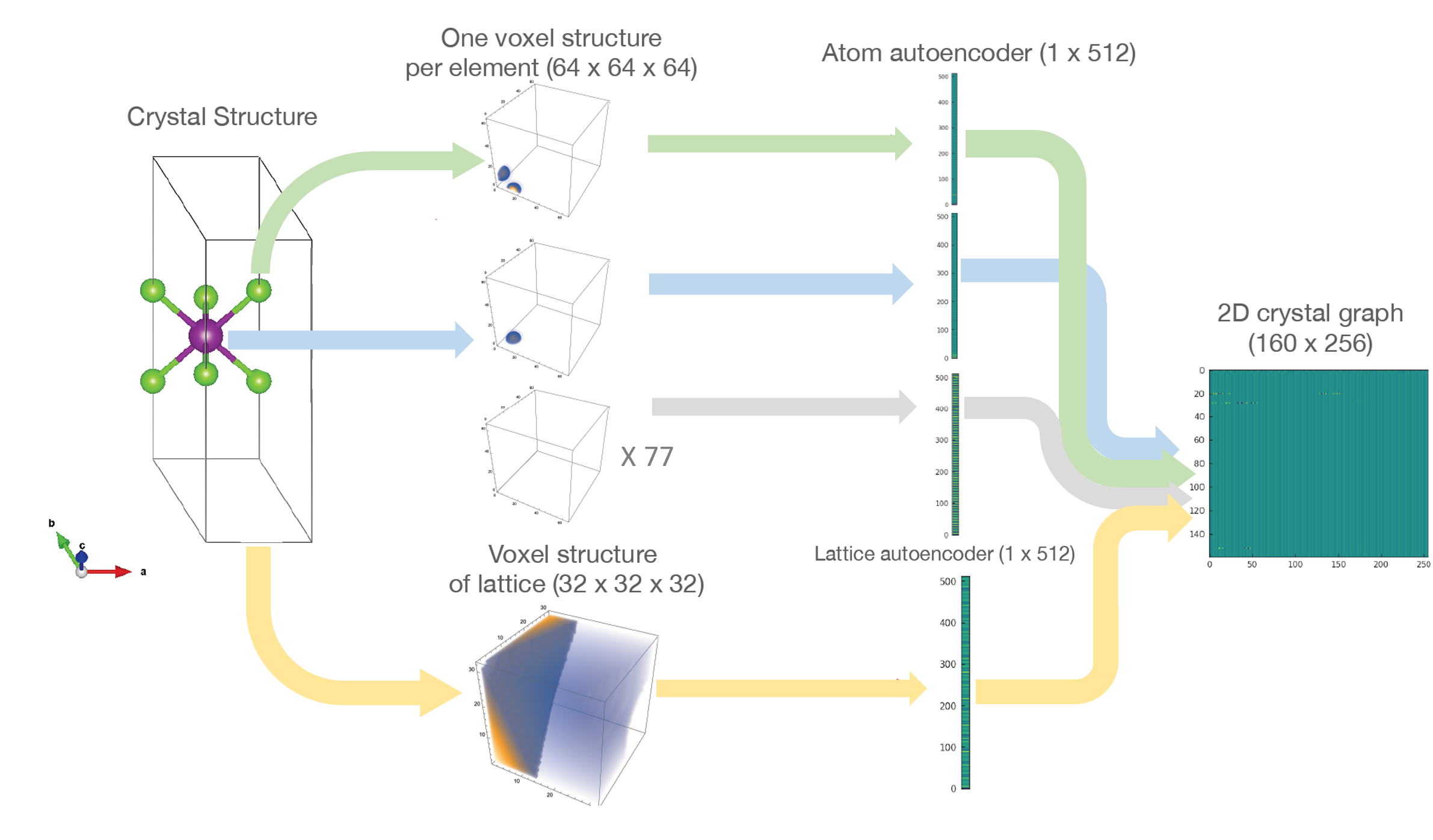}

\caption{\textbf{Construction of crystal graphs:} Schematic detailing the construction of two-dimensional crystal graphs for each compound in the training set. This process ensures a continuous representation of the structure that can be presented to the neural network.}
\label{fig:datasetdetails}
\end{figure*}

\par
It has been shown that for a spin-Hall insulator supporting $|C_{s,G}|=N$, when flux is tuned to $\phi=hc/(2e)$ ($\pi$-flux) there exists $2N$ mid-gap vortex bound modes (VBMs)\cite{QiSpinCharge,SpinChargeVishwanath,Lin2022Spin,tynerbismuthene}. The spin-Chern number can then determined by computing quantized induced charge on the vortex. If the VBMs are half-filled, the vortex acquires induced spin but no induced charge. If we dope by $N_{e} \in [-N,+N]$ electrons away from half-filling, occupying all VBMs, the vortex acquires induced charge $\delta Q= N_{e}\times e$. If this condition is satisfied, the spin Chern number can be directly calculated by fixing $N_{e}=N$ such that $\delta Q= |\mathcal{C}_{s,G}|\times e$. While the introduction of spin-orbit coupling causes the spin bound to the vortex to become finite but non-quantized, the quantization of bound charge remains robust. For this reason, quantization of bound charge was used in Ref. \cite{tyner2023solitons} as the criteria for topological classification. We choose to follow this procedure for construction of the training dataset in this work, screening all materials in the database of Mounet et al.\cite{mounet2018two} with less than 10 atoms in the unit cell which have not previously been labeled as topological via existing symmetry based methods.
\par
All first principles calculations based on density-functional theory (DFT) are carried out using the Quantum Espresso software package \cite{QE-2009,QE-2017,QE-2020}. Exchange-correlation potentials use the Perdew-Burke-Ernzerhof (PBE) parametrization of the generalized gradient approximation (GGA) \cite{Perdew1996}. We utilize norm-conserving pseudo-potentials\cite{Hamann2013} as obtained on the Pseudo-Dojo site\cite{van2018pseudodojo}. Spin-orbit coupling is considered in all calculations. The Wannier90\cite{Pizzi2020}, Z2pack\cite{Z2pack}, and BerryEasy\cite{tyner2023berryeasy} software packages were utilized in calculation of all topological invariants. In order to facilitate automated analysis of the bulk topology, Wannier tight-binding (WTB) models are constructed through use of the SCDM method introduced by Vitale et. al\cite{vitale2020automated}. Manipulation of Wannier tight-binding models for vortex insertion is done with a custom python program which will be made publicly available upon being developed into a stand-alone package. The result is a database of 246 symmetry non-indicative compounds with the full results available at Ref. \cite{github}. Despite this expanded search, in each system analyzed the bulk invariant is found to be either zero or an even spin-Chern number. A two-dimensional insulator supporting \emph{odd spin-Chern number and trivial $\mathbb{Z}_{2}$ index} remained elusive.

\par 
To expand the dataset, all insulators supporting a non-trivial mirror Chern number or $\mathcal{Z}_{2}$ index identified in Refs. \cite{2dsymmind,C2DBTIs,marrazzo2019relative} are incorporated into the training set. This is possible due to the fact that a non-trivial $\mathcal{Z}_{2}$  index or mirror Chern number can be considered as a special case of the (pseudo)spin-Chern number. We remark that we cannot include any materials labelled trivial from Refs. \cite{2dsymmind,C2DBTIs,marrazzo2019relative} in the training data as this may cause mislabelling of SNIPs. The resulting aggregate dataset consists of 443 two-dimensional materials. Further information regarding the composition of the dataset is visible in Fig. \eqref{fig:datasetdetails}.

\par

\begin{figure*}[t]
\centering

\includegraphics[scale=0.75]{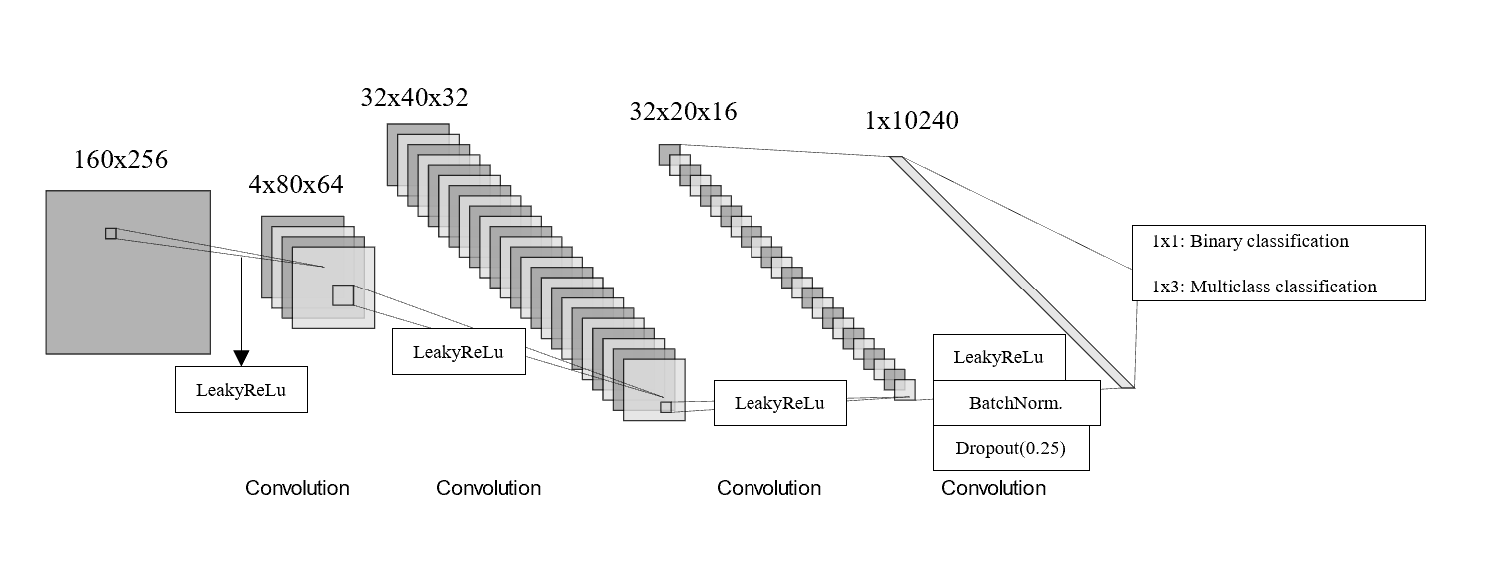}

\caption{\textbf{Convolutional neural network architecture:} Architecture of convolutional neural network for binary and multiclass classification. The binary and multiclass classification utilize a sigmoid and softmax activation function respectively for the final dense layer. The leakyReLu function is fixed to $\alpha=0.2$ in each case.}
\label{fig:NNA}
\end{figure*}

\begin{figure*}[t]
\centering
\subfigure[]{
\includegraphics[scale=0.45]{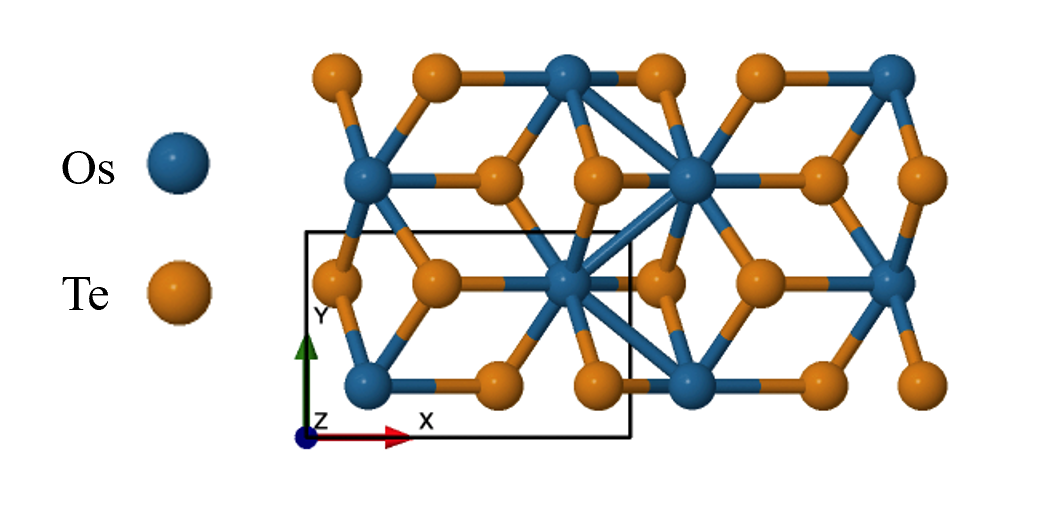}
\label{fig:}}
\subfigure[]{
\includegraphics[scale=0.4]{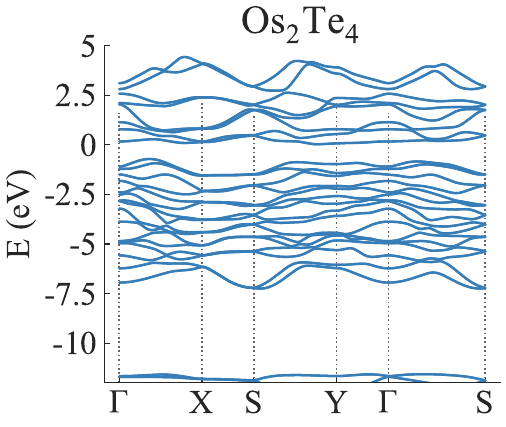}
\label{fig:}}
\subfigure[]{
\includegraphics[scale=0.2]{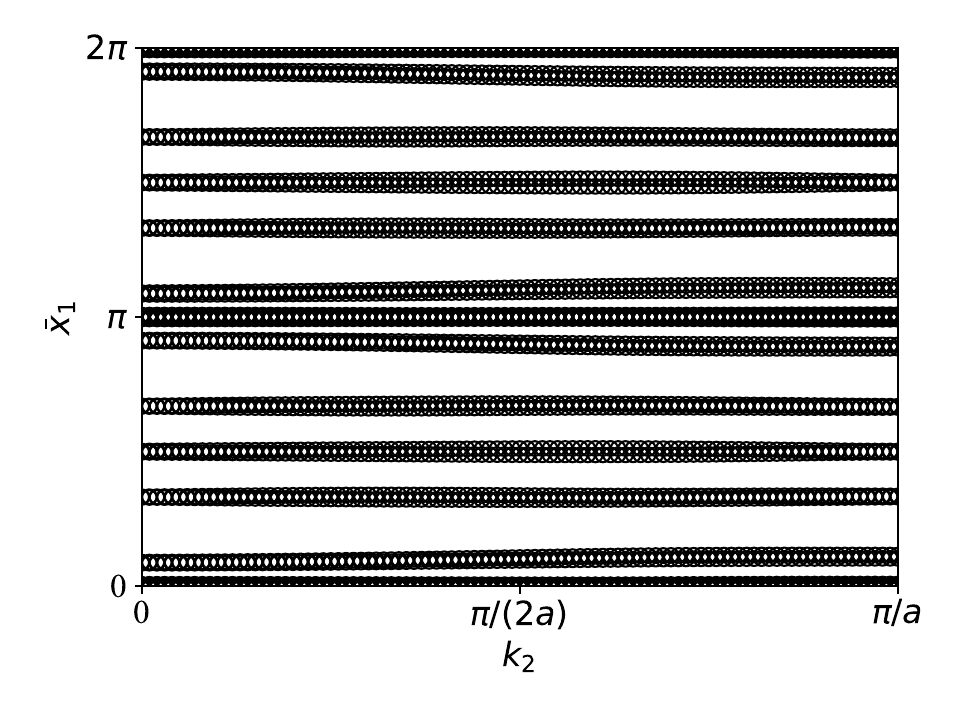}
\label{fig:}}
\subfigure[]{
\includegraphics[scale=0.4]{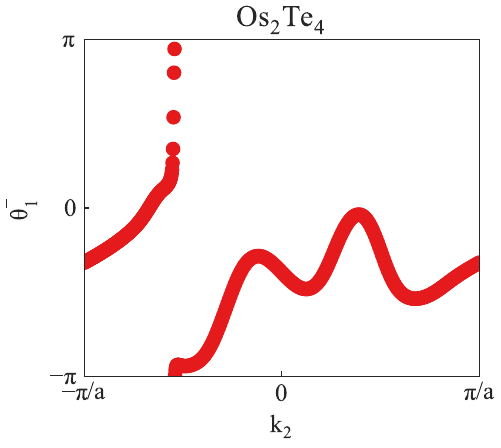}
\label{fig:}}
\subfigure[]{
\includegraphics[scale=0.45]{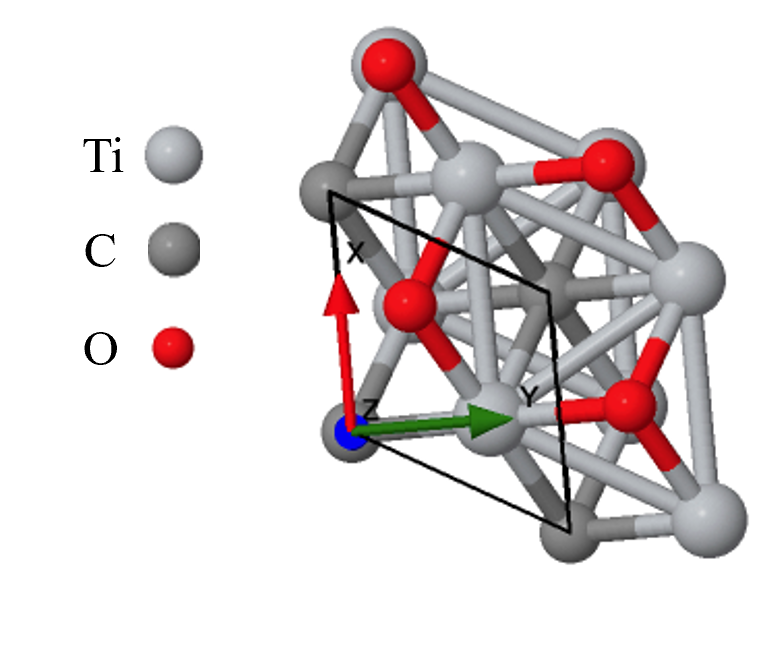}
\label{fig:}}
\subfigure[]{
\includegraphics[scale=0.4]{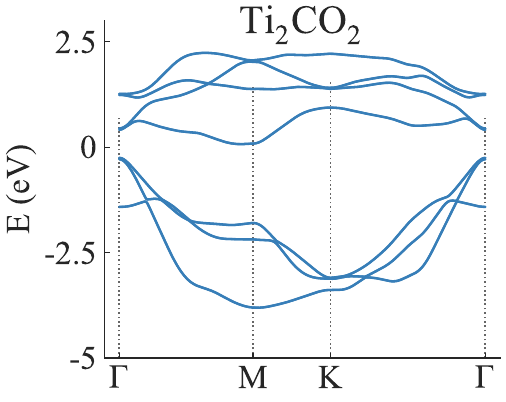}
\label{fig:}}
\subfigure[]{
\includegraphics[scale=0.4]{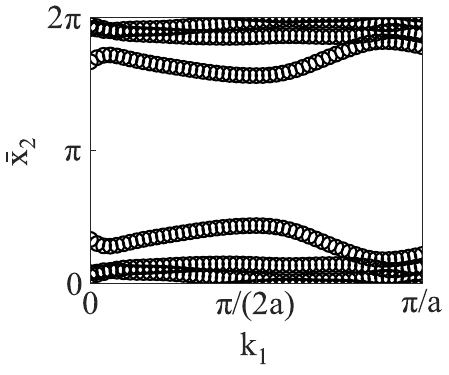}
\label{fig:}}
\subfigure[]{
\includegraphics[scale=0.4]{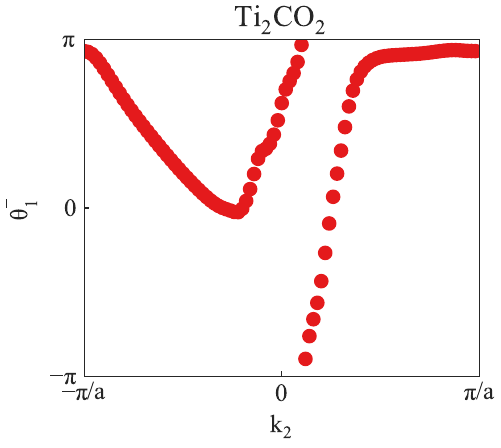}
\label{fig:}}

\caption{\textbf{Analysis of spin-resolved topological insulator candidates:} Structure of (a) Os$_{2}$Te$_{4}$ and (e) Ti$_{2}$CO$_{2}$ as given via the C2DB database \cite{rasmussen2015computational,haastrup2018computational}. The computed band structure along a high-symmetry path in the Brillouin zone detailing the bands nearest to the Fermi energy can be seen in (b) and (f) for Os$_{2}$Te$_{4}$ and Ti$_{2}$CO$_{2}$ respectively. The gapped wannier center spectra of both compounds, seen in (c) and (g) indicates these systems are trivial under the $\mathbb{Z}_{2}$ index. The ground state spin-Chern number is computed via spin-resolved Wilson loop, detailing a single winding for both compounds, confirming $|\mathcal{C}_{s}|=1$.}
\label{fig:Mats}
\end{figure*}

\par 
\section{Data augmentation and processing}
A number of strategies for presentation of lattice structure to CNNs have been developed and tested in recent years\cite{cgcnn,nouira2018crystalgan,hoffmann2019data,de2016comparing,kaufmann2020crystal,kim2020generative}. In this work, we utilize the strategy of forming a continuous representation by autoencoding voxel images of the crystal structure to create a 2D crystal graph. We account for the possibility of 79 different elements in the crystal structure, specifically atomic numbers 1-84 removing the noble gases. Details of the training set are shown in Fig. \eqref{fig:datasetdetails}. As a result, regardless of the number of elements in a single crystal structure, 80 voxel images will be produced. An autoencoder then translates each image into a vector. A similar process is done to form a voxel image of the lattice which again is translated into a one-dimensional array through use of an autoencoder. These one-dimensional arrays are then reshaped into 2D crytal images which can be presented to the CNN. In this way the crystal structure obtains a continuous and reversible representation. 
\par 
Importantly, this process allows for implementation of a data-augmentation strategy, expanding the dataset to include $\approx 10^3$ data-points such that the convolutional neural network may achieve a level of accuracy sufficient to benefit our materials search\cite{cgcnn}. Namely, we permute the primitive lattice vectors, altering the atomic positions accordingly and generating new 2D crystal images. This process allows for training data to be augmented by a factor of two if only the $a$ and $b$ primitive vectors are permuted, and up to a factor of six if all are permuted. For further details of voxel image production and auto-encoding please consult the supplementary material\cite{suppl}. Details of the CNNs architecture for binary and multiclass classification are visible in Fig. \eqref{fig:NNA}. 

\par 
\section{Neural network performance}
We begin with the CNN for binary classification, given the limited training data available it is expected that this model will achieve higher accuracy. The data set is randomized and an 80\%/10\%/10\% train/validate/test split is utilized. Early stopping based on validation loss is implemented and the batch size is set to 128. The CNN reaches a train/validation/test accuracy of 96\%/88.7\%/88.5\%. While extraordinary accuracy for neural networks has become commonplace, we note that these values are quite high given the limited training dataset.  {\color{black}For context, we compare these values with those produced utilizing a convolutional crystal graph neural network (CGCNN) in the supplementary material\cite{cgcnn,cgcnn2,suppl}.The CGCNN architecture is commonly regarded as the state-of-the-art for machine-learning based property prediction in computational materials science. It is important to note that CGCNN has not been selected as the primary approach in this work as it disallows the use of the data-augmentation technique of exchanging lattice parameters. This limits the size of the total training data set. For a larger dataset, the CGCNN method could be advantageous. Other common machine learning architectures for performing topological classification of crystal structures rely primarily on generation of an input vector constructed from features of the constituent atoms as well properties of the nearest neighbors rather than a direct continuous image of the crystal structure\cite{NRML,ma2023topogivity,2dmagml}. In models of this type, details of the crystal structure have also been incorporated by dividing the training dataset such that all constituent compounds correspond to a selected prototypical crystal structure\cite{schleder2021machine}. This approach is again problematic for the purposes of this work as it would result in severely limiting the size of the training dataset. We place emphasis on the use of a continuous and reversible image of the crystal structure as input to lay a foundation for the future incorporation of the model in a generative artificial intelligence architecture\cite{long2021constrained}.} 
\par 
For the multi-class classification model, we adjust the CNN architecture to the form seen in Fig. \eqref{fig:datasetdetails}. We again employ the same split for training, test and validation data and find a train/validation/test categorical accuracy of 95\%/89.5\%/88\%. This represents a significant step towards isolating optimal candidate SNIPs. 
\par 
{\color{black} At this point it is important to discuss potential sources of bias in the neural network. The most prominent source of bias is likely to be the limited quantity of training data available. This bias is expected as not all elements in the training data set will appear with the same frequency. Those which appear sparsely have the potential to have a significant impact on the model, particularly if all compounds in which the element is found have the same topological classification. Such biases and trends are explored in detail in the supplementary material\cite{suppl}.}

\par 
\section{Application of CNN}
We begin by selecting candidate compounds from the computational two-dimensional materials database (C2DB)\cite{rasmussen2015computational,haastrup2018computational} with the criteria that a compound must not be included in the training database or previously identified as topological via symmetry-based methods. Each compound must also admit a bulk band gap greater than 0.1 $eV$ and be dynamically stable. Two-dimensional crystal graphs are then constructed and analyzed using the binary classification CNN. Compounds labelled as non-trivial are isolated and subsequently fed to the multi-class CNN. We filter this final list by number of atoms in the unit cell, selecting two compounds with less than 10 atoms in the unit cell predicted to support $|\mathcal{C}_{s}|=1$, Os$_{2}$Te$_{4}$ and Ti$_{2}$CO$_{2}$.
\par 
Plots of the crystal structure and band structure for both compounds are visible in Fig. \eqref{fig:Mats}. In order to perform topological analysis of the structures, a Wannier tight-binding model is produced\cite{Pizzi2020}, exactly replicating only the Kramers-degenerate bands nearest to the Fermi energy using carefully selected orbitals. As these bands are significantly energetically separated from all other bands, the TB model reproduces the DFT data precisely. For more computational details please see the supplementary material\cite{suppl}. The spin Chern number is then computed directly via the method established by Prodan\cite{Prodan2009,tyner2023berryeasy}. This procedure requires defining the projected spin operator (PSO), $P(\mathbf{k})\hat{s}P(\mathbf{k})$, where $P(\mathbf{k})$ is the projector onto occupied bands and $\hat{s}$ is the preferred spin axis. We identify the preferred spin axis supporting a spin-gap through a computationally expensive trial and error procedure. However, once identified we are able to produce the results shown in Fig. \eqref{fig:Mats}, displaying calculation of the spin-Chern number via spin-resolved Wilson loop as detailed in Lin et. al\cite{Lin2022Spin}. Remarkably both compounds demonstrate proper labeling by the neural network and support of $|C_{s}|=1$.
\par 
These results are of significance due to the robust nature of the bulk invariant. Unlike other SNIPs which require the preservation of the bulk spin-gap, a challenging task experimentally, the bulk topological invariant of these systems requires only preservation of the bulk energetic gap to remain intact. As a result, the bulk-invariant is robust to the influence of disorder\cite{Prodan2009} and other perturbations, much like traditional $\mathbb{Z}_{2}$ topological insulators. Unlike existing topological insulators, both systems support sizeable bulk energetic gaps, $~0.325 eV$ and $~0.713eV$ for Ti$_{2}$CO$_{2}$ and Os$_{2}$Te$_{4}$ respectively. These large energetic gaps make both materials primary candidates for experiments and quantum devices as we explore in the following section. {\color{black} Furthermore, synthesis of Ti$_{2}$CO$_{2}$ has been reported in the literature\cite{naguib2012mxene} as well as single crystal Os$_{2}$Te$_{4}$ \cite{muller1991single}. In the case of Ti$_{2}$CO$_{2}$, synthesis is accelerated by the commercial availability of TiC.}  
\begin{figure}
    \subfigure[]{
    \includegraphics[width=3.7cm]{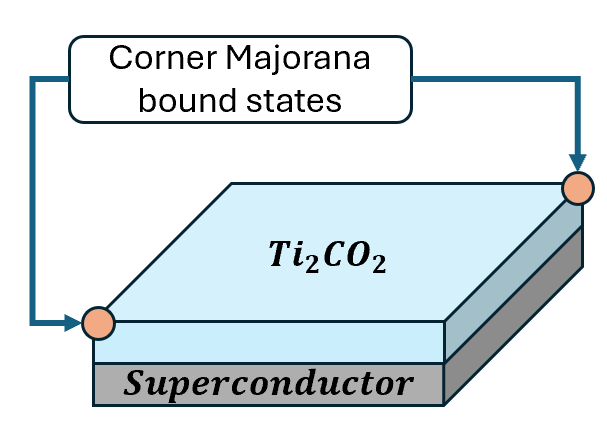}
    \label{fig:ProxSchem}}
    \subfigure[]{
    \includegraphics[width=3.7cm]{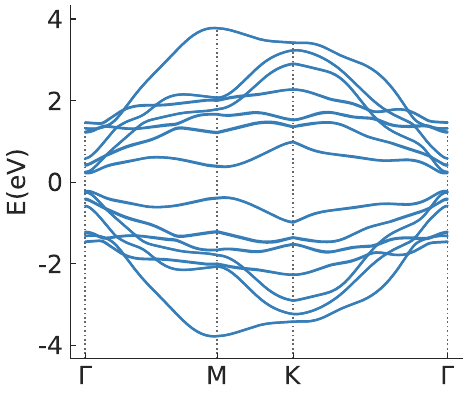}
    \label{fig:BdGBands}}
    \vfill
    \subfigure[]{
    \includegraphics[width=3.7cm]{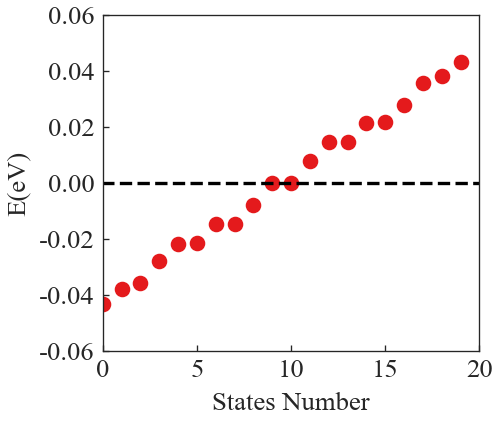}
    \label{fig:BdGStates}}
    \subfigure[]{
    \includegraphics[width=3.7cm]{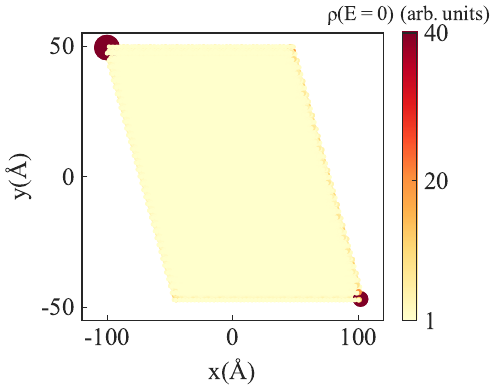}
    \label{fig:BdGCorner}}
    \caption{\textbf{Majorana corner modes via superconducting proximity effect:} (a) Schematic of setup for generating superconducting proximity effect in Ti$_{2}$CO$_{2}$. (b) Band structure of BdG Hamiltonian, $H_{eff}$. (c) States nearest zero energy for slab of $H_{eff}$ shown in (d) implementing open boundary conditions along both principal axes. (d) Localization of zero energy states shown in (c), demonstrating that they are corner bound. Local density of states on a given site is displayed as a function of color and site size for clarity. A larger site size indicates increased local density of states.}
    \label{fig:ProximityTiClO}
\end{figure}
\par 
\section{Accessing Majorana corner modes via superconducting proximity effect}
\par 
Prior works have explored the superconducting proximity effect in topological insulators\cite{FuProximity,JQIMajorana,cook1,cook2,prlProximity,reeg2018proximity,MCSHTC,LiuMBS}. In particular such proposals have focused on use of the superconducting proximity effect to realize Majorana bound states (MBSs) by layering a two-dimensional topological insulator on the surface of a superconductor. The experimental realization of MBSs is important due to their proposed utilization as a platform for topological quantum computing\cite{RevModPhys.80.1083}. A common issue in the proposed platforms for realizing MBSs via the superconducting proximity effect in topological insulators is an extremely small topological gap. Experimental realizations generally rely on known two-dimensional topological insulators\cite{MCSHTC,LiuMBS} such as 1T'-WTe$_{2}$\cite{qian2014quantum} and PbTe\cite{hsieh2012topological,liu2015crystal} which support normal state topological band gaps on the order of tens of $meV$. 
\par 
Similarly, proposals of platforms for the realization of Majorana corner states generally rely on an induced d-wave pairing. Experimental estimates for the induced pairing gap are on the order of several $meV$\cite{lupke2020proximity,shimamura2018ultrathin,ProxZhao}. Control of this small topological gap poses a significant experimental challenge, particularly as it requires extremely small levels of disorder to be present in a given sample.
\par 
In this section we briefly investigate the superconducting proximity effect in one of the proposed spin-resolved topological insulators listed above, Ti$_{2}$CO$_{2}$. In order to investigate the superconducting proximity effect we consider a device constructed from a two-dimensional slab of Ti$_{2}$CO$_{2}$ placed on top of a superconductor as shown schematically in Fig. \eqref{fig:ProxSchem}. To model the system we construct an effective Bouligobov-de-Gennes (BdG) Hamiltonian, $H_{eff}$, of the form, 
\begin{equation}
    H_{eff}(\mathbf{k})= \begin{bmatrix}
    H_{WTB}(\mathbf{k}) & \Delta\\
    \Delta^{\dagger} & -H_{WTB}^{*}(-\mathbf{k}),
    \end{bmatrix}
\end{equation}
where $H_{WTB}$ is the Wannier tight-binding model constructed in the previous section for analysis of Ti$_{2}$CO$_{2}$. Following Ref. \cite{prlProximity}, we include the superconducting proximity effect by introducing a simplistic nondissipative pairing term, $\Delta$, which we fix as a constant. The resulting band structure of $H_{eff}$ is shown in Fig. \eqref{fig:BdGBands}.
\par 
We investigate the presence of MBSs by performing exact diagonalization of $H_{eff}$ for a system of $49 \times 37$ unit cells, and $\Delta=180 \mu eV$. Fascinatingly, examining the states nearest to zero energy, the results in Fig. \eqref{fig:BdGStates} demonstrate the presence of two zero energy MBSs. Investigating the localization of the MBSs we find the results in Fig. \eqref{fig:BdGCorner} demonstrating that these states are corner bound and exist on opposite sides of the sample. Importantly, as seen in Fig. \eqref{fig:BdGBands}, the bulk energetic gap, $\Delta E_{gap}$, is on the order of $\approx 0.45 eV$. {\color{black}This is significantly larger than the typical value for known alternatives and underscores the implications of identifying large band-gap SNIPs. Beyond the intriguing properties which they support on a theoretical level, SNIPs provide a viable route to overcome current practical experimental issues in the identification of MBSs.}
\par
Furthermore, we highlight that in this setup, Majorana corner modes survive when the pairing term is suppressed. Such pairing is crucial for the existence of corner modes in alternative proposals based on the proximity effect in two-dimensional FOTIs\cite{PhysRevB.98.245413,liu2015crystal,MCSHTC}. This result underscores the potential utility of spin-resolved topological insulators in experimental platforms. 
\par 
\section{Summary}
In this work, machine learning guided discovery of symmetry non-indicative topological phases in two-dimensions has been shown to be a promising route to the identification of large band-gap quantum spin-Hall insulators. Furthermore, the resulting network can leverage a limited amount of training data through augmentation techniques. In the past, machine learning techniques for identification of non-trivial topology were dismissed due to the computational efficiency of directly calculating the bulk invariant using symmetry indicator techniques. The immense computational expense associated with direct calculation of band topology in SNIPs warrants the use of a novel machine learning approach. {\color{black} By limiting the associated computational expense, it is now possible to go beyond analysis of known compounds that have been successfully synthesized and explore large datasets such as the virtual two-dimensional material database (V2DB)\cite{sorkun2020artificial}. The neural network produced in this work can also serve as a building block in the future development of generative artificial intelligence models for inverse design of topological materials.} Finally, we expect that such a network can find extensive use in analyzing the growing number of two-dimensional hetero-structures and twisted architectures which have attracted experimental interest but where the normal state band topology is unknown due to the complexity of a many-atom unit cell which will be explored in a future work. 

\acknowledgements{}

I am grateful to Alexander V. Balatsky, Pallab Goswami, Fazel Tafti and Bianca Baldassarri for stimulating discussions. Nordita is supported in part by NordForsk. The computations were enabled by resources provided by the National Academic Infrastructure for Supercomputing in Sweden (NAISS), partially funded by the Swedish Research Council through grant agreement no. 2022-06725.

\bibliographystyle{apsrev4-1}
\nocite{apsrev41Control}
\bibliography{ref.bib}

\begin{thebibliography}{79}%
\makeatletter
\providecommand \@ifxundefined [1]{%
 \@ifx{#1\undefined}
}%
\providecommand \@ifnum [1]{%
 \ifnum #1\expandafter \@firstoftwo
 \else \expandafter \@secondoftwo
 \fi
}%
\providecommand \@ifx [1]{%
 \ifx #1\expandafter \@firstoftwo
 \else \expandafter \@secondoftwo
 \fi
}%
\providecommand \natexlab [1]{#1}%
\providecommand \enquote  [1]{``#1''}%
\providecommand \bibnamefont  [1]{#1}%
\providecommand \bibfnamefont [1]{#1}%
\providecommand \citenamefont [1]{#1}%
\providecommand \href@noop [0]{\@secondoftwo}%
\providecommand \href [0]{\begingroup \@sanitize@url \@href}%
\providecommand \@href[1]{\@@startlink{#1}\@@href}%
\providecommand \@@href[1]{\endgroup#1\@@endlink}%
\providecommand \@sanitize@url [0]{\catcode `\\12\catcode `\$12\catcode `\&12\catcode `\#12\catcode `\^12\catcode `\_12\catcode `\%12\relax}%
\providecommand \@@startlink[1]{}%
\providecommand \@@endlink[0]{}%
\providecommand \url  [0]{\begingroup\@sanitize@url \@url }%
\providecommand \@url [1]{\endgroup\@href {#1}{\urlprefix }}%
\providecommand \urlprefix  [0]{URL }%
\providecommand \Eprint [0]{\href }%
\providecommand \doibase [0]{http://dx.doi.org/}%
\providecommand \selectlanguage [0]{\@gobble}%
\providecommand \bibinfo  [0]{\@secondoftwo}%
\providecommand \bibfield  [0]{\@secondoftwo}%
\providecommand \translation [1]{[#1]}%
\providecommand \BibitemOpen [0]{}%
\providecommand \bibitemStop [0]{}%
\providecommand \bibitemNoStop [0]{.\EOS\space}%
\providecommand \EOS [0]{\spacefactor3000\relax}%
\providecommand \BibitemShut  [1]{\csname bibitem#1\endcsname}%
\let\auto@bib@innerbib\@empty
\bibitem [{\citenamefont {Tang}\ \emph {et~al.}(2019{\natexlab{a}})\citenamefont {Tang}, \citenamefont {Po}, \citenamefont {Vishwanath},\ and\ \citenamefont {Wan}}]{tang2019efficient}%
  \BibitemOpen
  \bibfield  {author} {\bibinfo {author} {\bibnamefont {Tang}, \bibfnamefont {F.}}, \bibinfo {author} {\bibnamefont {Po}, \bibfnamefont {H.~C.}}, \bibinfo {author} {\bibnamefont {Vishwanath}, \bibfnamefont {A.}}, \ and\ \bibinfo {author} {\bibnamefont {Wan}, \bibfnamefont {X.}},\ }\bibfield  {title} {\enquote {\bibinfo {title} {Efficient topological materials discovery using symmetry indicators},}\ }\href {\doibase 10.1038/s41586-019-0937-5} {\bibfield  {journal} {\bibinfo  {journal} {Nat. Phys.}\ }\textbf {\bibinfo {volume} {15}},\ \bibinfo {pages} {470--476} (\bibinfo {year} {2019}{\natexlab{a}})}\BibitemShut {NoStop}%
\bibitem [{\citenamefont {Zhang}\ \emph {et~al.}(2019)\citenamefont {Zhang}, \citenamefont {Jiang}, \citenamefont {Song}, \citenamefont {Huang}, \citenamefont {He}, \citenamefont {Fang}, \citenamefont {Weng},\ and\ \citenamefont {Fang}}]{zhang2019catalogue}%
  \BibitemOpen
  \bibfield  {author} {\bibinfo {author} {\bibnamefont {Zhang}, \bibfnamefont {T.}}, \bibinfo {author} {\bibnamefont {Jiang}, \bibfnamefont {Y.}}, \bibinfo {author} {\bibnamefont {Song}, \bibfnamefont {Z.}}, \bibinfo {author} {\bibnamefont {Huang}, \bibfnamefont {H.}}, \bibinfo {author} {\bibnamefont {He}, \bibfnamefont {Y.}}, \bibinfo {author} {\bibnamefont {Fang}, \bibfnamefont {Z.}}, \bibinfo {author} {\bibnamefont {Weng}, \bibfnamefont {H.}}, \ and\ \bibinfo {author} {\bibnamefont {Fang}, \bibfnamefont {C.}},\ }\bibfield  {title} {\enquote {\bibinfo {title} {Catalogue of topological electronic materials},}\ }\href {\doibase 10.1038/s41586-019-0944-6} {\bibfield  {journal} {\bibinfo  {journal} {Nature}\ }\textbf {\bibinfo {volume} {566}},\ \bibinfo {pages} {475--479} (\bibinfo {year} {2019})}\BibitemShut {NoStop}%
\bibitem [{\citenamefont {Vergniory}\ \emph {et~al.}(2019)\citenamefont {Vergniory}, \citenamefont {Elcoro}, \citenamefont {Felser}, \citenamefont {Regnault}, \citenamefont {Bernevig},\ and\ \citenamefont {Wang}}]{vergniory2019complete}%
  \BibitemOpen
  \bibfield  {author} {\bibinfo {author} {\bibnamefont {Vergniory}, \bibfnamefont {M.}}, \bibinfo {author} {\bibnamefont {Elcoro}, \bibfnamefont {L.}}, \bibinfo {author} {\bibnamefont {Felser}, \bibfnamefont {C.}}, \bibinfo {author} {\bibnamefont {Regnault}, \bibfnamefont {N.}}, \bibinfo {author} {\bibnamefont {Bernevig}, \bibfnamefont {B.~A.}}, \ and\ \bibinfo {author} {\bibnamefont {Wang}, \bibfnamefont {Z.}},\ }\bibfield  {title} {\enquote {\bibinfo {title} {A complete catalogue of high-quality topological materials},}\ }\href {\doibase 10.1038/s41586-019-0954-4} {\bibfield  {journal} {\bibinfo  {journal} {Nature}\ }\textbf {\bibinfo {volume} {566}},\ \bibinfo {pages} {480--485} (\bibinfo {year} {2019})}\BibitemShut {NoStop}%
\bibitem [{\citenamefont {Tang}\ \emph {et~al.}(2019{\natexlab{b}})\citenamefont {Tang}, \citenamefont {Po}, \citenamefont {Vishwanath},\ and\ \citenamefont {Wan}}]{tang2019comprehensive}%
  \BibitemOpen
  \bibfield  {author} {\bibinfo {author} {\bibnamefont {Tang}, \bibfnamefont {F.}}, \bibinfo {author} {\bibnamefont {Po}, \bibfnamefont {H.~C.}}, \bibinfo {author} {\bibnamefont {Vishwanath}, \bibfnamefont {A.}}, \ and\ \bibinfo {author} {\bibnamefont {Wan}, \bibfnamefont {X.}},\ }\bibfield  {title} {\enquote {\bibinfo {title} {Comprehensive search for topological materials using symmetry indicators},}\ }\href {\doibase 10.1038/s41586-019-0937-5} {\bibfield  {journal} {\bibinfo  {journal} {Nature}\ }\textbf {\bibinfo {volume} {566}},\ \bibinfo {pages} {486--489} (\bibinfo {year} {2019}{\natexlab{b}})}\BibitemShut {NoStop}%
\bibitem [{\citenamefont {Xu}\ \emph {et~al.}(2020)\citenamefont {Xu}, \citenamefont {Elcoro}, \citenamefont {Song}, \citenamefont {Wieder}, \citenamefont {Vergniory}, \citenamefont {Regnault}, \citenamefont {Chen}, \citenamefont {Felser},\ and\ \citenamefont {Bernevig}}]{xu2020high}%
  \BibitemOpen
  \bibfield  {author} {\bibinfo {author} {\bibnamefont {Xu}, \bibfnamefont {Y.}}, \bibinfo {author} {\bibnamefont {Elcoro}, \bibfnamefont {L.}}, \bibinfo {author} {\bibnamefont {Song}, \bibfnamefont {Z.-D.}}, \bibinfo {author} {\bibnamefont {Wieder}, \bibfnamefont {B.~J.}}, \bibinfo {author} {\bibnamefont {Vergniory}, \bibfnamefont {M.}}, \bibinfo {author} {\bibnamefont {Regnault}, \bibfnamefont {N.}}, \bibinfo {author} {\bibnamefont {Chen}, \bibfnamefont {Y.}}, \bibinfo {author} {\bibnamefont {Felser}, \bibfnamefont {C.}}, \ and\ \bibinfo {author} {\bibnamefont {Bernevig}, \bibfnamefont {B.~A.}},\ }\bibfield  {title} {\enquote {\bibinfo {title} {High-throughput calculations of magnetic topological materials},}\ }\href {\doibase 10.1038/s41586-020-2837-0} {\bibfield  {journal} {\bibinfo  {journal} {Nature}\ }\textbf {\bibinfo {volume} {586}},\ \bibinfo {pages} {702--707} (\bibinfo {year} {2020})}\BibitemShut {NoStop}%
\bibitem [{\citenamefont {Prodan}(2009)}]{Prodan2009}%
  \BibitemOpen
  \bibfield  {author} {\bibinfo {author} {\bibnamefont {Prodan}, \bibfnamefont {E.}},\ }\bibfield  {title} {\enquote {\bibinfo {title} {Robustness of the spin-chern number},}\ }\href {\doibase 10.1103/PhysRevB.80.125327} {\bibfield  {journal} {\bibinfo  {journal} {Phys. Rev. B}\ }\textbf {\bibinfo {volume} {80}},\ \bibinfo {pages} {125327} (\bibinfo {year} {2009})}\BibitemShut {NoStop}%
\bibitem [{\citenamefont {Moore}\ \emph {et~al.}(2008)\citenamefont {Moore}, \citenamefont {Ran},\ and\ \citenamefont {Wen}}]{MooreHopf}%
  \BibitemOpen
  \bibfield  {author} {\bibinfo {author} {\bibnamefont {Moore}, \bibfnamefont {J.~E.}}, \bibinfo {author} {\bibnamefont {Ran}, \bibfnamefont {Y.}}, \ and\ \bibinfo {author} {\bibnamefont {Wen}, \bibfnamefont {X.-G.}},\ }\bibfield  {title} {\enquote {\bibinfo {title} {Topological surface states in three-dimensional magnetic insulators},}\ }\href {\doibase 10.1103/PhysRevLett.101.186805} {\bibfield  {journal} {\bibinfo  {journal} {Phys. Rev. Lett.}\ }\textbf {\bibinfo {volume} {101}},\ \bibinfo {pages} {186805} (\bibinfo {year} {2008})}\BibitemShut {NoStop}%
\bibitem [{\citenamefont {Cano}\ \emph {et~al.}(2022)\citenamefont {Cano}, \citenamefont {Elcoro}, \citenamefont {Aroyo}, \citenamefont {Bernevig},\ and\ \citenamefont {Bradlyn}}]{PhysRevB.105.125115}%
  \BibitemOpen
  \bibfield  {author} {\bibinfo {author} {\bibnamefont {Cano}, \bibfnamefont {J.}}, \bibinfo {author} {\bibnamefont {Elcoro}, \bibfnamefont {L.}}, \bibinfo {author} {\bibnamefont {Aroyo}, \bibfnamefont {M.~I.}}, \bibinfo {author} {\bibnamefont {Bernevig}, \bibfnamefont {B.~A.}}, \ and\ \bibinfo {author} {\bibnamefont {Bradlyn}, \bibfnamefont {B.}},\ }\bibfield  {title} {\enquote {\bibinfo {title} {Topology invisible to eigenvalues in obstructed atomic insulators},}\ }\href {\doibase 10.1103/PhysRevB.105.125115} {\bibfield  {journal} {\bibinfo  {journal} {Phys. Rev. B}\ }\textbf {\bibinfo {volume} {105}},\ \bibinfo {pages} {125115} (\bibinfo {year} {2022})}\BibitemShut {NoStop}%
\bibitem [{\citenamefont {Nelson}\ \emph {et~al.}(2021)\citenamefont {Nelson}, \citenamefont {Neupert}, \citenamefont {Bzdu\ifmmode~\check{s}\else \v{s}\fi{}ek},\ and\ \citenamefont {Alexandradinata}}]{PhysRevLett.126.216404}%
  \BibitemOpen
  \bibfield  {author} {\bibinfo {author} {\bibnamefont {Nelson}, \bibfnamefont {A.}}, \bibinfo {author} {\bibnamefont {Neupert}, \bibfnamefont {T.}}, \bibinfo {author} {\bibnamefont {Bzdu\ifmmode~\check{s}\else \v{s}\fi{}ek}, \bibfnamefont {T.~c.~v.}}, \ and\ \bibinfo {author} {\bibnamefont {Alexandradinata}, \bibfnamefont {A.}},\ }\bibfield  {title} {\enquote {\bibinfo {title} {Multicellularity of delicate topological insulators},}\ }\href {\doibase 10.1103/PhysRevLett.126.216404} {\bibfield  {journal} {\bibinfo  {journal} {Phys. Rev. Lett.}\ }\textbf {\bibinfo {volume} {126}},\ \bibinfo {pages} {216404} (\bibinfo {year} {2021})}\BibitemShut {NoStop}%
\bibitem [{\citenamefont {Lapierre}\ \emph {et~al.}(2021)\citenamefont {Lapierre}, \citenamefont {Neupert},\ and\ \citenamefont {Trifunovic}}]{PhysRevResearch.3.033045}%
  \BibitemOpen
  \bibfield  {author} {\bibinfo {author} {\bibnamefont {Lapierre}, \bibfnamefont {B.}}, \bibinfo {author} {\bibnamefont {Neupert}, \bibfnamefont {T.}}, \ and\ \bibinfo {author} {\bibnamefont {Trifunovic}, \bibfnamefont {L.}},\ }\bibfield  {title} {\enquote {\bibinfo {title} {$n$-band hopf insulator},}\ }\href {\doibase 10.1103/PhysRevResearch.3.033045} {\bibfield  {journal} {\bibinfo  {journal} {Phys. Rev. Res.}\ }\textbf {\bibinfo {volume} {3}},\ \bibinfo {pages} {033045} (\bibinfo {year} {2021})}\BibitemShut {NoStop}%
\bibitem [{\citenamefont {Lange}\ \emph {et~al.}(2023)\citenamefont {Lange}, \citenamefont {Bouhon},\ and\ \citenamefont {Slager}}]{lange2023}%
  \BibitemOpen
  \bibfield  {author} {\bibinfo {author} {\bibnamefont {Lange}, \bibfnamefont {G.~F.}}, \bibinfo {author} {\bibnamefont {Bouhon}, \bibfnamefont {A.}}, \ and\ \bibinfo {author} {\bibnamefont {Slager}, \bibfnamefont {R.-J.}},\ }\bibfield  {title} {\enquote {\bibinfo {title} {Spin texture as a bulk indicator of fragile topology},}\ }\href {\doibase 10.1103/PhysRevResearch.5.033013} {\bibfield  {journal} {\bibinfo  {journal} {Phys. Rev. Res.}\ }\textbf {\bibinfo {volume} {5}},\ \bibinfo {pages} {033013} (\bibinfo {year} {2023})}\BibitemShut {NoStop}%
\bibitem [{\citenamefont {Bai}\ \emph {et~al.}(2022)\citenamefont {Bai}, \citenamefont {Cai}, \citenamefont {Mao}, \citenamefont {Li}, \citenamefont {Dai}, \citenamefont {Huang},\ and\ \citenamefont {Niu}}]{Bai2022Doubled}%
  \BibitemOpen
  \bibfield  {author} {\bibinfo {author} {\bibnamefont {Bai}, \bibfnamefont {Y.}}, \bibinfo {author} {\bibnamefont {Cai}, \bibfnamefont {L.}}, \bibinfo {author} {\bibnamefont {Mao}, \bibfnamefont {N.}}, \bibinfo {author} {\bibnamefont {Li}, \bibfnamefont {R.}}, \bibinfo {author} {\bibnamefont {Dai}, \bibfnamefont {Y.}}, \bibinfo {author} {\bibnamefont {Huang}, \bibfnamefont {B.}}, \ and\ \bibinfo {author} {\bibnamefont {Niu}, \bibfnamefont {C.}},\ }\bibfield  {title} {\enquote {\bibinfo {title} {Doubled quantum spin hall effect with high-spin chern number in $\ensuremath{\alpha}$-antimonene and $\ensuremath{\alpha}$-bismuthene},}\ }\href {\doibase 10.1103/PhysRevB.105.195142} {\bibfield  {journal} {\bibinfo  {journal} {Phys. Rev. B}\ }\textbf {\bibinfo {volume} {105}},\ \bibinfo {pages} {195142} (\bibinfo {year} {2022})}\BibitemShut {NoStop}%
\bibitem [{\citenamefont {Wang}\ \emph {et~al.}(2022)\citenamefont {Wang}, \citenamefont {Zhou}, \citenamefont {Lin}, \citenamefont {Lin},\ and\ \citenamefont {Bansil}}]{bansilspin}%
  \BibitemOpen
  \bibfield  {author} {\bibinfo {author} {\bibnamefont {Wang}, \bibfnamefont {B.}}, \bibinfo {author} {\bibnamefont {Zhou}, \bibfnamefont {X.}}, \bibinfo {author} {\bibnamefont {Lin}, \bibfnamefont {Y.-C.}}, \bibinfo {author} {\bibnamefont {Lin}, \bibfnamefont {H.}}, \ and\ \bibinfo {author} {\bibnamefont {Bansil}, \bibfnamefont {A.}},\ }\bibfield  {title} {\enquote {\bibinfo {title} {High spin-chern-number insulator in $\alpha$-antimonene with a hidden topological phase},}\ }\href {https://doi.org/10.48550/arXiv.2202.04162} {\bibfield  {journal} {\bibinfo  {journal} {arXiv:2202.04162}\ } (\bibinfo {year} {2022})}\BibitemShut {NoStop}%
\bibitem [{\citenamefont {Tyner}\ \emph {et~al.}(2023)\citenamefont {Tyner}, \citenamefont {Sur}, \citenamefont {Puggioni}, \citenamefont {Rondinelli},\ and\ \citenamefont {Goswami}}]{tyner2020topology}%
  \BibitemOpen
  \bibfield  {author} {\bibinfo {author} {\bibnamefont {Tyner}, \bibfnamefont {A.~C.}}, \bibinfo {author} {\bibnamefont {Sur}, \bibfnamefont {S.}}, \bibinfo {author} {\bibnamefont {Puggioni}, \bibfnamefont {D.}}, \bibinfo {author} {\bibnamefont {Rondinelli}, \bibfnamefont {J.~M.}}, \ and\ \bibinfo {author} {\bibnamefont {Goswami}, \bibfnamefont {P.}},\ }\bibfield  {title} {\enquote {\bibinfo {title} {Topology of three-dimensional dirac semimetals and quantum spin hall systems without gapless edge modes},}\ }\href {\doibase 10.1103/PhysRevResearch.5.L012019} {\bibfield  {journal} {\bibinfo  {journal} {Phys. Rev. Res.}\ }\textbf {\bibinfo {volume} {5}},\ \bibinfo {pages} {L012019} (\bibinfo {year} {2023})}\BibitemShut {NoStop}%
\bibitem [{\citenamefont {Tyner}\ and\ \citenamefont {Goswami}(2023{\natexlab{a}})}]{tynerbismuthene}%
  \BibitemOpen
  \bibfield  {author} {\bibinfo {author} {\bibnamefont {Tyner}, \bibfnamefont {A.~C.}}\ and\ \bibinfo {author} {\bibnamefont {Goswami}, \bibfnamefont {P.}},\ }\bibfield  {title} {\enquote {\bibinfo {title} {Spin-charge separation and quantum spin hall effect of $\beta$-bismuthene},}\ }\href {\doibase https://doi.org/10.1038/s41598-023-38491-1} {\bibfield  {journal} {\bibinfo  {journal} {Sci. Rep.}\ }\textbf {\bibinfo {volume} {13}},\ \bibinfo {pages} {11393} (\bibinfo {year} {2023}{\natexlab{a}})}\BibitemShut {NoStop}%
\bibitem [{\citenamefont {Lin}\ \emph {et~al.}(2024)\citenamefont {Lin}, \citenamefont {Palumbo}, \citenamefont {Guo}, \citenamefont {Hwang}, \citenamefont {Blackburn}, \citenamefont {Shoemaker}, \citenamefont {Mahmood}, \citenamefont {Wang}, \citenamefont {Fiete}, \citenamefont {Wieder} \emph {et~al.}}]{Lin2022Spin}%
  \BibitemOpen
  \bibfield  {author} {\bibinfo {author} {\bibnamefont {Lin}, \bibfnamefont {K.-S.}}, \bibinfo {author} {\bibnamefont {Palumbo}, \bibfnamefont {G.}}, \bibinfo {author} {\bibnamefont {Guo}, \bibfnamefont {Z.}}, \bibinfo {author} {\bibnamefont {Hwang}, \bibfnamefont {Y.}}, \bibinfo {author} {\bibnamefont {Blackburn}, \bibfnamefont {J.}}, \bibinfo {author} {\bibnamefont {Shoemaker}, \bibfnamefont {D.~P.}}, \bibinfo {author} {\bibnamefont {Mahmood}, \bibfnamefont {F.}}, \bibinfo {author} {\bibnamefont {Wang}, \bibfnamefont {Z.}}, \bibinfo {author} {\bibnamefont {Fiete}, \bibfnamefont {G.~A.}}, \bibinfo {author} {\bibnamefont {Wieder}, \bibfnamefont {B.~J.}},  \emph {et~al.},\ }\bibfield  {title} {\enquote {\bibinfo {title} {Spin-resolved topology and partial axion angles in three-dimensional insulators},}\ }\href {\doibase https://doi.org/10.1038/s41467-024-44762-w} {\bibfield  {journal} {\bibinfo  {journal} {Nat. Commun.}\ }\textbf {\bibinfo {volume} {15}},\ \bibinfo {pages} {550} (\bibinfo {year}
  {2024})}\BibitemShut {NoStop}%
\bibitem [{\citenamefont {Tyner}\ and\ \citenamefont {Goswami}(2023{\natexlab{b}})}]{tyner2023solitons}%
  \BibitemOpen
  \bibfield  {author} {\bibinfo {author} {\bibnamefont {Tyner}, \bibfnamefont {A.~C.}}\ and\ \bibinfo {author} {\bibnamefont {Goswami}, \bibfnamefont {P.}},\ }\bibfield  {title} {\enquote {\bibinfo {title} {Solitons and real-space screening of bulk topology of quantum materials},}\ }\href {\doibase https://doi.org/10.48550/arXiv.2304.05424} {\bibfield  {journal} {\bibinfo  {journal} {arXiv:2304.05424}\ } (\bibinfo {year} {2023}{\natexlab{b}}),\ https://doi.org/10.48550/arXiv.2304.05424}\BibitemShut {NoStop}%
\bibitem [{\citenamefont {Taherinejad}\ \emph {et~al.}(2014)\citenamefont {Taherinejad}, \citenamefont {Garrity},\ and\ \citenamefont {Vanderbilt}}]{Taherinejad2014}%
  \BibitemOpen
  \bibfield  {author} {\bibinfo {author} {\bibnamefont {Taherinejad}, \bibfnamefont {M.}}, \bibinfo {author} {\bibnamefont {Garrity}, \bibfnamefont {K.~F.}}, \ and\ \bibinfo {author} {\bibnamefont {Vanderbilt}, \bibfnamefont {D.}},\ }\bibfield  {title} {\enquote {\bibinfo {title} {{Wannier center sheets in topological insulators}},}\ }\href {\doibase 10.1103/PhysRevB.89.115102} {\bibfield  {journal} {\bibinfo  {journal} {Phys. Rev. B}\ }\textbf {\bibinfo {volume} {89}},\ \bibinfo {pages} {1--14} (\bibinfo {year} {2014})},\ \Eprint {http://arxiv.org/abs/1312.6940} {1312.6940} \BibitemShut {NoStop}%
\bibitem [{\citenamefont {Pizzi}\ \emph {et~al.}(2020)\citenamefont {Pizzi}, \citenamefont {Vitale}, \citenamefont {Arita}, \citenamefont {Blugel}, \citenamefont {Freimuth}, \citenamefont {G{\'{e}}ranton}, \citenamefont {Gibertini}, \citenamefont {Gresch}, \citenamefont {Johnson}, \citenamefont {Koretsune}, \citenamefont {Iba{\~{n}}ez-Azpiroz}, \citenamefont {Lee}, \citenamefont {Lihm}, \citenamefont {Marchand}, \citenamefont {Marrazzo}, \citenamefont {Mokrousov}, \citenamefont {Mustafa}, \citenamefont {Nohara}, \citenamefont {Nomura}, \citenamefont {Paulatto}, \citenamefont {Ponc{\'{e}}}, \citenamefont {Ponweiser}, \citenamefont {Qiao}, \citenamefont {Thole}, \citenamefont {Tsirkin}, \citenamefont {Wierzbowska}, \citenamefont {Marzari}, \citenamefont {Vanderbilt}, \citenamefont {Souza}, \citenamefont {Mostofi},\ and\ \citenamefont {Yates}}]{Pizzi2020}%
  \BibitemOpen
  \bibfield  {author} {\bibinfo {author} {\bibnamefont {Pizzi}, \bibfnamefont {G.}}, \bibinfo {author} {\bibnamefont {Vitale}, \bibfnamefont {V.}}, \bibinfo {author} {\bibnamefont {Arita}, \bibfnamefont {R.}}, \bibinfo {author} {\bibnamefont {Blugel}, \bibfnamefont {S.}}, \bibinfo {author} {\bibnamefont {Freimuth}, \bibfnamefont {F.}}, \bibinfo {author} {\bibnamefont {G{\'{e}}ranton}, \bibfnamefont {G.}}, \bibinfo {author} {\bibnamefont {Gibertini}, \bibfnamefont {M.}}, \bibinfo {author} {\bibnamefont {Gresch}, \bibfnamefont {D.}}, \bibinfo {author} {\bibnamefont {Johnson}, \bibfnamefont {C.}}, \bibinfo {author} {\bibnamefont {Koretsune}, \bibfnamefont {T.}}, \bibinfo {author} {\bibnamefont {Iba{\~{n}}ez-Azpiroz}, \bibfnamefont {J.}}, \bibinfo {author} {\bibnamefont {Lee}, \bibfnamefont {H.}}, \bibinfo {author} {\bibnamefont {Lihm}, \bibfnamefont {J.-M.}}, \bibinfo {author} {\bibnamefont {Marchand}, \bibfnamefont {D.}}, \bibinfo {author} {\bibnamefont {Marrazzo}, \bibfnamefont {A.}}, \bibinfo {author}
  {\bibnamefont {Mokrousov}, \bibfnamefont {Y.}}, \bibinfo {author} {\bibnamefont {Mustafa}, \bibfnamefont {J.~I.}}, \bibinfo {author} {\bibnamefont {Nohara}, \bibfnamefont {Y.}}, \bibinfo {author} {\bibnamefont {Nomura}, \bibfnamefont {Y.}}, \bibinfo {author} {\bibnamefont {Paulatto}, \bibfnamefont {L.}}, \bibinfo {author} {\bibnamefont {Ponc{\'{e}}}, \bibfnamefont {S.}}, \bibinfo {author} {\bibnamefont {Ponweiser}, \bibfnamefont {T.}}, \bibinfo {author} {\bibnamefont {Qiao}, \bibfnamefont {J.}}, \bibinfo {author} {\bibnamefont {Thole}, \bibfnamefont {F.}}, \bibinfo {author} {\bibnamefont {Tsirkin}, \bibfnamefont {S.~S.}}, \bibinfo {author} {\bibnamefont {Wierzbowska}, \bibfnamefont {M.}}, \bibinfo {author} {\bibnamefont {Marzari}, \bibfnamefont {N.}}, \bibinfo {author} {\bibnamefont {Vanderbilt}, \bibfnamefont {D.}}, \bibinfo {author} {\bibnamefont {Souza}, \bibfnamefont {I.}}, \bibinfo {author} {\bibnamefont {Mostofi}, \bibfnamefont {A.~A.}}, \ and\ \bibinfo {author} {\bibnamefont {Yates}, \bibfnamefont
  {J.~R.}},\ }\bibfield  {title} {\enquote {\bibinfo {title} {Wannier90 as a community code: new features and applications},}\ }\href {\doibase 10.1088/1361-648x/ab51ff} {\bibfield  {journal} {\bibinfo  {journal} {J. Phys. Condens. Matter}\ }\textbf {\bibinfo {volume} {32}},\ \bibinfo {pages} {165902} (\bibinfo {year} {2020})}\BibitemShut {NoStop}%
\bibitem [{\citenamefont {Gresch}\ \emph {et~al.}(2017)\citenamefont {Gresch}, \citenamefont {Aut\`es}, \citenamefont {Yazyev}, \citenamefont {Troyer}, \citenamefont {Vanderbilt}, \citenamefont {Bernevig},\ and\ \citenamefont {Soluyanov}}]{Z2pack}%
  \BibitemOpen
  \bibfield  {author} {\bibinfo {author} {\bibnamefont {Gresch}, \bibfnamefont {D.}}, \bibinfo {author} {\bibnamefont {Aut\`es}, \bibfnamefont {G.}}, \bibinfo {author} {\bibnamefont {Yazyev}, \bibfnamefont {O.~V.}}, \bibinfo {author} {\bibnamefont {Troyer}, \bibfnamefont {M.}}, \bibinfo {author} {\bibnamefont {Vanderbilt}, \bibfnamefont {D.}}, \bibinfo {author} {\bibnamefont {Bernevig}, \bibfnamefont {B.~A.}}, \ and\ \bibinfo {author} {\bibnamefont {Soluyanov}, \bibfnamefont {A.~A.}},\ }\bibfield  {title} {\enquote {\bibinfo {title} {Z2pack: Numerical implementation of hybrid wannier centers for identifying topological materials},}\ }\href {\doibase 10.1103/PhysRevB.95.075146} {\bibfield  {journal} {\bibinfo  {journal} {Phys. Rev. B}\ }\textbf {\bibinfo {volume} {95}},\ \bibinfo {pages} {075146} (\bibinfo {year} {2017})}\BibitemShut {NoStop}%
\bibitem [{\citenamefont {Wu}\ \emph {et~al.}(2018)\citenamefont {Wu}, \citenamefont {Zhang}, \citenamefont {Song}, \citenamefont {Troyer},\ and\ \citenamefont {Soluyanov}}]{WU2017}%
  \BibitemOpen
  \bibfield  {author} {\bibinfo {author} {\bibnamefont {Wu}, \bibfnamefont {Q.}}, \bibinfo {author} {\bibnamefont {Zhang}, \bibfnamefont {S.}}, \bibinfo {author} {\bibnamefont {Song}, \bibfnamefont {H.-F.}}, \bibinfo {author} {\bibnamefont {Troyer}, \bibfnamefont {M.}}, \ and\ \bibinfo {author} {\bibnamefont {Soluyanov}, \bibfnamefont {A.~A.}},\ }\bibfield  {title} {\enquote {\bibinfo {title} {Wanniertools : An open-source software package for novel topological materials},}\ }\href {\doibase https://doi.org/10.1016/j.cpc.2017.09.033} {\bibfield  {journal} {\bibinfo  {journal} {Computer Physics Communications}\ }\textbf {\bibinfo {volume} {224}},\ \bibinfo {pages} {405 -- 416} (\bibinfo {year} {2018})}\BibitemShut {NoStop}%
\bibitem [{\citenamefont {Benalcazar}\ \emph {et~al.}(2017)\citenamefont {Benalcazar}, \citenamefont {Bernevig},\ and\ \citenamefont {Hughes}}]{Benalcazar61}%
  \BibitemOpen
  \bibfield  {author} {\bibinfo {author} {\bibnamefont {Benalcazar}, \bibfnamefont {W.~A.}}, \bibinfo {author} {\bibnamefont {Bernevig}, \bibfnamefont {B.~A.}}, \ and\ \bibinfo {author} {\bibnamefont {Hughes}, \bibfnamefont {T.~L.}},\ }\bibfield  {title} {\enquote {\bibinfo {title} {Quantized electric multipole insulators},}\ }\href {\doibase 10.1126/science.aah6442} {\bibfield  {journal} {\bibinfo  {journal} {Science}\ }\textbf {\bibinfo {volume} {357}},\ \bibinfo {pages} {61--66} (\bibinfo {year} {2017})}\BibitemShut {NoStop}%
\bibitem [{\citenamefont {Benalcazar}\ \emph {et~al.}(2019)\citenamefont {Benalcazar}, \citenamefont {Li},\ and\ \citenamefont {Hughes}}]{BenalacazarCn}%
  \BibitemOpen
  \bibfield  {author} {\bibinfo {author} {\bibnamefont {Benalcazar}, \bibfnamefont {W.~A.}}, \bibinfo {author} {\bibnamefont {Li}, \bibfnamefont {T.}}, \ and\ \bibinfo {author} {\bibnamefont {Hughes}, \bibfnamefont {T.~L.}},\ }\bibfield  {title} {\enquote {\bibinfo {title} {Quantization of fractional corner charge in ${C}_{n}$-symmetric higher-order topological crystalline insulators},}\ }\href {\doibase 10.1103/PhysRevB.99.245151} {\bibfield  {journal} {\bibinfo  {journal} {Phys. Rev. B}\ }\textbf {\bibinfo {volume} {99}},\ \bibinfo {pages} {245151} (\bibinfo {year} {2019})}\BibitemShut {NoStop}%
\bibitem [{\citenamefont {Schindler}\ \emph {et~al.}(2018{\natexlab{a}})\citenamefont {Schindler}, \citenamefont {Wang}, \citenamefont {Vergniory}, \citenamefont {Cook}, \citenamefont {Murani}, \citenamefont {Sengupta}, \citenamefont {Kasumov}, \citenamefont {Deblock}, \citenamefont {Jeon}, \citenamefont {Drozdov} \emph {et~al.}}]{schindler2018higher}%
  \BibitemOpen
  \bibfield  {author} {\bibinfo {author} {\bibnamefont {Schindler}, \bibfnamefont {F.}}, \bibinfo {author} {\bibnamefont {Wang}, \bibfnamefont {Z.}}, \bibinfo {author} {\bibnamefont {Vergniory}, \bibfnamefont {M.~G.}}, \bibinfo {author} {\bibnamefont {Cook}, \bibfnamefont {A.~M.}}, \bibinfo {author} {\bibnamefont {Murani}, \bibfnamefont {A.}}, \bibinfo {author} {\bibnamefont {Sengupta}, \bibfnamefont {S.}}, \bibinfo {author} {\bibnamefont {Kasumov}, \bibfnamefont {A.~Y.}}, \bibinfo {author} {\bibnamefont {Deblock}, \bibfnamefont {R.}}, \bibinfo {author} {\bibnamefont {Jeon}, \bibfnamefont {S.}}, \bibinfo {author} {\bibnamefont {Drozdov}, \bibfnamefont {I.}},  \emph {et~al.},\ }\bibfield  {title} {\enquote {\bibinfo {title} {Higher-order topology in bismuth},}\ }\href {\doibase 10.1038/s41567-020-0902-0} {\bibfield  {journal} {\bibinfo  {journal} {Nat. Phys.}\ }\textbf {\bibinfo {volume} {14}},\ \bibinfo {pages} {918--924} (\bibinfo {year} {2018}{\natexlab{a}})}\BibitemShut {NoStop}%
\bibitem [{\citenamefont {Schindler}\ \emph {et~al.}(2018{\natexlab{b}})\citenamefont {Schindler}, \citenamefont {Cook}, \citenamefont {Vergniory}, \citenamefont {Wang}, \citenamefont {Parkin}, \citenamefont {Bernevig},\ and\ \citenamefont {Neupert}}]{Schindlereaat0346}%
  \BibitemOpen
  \bibfield  {author} {\bibinfo {author} {\bibnamefont {Schindler}, \bibfnamefont {F.}}, \bibinfo {author} {\bibnamefont {Cook}, \bibfnamefont {A.~M.}}, \bibinfo {author} {\bibnamefont {Vergniory}, \bibfnamefont {M.~G.}}, \bibinfo {author} {\bibnamefont {Wang}, \bibfnamefont {Z.}}, \bibinfo {author} {\bibnamefont {Parkin}, \bibfnamefont {S.~S.~P.}}, \bibinfo {author} {\bibnamefont {Bernevig}, \bibfnamefont {B.~A.}}, \ and\ \bibinfo {author} {\bibnamefont {Neupert}, \bibfnamefont {T.}},\ }\bibfield  {title} {\enquote {\bibinfo {title} {Higher-order topological insulators},}\ }\href {\doibase 10.1126/sciadv.aat0346} {\bibfield  {journal} {\bibinfo  {journal} {Sci. Adv.}\ }\textbf {\bibinfo {volume} {4}} (\bibinfo {year} {2018}{\natexlab{b}}),\ 10.1126/sciadv.aat0346}\BibitemShut {NoStop}%
\bibitem [{\citenamefont {S{\o}dequist}\ \emph {et~al.}(2022)\citenamefont {S{\o}dequist}, \citenamefont {Petralanda},\ and\ \citenamefont {Olsen}}]{sodequist2022abundance}%
  \BibitemOpen
  \bibfield  {author} {\bibinfo {author} {\bibnamefont {S{\o}dequist}, \bibfnamefont {J.}}, \bibinfo {author} {\bibnamefont {Petralanda}, \bibfnamefont {U.}}, \ and\ \bibinfo {author} {\bibnamefont {Olsen}, \bibfnamefont {T.}},\ }\bibfield  {title} {\enquote {\bibinfo {title} {Abundance of second order topology in c3 symmetric two-dimensional insulators},}\ }\href {\doibase 10.1088/2053-1583/ac9fe2} {\bibfield  {journal} {\bibinfo  {journal} {2D Mat.}\ }\textbf {\bibinfo {volume} {10}},\ \bibinfo {pages} {015009} (\bibinfo {year} {2022})}\BibitemShut {NoStop}%
\bibitem [{\citenamefont {Qi}\ and\ \citenamefont {Zhang}(2008)}]{QiSpinCharge}%
  \BibitemOpen
  \bibfield  {author} {\bibinfo {author} {\bibnamefont {Qi}, \bibfnamefont {X.-L.}}\ and\ \bibinfo {author} {\bibnamefont {Zhang}, \bibfnamefont {S.-C.}},\ }\bibfield  {title} {\enquote {\bibinfo {title} {Spin-charge separation in the quantum spin hall state},}\ }\href {\doibase 10.1103/PhysRevLett.101.086802} {\bibfield  {journal} {\bibinfo  {journal} {Phys. Rev. Lett.}\ }\textbf {\bibinfo {volume} {101}},\ \bibinfo {pages} {086802} (\bibinfo {year} {2008})}\BibitemShut {NoStop}%
\bibitem [{\citenamefont {Ran}\ \emph {et~al.}(2008)\citenamefont {Ran}, \citenamefont {Vishwanath},\ and\ \citenamefont {Lee}}]{SpinChargeVishwanath}%
  \BibitemOpen
  \bibfield  {author} {\bibinfo {author} {\bibnamefont {Ran}, \bibfnamefont {Y.}}, \bibinfo {author} {\bibnamefont {Vishwanath}, \bibfnamefont {A.}}, \ and\ \bibinfo {author} {\bibnamefont {Lee}, \bibfnamefont {D.-H.}},\ }\bibfield  {title} {\enquote {\bibinfo {title} {Spin-charge separated solitons in a topological band insulator},}\ }\href {\doibase 10.1103/PhysRevLett.101.086801} {\bibfield  {journal} {\bibinfo  {journal} {Phys. Rev. Lett.}\ }\textbf {\bibinfo {volume} {101}},\ \bibinfo {pages} {086801} (\bibinfo {year} {2008})}\BibitemShut {NoStop}%
\bibitem [{\citenamefont {Noh}\ \emph {et~al.}(2019)\citenamefont {Noh}, \citenamefont {Kim}, \citenamefont {Stein}, \citenamefont {Sanchez-Lengeling}, \citenamefont {Gregoire}, \citenamefont {Aspuru-Guzik},\ and\ \citenamefont {Jung}}]{noh2019inverse}%
  \BibitemOpen
  \bibfield  {author} {\bibinfo {author} {\bibnamefont {Noh}, \bibfnamefont {J.}}, \bibinfo {author} {\bibnamefont {Kim}, \bibfnamefont {J.}}, \bibinfo {author} {\bibnamefont {Stein}, \bibfnamefont {H.~S.}}, \bibinfo {author} {\bibnamefont {Sanchez-Lengeling}, \bibfnamefont {B.}}, \bibinfo {author} {\bibnamefont {Gregoire}, \bibfnamefont {J.~M.}}, \bibinfo {author} {\bibnamefont {Aspuru-Guzik}, \bibfnamefont {A.}}, \ and\ \bibinfo {author} {\bibnamefont {Jung}, \bibfnamefont {Y.}},\ }\bibfield  {title} {\enquote {\bibinfo {title} {Inverse design of solid-state materials via a continuous representation},}\ }\href {\doibase https://doi.org/10.1016/j.matt.2019.08.017} {\bibfield  {journal} {\bibinfo  {journal} {Matter}\ }\textbf {\bibinfo {volume} {1}},\ \bibinfo {pages} {1370--1384} (\bibinfo {year} {2019})}\BibitemShut {NoStop}%
\bibitem [{\citenamefont {Long}\ \emph {et~al.}(2021)\citenamefont {Long}, \citenamefont {Fortunato}, \citenamefont {Opahle}, \citenamefont {Zhang}, \citenamefont {Samathrakis}, \citenamefont {Shen}, \citenamefont {Gutfleisch},\ and\ \citenamefont {Zhang}}]{long2021constrained}%
  \BibitemOpen
  \bibfield  {author} {\bibinfo {author} {\bibnamefont {Long}, \bibfnamefont {T.}}, \bibinfo {author} {\bibnamefont {Fortunato}, \bibfnamefont {N.~M.}}, \bibinfo {author} {\bibnamefont {Opahle}, \bibfnamefont {I.}}, \bibinfo {author} {\bibnamefont {Zhang}, \bibfnamefont {Y.}}, \bibinfo {author} {\bibnamefont {Samathrakis}, \bibfnamefont {I.}}, \bibinfo {author} {\bibnamefont {Shen}, \bibfnamefont {C.}}, \bibinfo {author} {\bibnamefont {Gutfleisch}, \bibfnamefont {O.}}, \ and\ \bibinfo {author} {\bibnamefont {Zhang}, \bibfnamefont {H.}},\ }\bibfield  {title} {\enquote {\bibinfo {title} {Constrained crystals deep convolutional generative adversarial network for the inverse design of crystal structures},}\ }\href {\doibase https://doi.org/10.1038/s41524-021-00526-4} {\bibfield  {journal} {\bibinfo  {journal} {npj Computational Materials}\ }\textbf {\bibinfo {volume} {7}},\ \bibinfo {pages} {66} (\bibinfo {year} {2021})}\BibitemShut {NoStop}%
\bibitem [{\citenamefont {Rasmussen}\ and\ \citenamefont {Thygesen}(2015)}]{rasmussen2015computational}%
  \BibitemOpen
  \bibfield  {author} {\bibinfo {author} {\bibnamefont {Rasmussen}, \bibfnamefont {F.~A.}}\ and\ \bibinfo {author} {\bibnamefont {Thygesen}, \bibfnamefont {K.~S.}},\ }\bibfield  {title} {\enquote {\bibinfo {title} {Computational 2d materials database: electronic structure of transition-metal dichalcogenides and oxides},}\ }\href {\doibase 10.1021/acs.jpcc.5b02950} {\bibfield  {journal} {\bibinfo  {journal} {J. Phys. Chem. C}\ }\textbf {\bibinfo {volume} {119}},\ \bibinfo {pages} {13169--13183} (\bibinfo {year} {2015})}\BibitemShut {NoStop}%
\bibitem [{\citenamefont {Haastrup}\ \emph {et~al.}(2018)\citenamefont {Haastrup}, \citenamefont {Strange}, \citenamefont {Pandey}, \citenamefont {Deilmann}, \citenamefont {Schmidt}, \citenamefont {Hinsche}, \citenamefont {Gjerding}, \citenamefont {Torelli}, \citenamefont {Larsen}, \citenamefont {Riis-Jensen} \emph {et~al.}}]{haastrup2018computational}%
  \BibitemOpen
  \bibfield  {author} {\bibinfo {author} {\bibnamefont {Haastrup}, \bibfnamefont {S.}}, \bibinfo {author} {\bibnamefont {Strange}, \bibfnamefont {M.}}, \bibinfo {author} {\bibnamefont {Pandey}, \bibfnamefont {M.}}, \bibinfo {author} {\bibnamefont {Deilmann}, \bibfnamefont {T.}}, \bibinfo {author} {\bibnamefont {Schmidt}, \bibfnamefont {P.~S.}}, \bibinfo {author} {\bibnamefont {Hinsche}, \bibfnamefont {N.~F.}}, \bibinfo {author} {\bibnamefont {Gjerding}, \bibfnamefont {M.~N.}}, \bibinfo {author} {\bibnamefont {Torelli}, \bibfnamefont {D.}}, \bibinfo {author} {\bibnamefont {Larsen}, \bibfnamefont {P.~M.}}, \bibinfo {author} {\bibnamefont {Riis-Jensen}, \bibfnamefont {A.~C.}},  \emph {et~al.},\ }\bibfield  {title} {\enquote {\bibinfo {title} {The computational 2d materials database: high-throughput modeling and discovery of atomically thin crystals},}\ }\href {\doibase 10.1088/2053-1583/aacfc1} {\bibfield  {journal} {\bibinfo  {journal} {2D Mat.}\ }\textbf {\bibinfo {volume} {5}},\ \bibinfo {pages} {042002}
  (\bibinfo {year} {2018})}\BibitemShut {NoStop}%
\bibitem [{\citenamefont {Mak}\ \emph {et~al.}(2014)\citenamefont {Mak}, \citenamefont {McGill}, \citenamefont {Park},\ and\ \citenamefont {McEuen}}]{mak2014valley}%
  \BibitemOpen
  \bibfield  {author} {\bibinfo {author} {\bibnamefont {Mak}, \bibfnamefont {K.~F.}}, \bibinfo {author} {\bibnamefont {McGill}, \bibfnamefont {K.~L.}}, \bibinfo {author} {\bibnamefont {Park}, \bibfnamefont {J.}}, \ and\ \bibinfo {author} {\bibnamefont {McEuen}, \bibfnamefont {P.~L.}},\ }\bibfield  {title} {\enquote {\bibinfo {title} {The valley hall effect in mos2 transistors},}\ }\href {\doibase 10.1126/science.1250140} {\bibfield  {journal} {\bibinfo  {journal} {Science}\ }\textbf {\bibinfo {volume} {344}},\ \bibinfo {pages} {1489--1492} (\bibinfo {year} {2014})}\BibitemShut {NoStop}%
\bibitem [{\citenamefont {Tao}\ \emph {et~al.}(2024)\citenamefont {Tao}, \citenamefont {Shen}, \citenamefont {Zhao}, \citenamefont {Hu}, \citenamefont {Li}, \citenamefont {Jiang}, \citenamefont {Li}, \citenamefont {Watanabe}, \citenamefont {Taniguchi}, \citenamefont {MacDonald} \emph {et~al.}}]{tao2024giant}%
  \BibitemOpen
  \bibfield  {author} {\bibinfo {author} {\bibnamefont {Tao}, \bibfnamefont {Z.}}, \bibinfo {author} {\bibnamefont {Shen}, \bibfnamefont {B.}}, \bibinfo {author} {\bibnamefont {Zhao}, \bibfnamefont {W.}}, \bibinfo {author} {\bibnamefont {Hu}, \bibfnamefont {N.~C.}}, \bibinfo {author} {\bibnamefont {Li}, \bibfnamefont {T.}}, \bibinfo {author} {\bibnamefont {Jiang}, \bibfnamefont {S.}}, \bibinfo {author} {\bibnamefont {Li}, \bibfnamefont {L.}}, \bibinfo {author} {\bibnamefont {Watanabe}, \bibfnamefont {K.}}, \bibinfo {author} {\bibnamefont {Taniguchi}, \bibfnamefont {T.}}, \bibinfo {author} {\bibnamefont {MacDonald}, \bibfnamefont {A.~H.}},  \emph {et~al.},\ }\bibfield  {title} {\enquote {\bibinfo {title} {Giant spin hall effect in ab-stacked mote2/wse2 bilayers},}\ }\href {\doibase https://doi.org/10.1038/s41565-023-01492-2} {\bibfield  {journal} {\bibinfo  {journal} {Nat. Nanotechnol.}\ }\textbf {\bibinfo {volume} {19}},\ \bibinfo {pages} {28--33} (\bibinfo {year} {2024})}\BibitemShut {NoStop}%
\bibitem [{\citenamefont {Mounet}\ \emph {et~al.}(2018)\citenamefont {Mounet}, \citenamefont {Gibertini}, \citenamefont {Schwaller}, \citenamefont {Campi}, \citenamefont {Merkys}, \citenamefont {Marrazzo}, \citenamefont {Sohier}, \citenamefont {Castelli}, \citenamefont {Cepellotti}, \citenamefont {Pizzi} \emph {et~al.}}]{mounet2018two}%
  \BibitemOpen
  \bibfield  {author} {\bibinfo {author} {\bibnamefont {Mounet}, \bibfnamefont {N.}}, \bibinfo {author} {\bibnamefont {Gibertini}, \bibfnamefont {M.}}, \bibinfo {author} {\bibnamefont {Schwaller}, \bibfnamefont {P.}}, \bibinfo {author} {\bibnamefont {Campi}, \bibfnamefont {D.}}, \bibinfo {author} {\bibnamefont {Merkys}, \bibfnamefont {A.}}, \bibinfo {author} {\bibnamefont {Marrazzo}, \bibfnamefont {A.}}, \bibinfo {author} {\bibnamefont {Sohier}, \bibfnamefont {T.}}, \bibinfo {author} {\bibnamefont {Castelli}, \bibfnamefont {I.~E.}}, \bibinfo {author} {\bibnamefont {Cepellotti}, \bibfnamefont {A.}}, \bibinfo {author} {\bibnamefont {Pizzi}, \bibfnamefont {G.}},  \emph {et~al.},\ }\bibfield  {title} {\enquote {\bibinfo {title} {Two-dimensional materials from high-throughput computational exfoliation of experimentally known compounds},}\ }\href {\doibase 10.1038/s41565-017-0035-5} {\bibfield  {journal} {\bibinfo  {journal} {Nat. nanotechnol.}\ }\textbf {\bibinfo {volume} {13}},\ \bibinfo {pages} {246--252}
  (\bibinfo {year} {2018})}\BibitemShut {NoStop}%
\bibitem [{\citenamefont {Giannozzi}\ \emph {et~al.}(2009)\citenamefont {Giannozzi}, \citenamefont {Baroni}, \citenamefont {Bonini}, \citenamefont {Calandra}, \citenamefont {Car}, \citenamefont {Cavazzoni}, \citenamefont {Ceresoli}, \citenamefont {Chiarotti}, \citenamefont {Cococcioni}, \citenamefont {Dabo}, \citenamefont {{Dal Corso}}, \citenamefont {de~Gironcoli}, \citenamefont {Fabris}, \citenamefont {Fratesi}, \citenamefont {Gebauer}, \citenamefont {Gerstmann}, \citenamefont {Gougoussis}, \citenamefont {Kokalj}, \citenamefont {Lazzeri}, \citenamefont {Martin-Samos}, \citenamefont {Marzari}, \citenamefont {Mauri}, \citenamefont {Mazzarello}, \citenamefont {Paolini}, \citenamefont {Pasquarello}, \citenamefont {Paulatto}, \citenamefont {Sbraccia}, \citenamefont {Scandolo}, \citenamefont {Sclauzero}, \citenamefont {Seitsonen}, \citenamefont {Smogunov}, \citenamefont {Umari},\ and\ \citenamefont {Wentzcovitch}}]{QE-2009}%
  \BibitemOpen
  \bibfield  {author} {\bibinfo {author} {\bibnamefont {Giannozzi}, \bibfnamefont {P.}}, \bibinfo {author} {\bibnamefont {Baroni}, \bibfnamefont {S.}}, \bibinfo {author} {\bibnamefont {Bonini}, \bibfnamefont {N.}}, \bibinfo {author} {\bibnamefont {Calandra}, \bibfnamefont {M.}}, \bibinfo {author} {\bibnamefont {Car}, \bibfnamefont {R.}}, \bibinfo {author} {\bibnamefont {Cavazzoni}, \bibfnamefont {C.}}, \bibinfo {author} {\bibnamefont {Ceresoli}, \bibfnamefont {D.}}, \bibinfo {author} {\bibnamefont {Chiarotti}, \bibfnamefont {G.~L.}}, \bibinfo {author} {\bibnamefont {Cococcioni}, \bibfnamefont {M.}}, \bibinfo {author} {\bibnamefont {Dabo}, \bibfnamefont {I.}}, \bibinfo {author} {\bibnamefont {{Dal Corso}}, \bibfnamefont {A.}}, \bibinfo {author} {\bibnamefont {de~Gironcoli}, \bibfnamefont {S.}}, \bibinfo {author} {\bibnamefont {Fabris}, \bibfnamefont {S.}}, \bibinfo {author} {\bibnamefont {Fratesi}, \bibfnamefont {G.}}, \bibinfo {author} {\bibnamefont {Gebauer}, \bibfnamefont {R.}}, \bibinfo {author}
  {\bibnamefont {Gerstmann}, \bibfnamefont {U.}}, \bibinfo {author} {\bibnamefont {Gougoussis}, \bibfnamefont {C.}}, \bibinfo {author} {\bibnamefont {Kokalj}, \bibfnamefont {A.}}, \bibinfo {author} {\bibnamefont {Lazzeri}, \bibfnamefont {M.}}, \bibinfo {author} {\bibnamefont {Martin-Samos}, \bibfnamefont {L.}}, \bibinfo {author} {\bibnamefont {Marzari}, \bibfnamefont {N.}}, \bibinfo {author} {\bibnamefont {Mauri}, \bibfnamefont {F.}}, \bibinfo {author} {\bibnamefont {Mazzarello}, \bibfnamefont {R.}}, \bibinfo {author} {\bibnamefont {Paolini}, \bibfnamefont {S.}}, \bibinfo {author} {\bibnamefont {Pasquarello}, \bibfnamefont {A.}}, \bibinfo {author} {\bibnamefont {Paulatto}, \bibfnamefont {L.}}, \bibinfo {author} {\bibnamefont {Sbraccia}, \bibfnamefont {C.}}, \bibinfo {author} {\bibnamefont {Scandolo}, \bibfnamefont {S.}}, \bibinfo {author} {\bibnamefont {Sclauzero}, \bibfnamefont {G.}}, \bibinfo {author} {\bibnamefont {Seitsonen}, \bibfnamefont {A.~P.}}, \bibinfo {author} {\bibnamefont {Smogunov},
  \bibfnamefont {A.}}, \bibinfo {author} {\bibnamefont {Umari}, \bibfnamefont {P.}}, \ and\ \bibinfo {author} {\bibnamefont {Wentzcovitch}, \bibfnamefont {R.~M.}},\ }\bibfield  {title} {\enquote {\bibinfo {title} {Quantum espresso: a modular and open-source software project for quantum simulations of materials},}\ }\href {http://www.quantum-espresso.org} {\bibfield  {journal} {\bibinfo  {journal} {J. Phys. Condens. Matter}\ }\textbf {\bibinfo {volume} {21}},\ \bibinfo {pages} {395502 (19pp)} (\bibinfo {year} {2009})}\BibitemShut {NoStop}%
\bibitem [{\citenamefont {Giannozzi}\ \emph {et~al.}(2017)\citenamefont {Giannozzi}, \citenamefont {Andreussi}, \citenamefont {Brumme}, \citenamefont {Bunau}, \citenamefont {Nardelli}, \citenamefont {Calandra}, \citenamefont {Car}, \citenamefont {Cavazzoni}, \citenamefont {Ceresoli}, \citenamefont {Cococcioni}, \citenamefont {Colonna}, \citenamefont {Carnimeo}, \citenamefont {Corso}, \citenamefont {de~Gironcoli}, \citenamefont {Delugas}, \citenamefont {Jr}, \citenamefont {Ferretti}, \citenamefont {Floris}, \citenamefont {Fratesi}, \citenamefont {Fugallo}, \citenamefont {Gebauer}, \citenamefont {Gerstmann}, \citenamefont {Giustino}, \citenamefont {Gorni}, \citenamefont {Jia}, \citenamefont {Kawamura}, \citenamefont {Ko}, \citenamefont {Kokalj}, \citenamefont {Küçükbenli}, \citenamefont {Lazzeri}, \citenamefont {Marsili}, \citenamefont {Marzari}, \citenamefont {Mauri}, \citenamefont {Nguyen}, \citenamefont {Nguyen}, \citenamefont {de-la Roza}, \citenamefont {Paulatto}, \citenamefont {Poncé}, \citenamefont
  {Rocca}, \citenamefont {Sabatini}, \citenamefont {Santra}, \citenamefont {Schlipf}, \citenamefont {Seitsonen}, \citenamefont {Smogunov}, \citenamefont {Timrov}, \citenamefont {Thonhauser}, \citenamefont {Umari}, \citenamefont {Vast}, \citenamefont {Wu},\ and\ \citenamefont {Baroni}}]{QE-2017}%
  \BibitemOpen
  \bibfield  {author} {\bibinfo {author} {\bibnamefont {Giannozzi}, \bibfnamefont {P.}}, \bibinfo {author} {\bibnamefont {Andreussi}, \bibfnamefont {O.}}, \bibinfo {author} {\bibnamefont {Brumme}, \bibfnamefont {T.}}, \bibinfo {author} {\bibnamefont {Bunau}, \bibfnamefont {O.}}, \bibinfo {author} {\bibnamefont {Nardelli}, \bibfnamefont {M.~B.}}, \bibinfo {author} {\bibnamefont {Calandra}, \bibfnamefont {M.}}, \bibinfo {author} {\bibnamefont {Car}, \bibfnamefont {R.}}, \bibinfo {author} {\bibnamefont {Cavazzoni}, \bibfnamefont {C.}}, \bibinfo {author} {\bibnamefont {Ceresoli}, \bibfnamefont {D.}}, \bibinfo {author} {\bibnamefont {Cococcioni}, \bibfnamefont {M.}}, \bibinfo {author} {\bibnamefont {Colonna}, \bibfnamefont {N.}}, \bibinfo {author} {\bibnamefont {Carnimeo}, \bibfnamefont {I.}}, \bibinfo {author} {\bibnamefont {Corso}, \bibfnamefont {A.~D.}}, \bibinfo {author} {\bibnamefont {de~Gironcoli}, \bibfnamefont {S.}}, \bibinfo {author} {\bibnamefont {Delugas}, \bibfnamefont {P.}}, \bibinfo {author}
  {\bibnamefont {Jr}, \bibfnamefont {R.~A.~D.}}, \bibinfo {author} {\bibnamefont {Ferretti}, \bibfnamefont {A.}}, \bibinfo {author} {\bibnamefont {Floris}, \bibfnamefont {A.}}, \bibinfo {author} {\bibnamefont {Fratesi}, \bibfnamefont {G.}}, \bibinfo {author} {\bibnamefont {Fugallo}, \bibfnamefont {G.}}, \bibinfo {author} {\bibnamefont {Gebauer}, \bibfnamefont {R.}}, \bibinfo {author} {\bibnamefont {Gerstmann}, \bibfnamefont {U.}}, \bibinfo {author} {\bibnamefont {Giustino}, \bibfnamefont {F.}}, \bibinfo {author} {\bibnamefont {Gorni}, \bibfnamefont {T.}}, \bibinfo {author} {\bibnamefont {Jia}, \bibfnamefont {J.}}, \bibinfo {author} {\bibnamefont {Kawamura}, \bibfnamefont {M.}}, \bibinfo {author} {\bibnamefont {Ko}, \bibfnamefont {H.-Y.}}, \bibinfo {author} {\bibnamefont {Kokalj}, \bibfnamefont {A.}}, \bibinfo {author} {\bibnamefont {Küçükbenli}, \bibfnamefont {E.}}, \bibinfo {author} {\bibnamefont {Lazzeri}, \bibfnamefont {M.}}, \bibinfo {author} {\bibnamefont {Marsili}, \bibfnamefont {M.}}, \bibinfo
  {author} {\bibnamefont {Marzari}, \bibfnamefont {N.}}, \bibinfo {author} {\bibnamefont {Mauri}, \bibfnamefont {F.}}, \bibinfo {author} {\bibnamefont {Nguyen}, \bibfnamefont {N.~L.}}, \bibinfo {author} {\bibnamefont {Nguyen}, \bibfnamefont {H.-V.}}, \bibinfo {author} {\bibnamefont {de-la Roza}, \bibfnamefont {A.~O.}}, \bibinfo {author} {\bibnamefont {Paulatto}, \bibfnamefont {L.}}, \bibinfo {author} {\bibnamefont {Poncé}, \bibfnamefont {S.}}, \bibinfo {author} {\bibnamefont {Rocca}, \bibfnamefont {D.}}, \bibinfo {author} {\bibnamefont {Sabatini}, \bibfnamefont {R.}}, \bibinfo {author} {\bibnamefont {Santra}, \bibfnamefont {B.}}, \bibinfo {author} {\bibnamefont {Schlipf}, \bibfnamefont {M.}}, \bibinfo {author} {\bibnamefont {Seitsonen}, \bibfnamefont {A.~P.}}, \bibinfo {author} {\bibnamefont {Smogunov}, \bibfnamefont {A.}}, \bibinfo {author} {\bibnamefont {Timrov}, \bibfnamefont {I.}}, \bibinfo {author} {\bibnamefont {Thonhauser}, \bibfnamefont {T.}}, \bibinfo {author} {\bibnamefont {Umari}, \bibfnamefont
  {P.}}, \bibinfo {author} {\bibnamefont {Vast}, \bibfnamefont {N.}}, \bibinfo {author} {\bibnamefont {Wu}, \bibfnamefont {X.}}, \ and\ \bibinfo {author} {\bibnamefont {Baroni}, \bibfnamefont {S.}},\ }\bibfield  {title} {\enquote {\bibinfo {title} {Advanced capabilities for materials modelling with quantum espresso},}\ }\href {http://stacks.iop.org/0953-8984/29/i=46/a=465901} {\bibfield  {journal} {\bibinfo  {journal} {J. Phys. Condens. Matter}\ }\textbf {\bibinfo {volume} {29}},\ \bibinfo {pages} {465901} (\bibinfo {year} {2017})}\BibitemShut {NoStop}%
\bibitem [{\citenamefont {Giannozzi}\ \emph {et~al.}(2020)\citenamefont {Giannozzi}, \citenamefont {Baseggio}, \citenamefont {Bonfà}, \citenamefont {Brunato}, \citenamefont {Car}, \citenamefont {Carnimeo}, \citenamefont {Cavazzoni}, \citenamefont {de~Gironcoli}, \citenamefont {Delugas}, \citenamefont {Ferrari~Ruffino}, \citenamefont {Ferretti}, \citenamefont {Marzari}, \citenamefont {Timrov}, \citenamefont {Urru},\ and\ \citenamefont {Baroni}}]{QE-2020}%
  \BibitemOpen
  \bibfield  {author} {\bibinfo {author} {\bibnamefont {Giannozzi}, \bibfnamefont {P.}}, \bibinfo {author} {\bibnamefont {Baseggio}, \bibfnamefont {O.}}, \bibinfo {author} {\bibnamefont {Bonfà}, \bibfnamefont {P.}}, \bibinfo {author} {\bibnamefont {Brunato}, \bibfnamefont {D.}}, \bibinfo {author} {\bibnamefont {Car}, \bibfnamefont {R.}}, \bibinfo {author} {\bibnamefont {Carnimeo}, \bibfnamefont {I.}}, \bibinfo {author} {\bibnamefont {Cavazzoni}, \bibfnamefont {C.}}, \bibinfo {author} {\bibnamefont {de~Gironcoli}, \bibfnamefont {S.}}, \bibinfo {author} {\bibnamefont {Delugas}, \bibfnamefont {P.}}, \bibinfo {author} {\bibnamefont {Ferrari~Ruffino}, \bibfnamefont {F.}}, \bibinfo {author} {\bibnamefont {Ferretti}, \bibfnamefont {A.}}, \bibinfo {author} {\bibnamefont {Marzari}, \bibfnamefont {N.}}, \bibinfo {author} {\bibnamefont {Timrov}, \bibfnamefont {I.}}, \bibinfo {author} {\bibnamefont {Urru}, \bibfnamefont {A.}}, \ and\ \bibinfo {author} {\bibnamefont {Baroni}, \bibfnamefont {S.}},\ }\bibfield  {title}
  {\enquote {\bibinfo {title} {Quantum espresso toward the exascale},}\ }\href {\doibase 10.1063/5.0005082} {\bibfield  {journal} {\bibinfo  {journal} {J. Chem. Phys.}\ }\textbf {\bibinfo {volume} {152}},\ \bibinfo {pages} {154105} (\bibinfo {year} {2020})}\BibitemShut {NoStop}%
\bibitem [{\citenamefont {Perdew}\ \emph {et~al.}(1997)\citenamefont {Perdew}, \citenamefont {Burke},\ and\ \citenamefont {Ernzerhof}}]{Perdew1996}%
  \BibitemOpen
  \bibfield  {author} {\bibinfo {author} {\bibnamefont {Perdew}, \bibfnamefont {J.~P.}}, \bibinfo {author} {\bibnamefont {Burke}, \bibfnamefont {K.}}, \ and\ \bibinfo {author} {\bibnamefont {Ernzerhof}, \bibfnamefont {M.}},\ }\bibfield  {title} {\enquote {\bibinfo {title} {Generalized gradient approximation made simple},}\ }\href {\doibase 10.1103/PhysRevLett.78.1396} {\bibfield  {journal} {\bibinfo  {journal} {Phys. Rev. Lett.}\ }\textbf {\bibinfo {volume} {78}},\ \bibinfo {pages} {1396--1396} (\bibinfo {year} {1997})}\BibitemShut {NoStop}%
\bibitem [{\citenamefont {Hamann}(2013)}]{Hamann2013}%
  \BibitemOpen
  \bibfield  {author} {\bibinfo {author} {\bibnamefont {Hamann}, \bibfnamefont {D.~R.}},\ }\bibfield  {title} {\enquote {\bibinfo {title} {Optimized norm-conserving vanderbilt pseudopotentials},}\ }\href {\doibase 10.1103/PhysRevB.88.085117} {\bibfield  {journal} {\bibinfo  {journal} {Phys. Rev. B}\ }\textbf {\bibinfo {volume} {88}},\ \bibinfo {pages} {085117} (\bibinfo {year} {2013})}\BibitemShut {NoStop}%
\bibitem [{\citenamefont {van Setten}\ \emph {et~al.}(2018)\citenamefont {van Setten}, \citenamefont {Giantomassi}, \citenamefont {Bousquet}, \citenamefont {Verstraete}, \citenamefont {Hamann}, \citenamefont {Gonze},\ and\ \citenamefont {Rignanese}}]{van2018pseudodojo}%
  \BibitemOpen
  \bibfield  {author} {\bibinfo {author} {\bibnamefont {van Setten}, \bibfnamefont {M.~J.}}, \bibinfo {author} {\bibnamefont {Giantomassi}, \bibfnamefont {M.}}, \bibinfo {author} {\bibnamefont {Bousquet}, \bibfnamefont {E.}}, \bibinfo {author} {\bibnamefont {Verstraete}, \bibfnamefont {M.~J.}}, \bibinfo {author} {\bibnamefont {Hamann}, \bibfnamefont {D.~R.}}, \bibinfo {author} {\bibnamefont {Gonze}, \bibfnamefont {X.}}, \ and\ \bibinfo {author} {\bibnamefont {Rignanese}, \bibfnamefont {G.-M.}},\ }\bibfield  {title} {\enquote {\bibinfo {title} {The pseudodojo: Training and grading a 85 element optimized norm-conserving pseudopotential table},}\ }\href {\doibase https://doi.org/10.1016/j.cpc.2018.01.012} {\bibfield  {journal} {\bibinfo  {journal} {Comput. Phys. Commun.}\ }\textbf {\bibinfo {volume} {226}},\ \bibinfo {pages} {39--54} (\bibinfo {year} {2018})}\BibitemShut {NoStop}%
\bibitem [{\citenamefont {Tyner}(2024)}]{tyner2023berryeasy}%
  \BibitemOpen
  \bibfield  {author} {\bibinfo {author} {\bibnamefont {Tyner}, \bibfnamefont {A.~C.}},\ }\bibfield  {title} {\enquote {\bibinfo {title} {Berryeasy: A gpu enabled python package for diagnosis of $ n $-th-order and spin-resolved topology in the presence of fields and effects},}\ }\href {\doibase 10.1088/1361-648X/ad475f} {\bibfield  {journal} {\bibinfo  {journal} {J. Condens. Matter Phys.}\ }\textbf {\bibinfo {volume} {36}},\ \bibinfo {pages} {325902} (\bibinfo {year} {2024})}\BibitemShut {NoStop}%
\bibitem [{\citenamefont {Vitale}\ \emph {et~al.}(2020)\citenamefont {Vitale}, \citenamefont {Pizzi}, \citenamefont {Marrazzo}, \citenamefont {Yates}, \citenamefont {Marzari},\ and\ \citenamefont {Mostofi}}]{vitale2020automated}%
  \BibitemOpen
  \bibfield  {author} {\bibinfo {author} {\bibnamefont {Vitale}, \bibfnamefont {V.}}, \bibinfo {author} {\bibnamefont {Pizzi}, \bibfnamefont {G.}}, \bibinfo {author} {\bibnamefont {Marrazzo}, \bibfnamefont {A.}}, \bibinfo {author} {\bibnamefont {Yates}, \bibfnamefont {J.~R.}}, \bibinfo {author} {\bibnamefont {Marzari}, \bibfnamefont {N.}}, \ and\ \bibinfo {author} {\bibnamefont {Mostofi}, \bibfnamefont {A.~A.}},\ }\bibfield  {title} {\enquote {\bibinfo {title} {Automated high-throughput wannierisation},}\ }\href {\doibase 10.1038/s41524-020-0312-y} {\bibfield  {journal} {\bibinfo  {journal} {npj Comput. Mater.}\ }\textbf {\bibinfo {volume} {6}},\ \bibinfo {pages} {1--18} (\bibinfo {year} {2020})}\BibitemShut {NoStop}%
\bibitem [{git()}]{github}%
  \BibitemOpen
  \href {\doibase https://github.com/actyner/MLSpinResolvedTopology} {\ https://github.com/actyner/MLSpinResolvedTopology}\BibitemShut {NoStop}%
\bibitem [{\citenamefont {Wang}\ \emph {et~al.}(2019)\citenamefont {Wang}, \citenamefont {Tang}, \citenamefont {Ji}, \citenamefont {Zhang}, \citenamefont {Vishwanath}, \citenamefont {Po},\ and\ \citenamefont {Wan}}]{2dsymmind}%
  \BibitemOpen
  \bibfield  {author} {\bibinfo {author} {\bibnamefont {Wang}, \bibfnamefont {D.}}, \bibinfo {author} {\bibnamefont {Tang}, \bibfnamefont {F.}}, \bibinfo {author} {\bibnamefont {Ji}, \bibfnamefont {J.}}, \bibinfo {author} {\bibnamefont {Zhang}, \bibfnamefont {W.}}, \bibinfo {author} {\bibnamefont {Vishwanath}, \bibfnamefont {A.}}, \bibinfo {author} {\bibnamefont {Po}, \bibfnamefont {H.~C.}}, \ and\ \bibinfo {author} {\bibnamefont {Wan}, \bibfnamefont {X.}},\ }\bibfield  {title} {\enquote {\bibinfo {title} {Two-dimensional topological materials discovery by symmetry-indicator method},}\ }\href {\doibase 10.1103/PhysRevB.100.195108} {\bibfield  {journal} {\bibinfo  {journal} {Phys. Rev. B}\ }\textbf {\bibinfo {volume} {100}},\ \bibinfo {pages} {195108} (\bibinfo {year} {2019})}\BibitemShut {NoStop}%
\bibitem [{\citenamefont {Olsen}\ \emph {et~al.}(2019)\citenamefont {Olsen}, \citenamefont {Andersen}, \citenamefont {Okugawa}, \citenamefont {Torelli}, \citenamefont {Deilmann},\ and\ \citenamefont {Thygesen}}]{C2DBTIs}%
  \BibitemOpen
  \bibfield  {author} {\bibinfo {author} {\bibnamefont {Olsen}, \bibfnamefont {T.}}, \bibinfo {author} {\bibnamefont {Andersen}, \bibfnamefont {E.}}, \bibinfo {author} {\bibnamefont {Okugawa}, \bibfnamefont {T.}}, \bibinfo {author} {\bibnamefont {Torelli}, \bibfnamefont {D.}}, \bibinfo {author} {\bibnamefont {Deilmann}, \bibfnamefont {T.}}, \ and\ \bibinfo {author} {\bibnamefont {Thygesen}, \bibfnamefont {K.~S.}},\ }\bibfield  {title} {\enquote {\bibinfo {title} {Discovering two-dimensional topological insulators from high-throughput computations},}\ }\href {\doibase 10.1103/PhysRevMaterials.3.024005} {\bibfield  {journal} {\bibinfo  {journal} {Phys. Rev. Mater.}\ }\textbf {\bibinfo {volume} {3}},\ \bibinfo {pages} {024005} (\bibinfo {year} {2019})}\BibitemShut {NoStop}%
\bibitem [{\citenamefont {Marrazzo}\ \emph {et~al.}(2019)\citenamefont {Marrazzo}, \citenamefont {Gibertini}, \citenamefont {Campi}, \citenamefont {Mounet},\ and\ \citenamefont {Marzari}}]{marrazzo2019relative}%
  \BibitemOpen
  \bibfield  {author} {\bibinfo {author} {\bibnamefont {Marrazzo}, \bibfnamefont {A.}}, \bibinfo {author} {\bibnamefont {Gibertini}, \bibfnamefont {M.}}, \bibinfo {author} {\bibnamefont {Campi}, \bibfnamefont {D.}}, \bibinfo {author} {\bibnamefont {Mounet}, \bibfnamefont {N.}}, \ and\ \bibinfo {author} {\bibnamefont {Marzari}, \bibfnamefont {N.}},\ }\bibfield  {title} {\enquote {\bibinfo {title} {Relative abundance of z 2 topological order in exfoliable two-dimensional insulators},}\ }\href {\doibase https://doi.org/10.1021/acs.nanolett.9b02689} {\bibfield  {journal} {\bibinfo  {journal} {Nano letters}\ }\textbf {\bibinfo {volume} {19}},\ \bibinfo {pages} {8431--8440} (\bibinfo {year} {2019})}\BibitemShut {NoStop}%
\bibitem [{\citenamefont {Xie}\ and\ \citenamefont {Grossman}(2018)}]{cgcnn}%
  \BibitemOpen
  \bibfield  {author} {\bibinfo {author} {\bibnamefont {Xie}, \bibfnamefont {T.}}\ and\ \bibinfo {author} {\bibnamefont {Grossman}, \bibfnamefont {J.~C.}},\ }\bibfield  {title} {\enquote {\bibinfo {title} {Crystal graph convolutional neural networks for an accurate and interpretable prediction of material properties},}\ }\href {\doibase 10.1103/PhysRevLett.120.145301} {\bibfield  {journal} {\bibinfo  {journal} {Phys. Rev. Lett.}\ }\textbf {\bibinfo {volume} {120}},\ \bibinfo {pages} {145301} (\bibinfo {year} {2018})}\BibitemShut {NoStop}%
\bibitem [{\citenamefont {Nouira}\ \emph {et~al.}(2018)\citenamefont {Nouira}, \citenamefont {Sokolovska},\ and\ \citenamefont {Crivello}}]{nouira2018crystalgan}%
  \BibitemOpen
  \bibfield  {author} {\bibinfo {author} {\bibnamefont {Nouira}, \bibfnamefont {A.}}, \bibinfo {author} {\bibnamefont {Sokolovska}, \bibfnamefont {N.}}, \ and\ \bibinfo {author} {\bibnamefont {Crivello}, \bibfnamefont {J.-C.}},\ }\bibfield  {title} {\enquote {\bibinfo {title} {Crystalgan: learning to discover crystallographic structures with generative adversarial networks},}\ }\href {\doibase https://doi.org/10.48550/arXiv.1810.11203} {\bibfield  {journal} {\bibinfo  {journal} {arXiv:1810.11203}\ } (\bibinfo {year} {2018}),\ https://doi.org/10.48550/arXiv.1810.11203}\BibitemShut {NoStop}%
\bibitem [{\citenamefont {Hoffmann}\ \emph {et~al.}(2019)\citenamefont {Hoffmann}, \citenamefont {Maestrati}, \citenamefont {Sawada}, \citenamefont {Tang}, \citenamefont {Sellier},\ and\ \citenamefont {Bengio}}]{hoffmann2019data}%
  \BibitemOpen
  \bibfield  {author} {\bibinfo {author} {\bibnamefont {Hoffmann}, \bibfnamefont {J.}}, \bibinfo {author} {\bibnamefont {Maestrati}, \bibfnamefont {L.}}, \bibinfo {author} {\bibnamefont {Sawada}, \bibfnamefont {Y.}}, \bibinfo {author} {\bibnamefont {Tang}, \bibfnamefont {J.}}, \bibinfo {author} {\bibnamefont {Sellier}, \bibfnamefont {J.~M.}}, \ and\ \bibinfo {author} {\bibnamefont {Bengio}, \bibfnamefont {Y.}},\ }\bibfield  {title} {\enquote {\bibinfo {title} {Data-driven approach to encoding and decoding 3-d crystal structures},}\ }\href {\doibase https://doi.org/10.48550/arXiv.1909.00949} {\bibfield  {journal} {\bibinfo  {journal} {arXiv:1909.00949}\ } (\bibinfo {year} {2019}),\ https://doi.org/10.48550/arXiv.1909.00949}\BibitemShut {NoStop}%
\bibitem [{\citenamefont {De}\ \emph {et~al.}(2016)\citenamefont {De}, \citenamefont {Bart{\'o}k}, \citenamefont {Cs{\'a}nyi},\ and\ \citenamefont {Ceriotti}}]{de2016comparing}%
  \BibitemOpen
  \bibfield  {author} {\bibinfo {author} {\bibnamefont {De}, \bibfnamefont {S.}}, \bibinfo {author} {\bibnamefont {Bart{\'o}k}, \bibfnamefont {A.~P.}}, \bibinfo {author} {\bibnamefont {Cs{\'a}nyi}, \bibfnamefont {G.}}, \ and\ \bibinfo {author} {\bibnamefont {Ceriotti}, \bibfnamefont {M.}},\ }\bibfield  {title} {\enquote {\bibinfo {title} {Comparing molecules and solids across structural and alchemical space},}\ }\href {\doibase 10.1039/C6CP00415F} {\bibfield  {journal} {\bibinfo  {journal} {Physical Chemistry Chemical Physics}\ }\textbf {\bibinfo {volume} {18}},\ \bibinfo {pages} {13754--13769} (\bibinfo {year} {2016})}\BibitemShut {NoStop}%
\bibitem [{\citenamefont {Kaufmann}\ \emph {et~al.}(2020)\citenamefont {Kaufmann}, \citenamefont {Zhu}, \citenamefont {Rosengarten}, \citenamefont {Maryanovsky}, \citenamefont {Harrington}, \citenamefont {Marin},\ and\ \citenamefont {Vecchio}}]{kaufmann2020crystal}%
  \BibitemOpen
  \bibfield  {author} {\bibinfo {author} {\bibnamefont {Kaufmann}, \bibfnamefont {K.}}, \bibinfo {author} {\bibnamefont {Zhu}, \bibfnamefont {C.}}, \bibinfo {author} {\bibnamefont {Rosengarten}, \bibfnamefont {A.~S.}}, \bibinfo {author} {\bibnamefont {Maryanovsky}, \bibfnamefont {D.}}, \bibinfo {author} {\bibnamefont {Harrington}, \bibfnamefont {T.~J.}}, \bibinfo {author} {\bibnamefont {Marin}, \bibfnamefont {E.}}, \ and\ \bibinfo {author} {\bibnamefont {Vecchio}, \bibfnamefont {K.~S.}},\ }\bibfield  {title} {\enquote {\bibinfo {title} {Crystal symmetry determination in electron diffraction using machine learning},}\ }\href {\doibase 10.1126/science.aay3062} {\bibfield  {journal} {\bibinfo  {journal} {Science}\ }\textbf {\bibinfo {volume} {367}},\ \bibinfo {pages} {564--568} (\bibinfo {year} {2020})}\BibitemShut {NoStop}%
\bibitem [{\citenamefont {Kim}\ \emph {et~al.}(2020)\citenamefont {Kim}, \citenamefont {Noh}, \citenamefont {Gu}, \citenamefont {Aspuru-Guzik},\ and\ \citenamefont {Jung}}]{kim2020generative}%
  \BibitemOpen
  \bibfield  {author} {\bibinfo {author} {\bibnamefont {Kim}, \bibfnamefont {S.}}, \bibinfo {author} {\bibnamefont {Noh}, \bibfnamefont {J.}}, \bibinfo {author} {\bibnamefont {Gu}, \bibfnamefont {G.~H.}}, \bibinfo {author} {\bibnamefont {Aspuru-Guzik}, \bibfnamefont {A.}}, \ and\ \bibinfo {author} {\bibnamefont {Jung}, \bibfnamefont {Y.}},\ }\bibfield  {title} {\enquote {\bibinfo {title} {Generative adversarial networks for crystal structure prediction},}\ }\href {\doibase https://doi.org/10.1021/acscentsci.0c00426} {\bibfield  {journal} {\bibinfo  {journal} {ACS central science}\ }\textbf {\bibinfo {volume} {6}},\ \bibinfo {pages} {1412--1420} (\bibinfo {year} {2020})}\BibitemShut {NoStop}%
\bibitem [{sup()}]{suppl}%
  \BibitemOpen
  \href@noop {} {\bibinfo  {journal} {See Supplemental Material at [URL will be inserted by publisher] for details of voxel image production, auto-encoding, comparison to CGCNN, analysis of model bias and details of spin-Chern number computation}\ }\BibitemShut {NoStop}%
\bibitem [{\citenamefont {Park}\ and\ \citenamefont {Wolverton}(2020)}]{cgcnn2}%
  \BibitemOpen
\bibfield  {journal} {  }\bibfield  {author} {\bibinfo {author} {\bibnamefont {Park}, \bibfnamefont {C.~W.}}\ and\ \bibinfo {author} {\bibnamefont {Wolverton}, \bibfnamefont {C.}},\ }\bibfield  {title} {\enquote {\bibinfo {title} {Developing an improved crystal graph convolutional neural network framework for accelerated materials discovery},}\ }\href {\doibase 10.1103/PhysRevMaterials.4.063801} {\bibfield  {journal} {\bibinfo  {journal} {Phys. Rev. Mater.}\ }\textbf {\bibinfo {volume} {4}},\ \bibinfo {pages} {063801} (\bibinfo {year} {2020})}\BibitemShut {NoStop}%
\bibitem [{\citenamefont {Claussen}\ \emph {et~al.}(2020)\citenamefont {Claussen}, \citenamefont {Bernevig},\ and\ \citenamefont {Regnault}}]{NRML}%
  \BibitemOpen
  \bibfield  {author} {\bibinfo {author} {\bibnamefont {Claussen}, \bibfnamefont {N.}}, \bibinfo {author} {\bibnamefont {Bernevig}, \bibfnamefont {B.~A.}}, \ and\ \bibinfo {author} {\bibnamefont {Regnault}, \bibfnamefont {N.}},\ }\bibfield  {title} {\enquote {\bibinfo {title} {Detection of topological materials with machine learning},}\ }\href {\doibase 10.1103/PhysRevB.101.245117} {\bibfield  {journal} {\bibinfo  {journal} {Phys. Rev. B}\ }\textbf {\bibinfo {volume} {101}},\ \bibinfo {pages} {245117} (\bibinfo {year} {2020})}\BibitemShut {NoStop}%
\bibitem [{\citenamefont {Ma}\ \emph {et~al.}(2023)\citenamefont {Ma}, \citenamefont {Zhang}, \citenamefont {Christensen}, \citenamefont {Po}, \citenamefont {Jing}, \citenamefont {Fu},\ and\ \citenamefont {Soljacic}}]{ma2023topogivity}%
  \BibitemOpen
  \bibfield  {author} {\bibinfo {author} {\bibnamefont {Ma}, \bibfnamefont {A.}}, \bibinfo {author} {\bibnamefont {Zhang}, \bibfnamefont {Y.}}, \bibinfo {author} {\bibnamefont {Christensen}, \bibfnamefont {T.}}, \bibinfo {author} {\bibnamefont {Po}, \bibfnamefont {H.~C.}}, \bibinfo {author} {\bibnamefont {Jing}, \bibfnamefont {L.}}, \bibinfo {author} {\bibnamefont {Fu}, \bibfnamefont {L.}}, \ and\ \bibinfo {author} {\bibnamefont {Soljacic}, \bibfnamefont {M.}},\ }\bibfield  {title} {\enquote {\bibinfo {title} {Topogivity: A machine-learned chemical rule for discovering topological materials},}\ }\href {\doibase https://doi.org/10.1021/acs.nanolett.2c03307} {\bibfield  {journal} {\bibinfo  {journal} {Nano Lett.}\ }\textbf {\bibinfo {volume} {23}},\ \bibinfo {pages} {772--778} (\bibinfo {year} {2023})}\BibitemShut {NoStop}%
\bibitem [{\citenamefont {Xu}\ \emph {et~al.}(2024)\citenamefont {Xu}, \citenamefont {Jiang}, \citenamefont {Wang},\ and\ \citenamefont {Wang}}]{2dmagml}%
  \BibitemOpen
  \bibfield  {author} {\bibinfo {author} {\bibnamefont {Xu}, \bibfnamefont {H.}}, \bibinfo {author} {\bibnamefont {Jiang}, \bibfnamefont {Y.}}, \bibinfo {author} {\bibnamefont {Wang}, \bibfnamefont {H.}}, \ and\ \bibinfo {author} {\bibnamefont {Wang}, \bibfnamefont {J.}},\ }\bibfield  {title} {\enquote {\bibinfo {title} {Discovering two-dimensional magnetic topological insulators by machine learning},}\ }\href {\doibase 10.1103/PhysRevB.109.035122} {\bibfield  {journal} {\bibinfo  {journal} {Phys. Rev. B}\ }\textbf {\bibinfo {volume} {109}},\ \bibinfo {pages} {035122} (\bibinfo {year} {2024})}\BibitemShut {NoStop}%
\bibitem [{\citenamefont {Schleder}\ \emph {et~al.}(2021)\citenamefont {Schleder}, \citenamefont {Focassio},\ and\ \citenamefont {Fazzio}}]{schleder2021machine}%
  \BibitemOpen
  \bibfield  {author} {\bibinfo {author} {\bibnamefont {Schleder}, \bibfnamefont {G.~R.}}, \bibinfo {author} {\bibnamefont {Focassio}, \bibfnamefont {B.}}, \ and\ \bibinfo {author} {\bibnamefont {Fazzio}, \bibfnamefont {A.}},\ }\bibfield  {title} {\enquote {\bibinfo {title} {Machine learning for materials discovery: Two-dimensional topological insulators},}\ }\href {\doibase https://doi.org/10.1063/5.0055035} {\bibfield  {journal} {\bibinfo  {journal} {Appl. Phys. Rev.}\ }\textbf {\bibinfo {volume} {8}} (\bibinfo {year} {2021}),\ https://doi.org/10.1063/5.0055035}\BibitemShut {NoStop}%
\bibitem [{\citenamefont {Naguib}\ \emph {et~al.}(2012)\citenamefont {Naguib}, \citenamefont {Come}, \citenamefont {Dyatkin}, \citenamefont {Presser}, \citenamefont {Taberna}, \citenamefont {Simon}, \citenamefont {Barsoum},\ and\ \citenamefont {Gogotsi}}]{naguib2012mxene}%
  \BibitemOpen
  \bibfield  {author} {\bibinfo {author} {\bibnamefont {Naguib}, \bibfnamefont {M.}}, \bibinfo {author} {\bibnamefont {Come}, \bibfnamefont {J.}}, \bibinfo {author} {\bibnamefont {Dyatkin}, \bibfnamefont {B.}}, \bibinfo {author} {\bibnamefont {Presser}, \bibfnamefont {V.}}, \bibinfo {author} {\bibnamefont {Taberna}, \bibfnamefont {P.-L.}}, \bibinfo {author} {\bibnamefont {Simon}, \bibfnamefont {P.}}, \bibinfo {author} {\bibnamefont {Barsoum}, \bibfnamefont {M.~W.}}, \ and\ \bibinfo {author} {\bibnamefont {Gogotsi}, \bibfnamefont {Y.}},\ }\bibfield  {title} {\enquote {\bibinfo {title} {Mxene: a promising transition metal carbide anode for lithium-ion batteries},}\ }\href {\doibase https://doi.org/10.1016/j.elecom.2012.01.002} {\bibfield  {journal} {\bibinfo  {journal} {Electrochem. commun.}\ }\textbf {\bibinfo {volume} {16}},\ \bibinfo {pages} {61--64} (\bibinfo {year} {2012})}\BibitemShut {NoStop}%
\bibitem [{\citenamefont {M{\"u}ller}\ and\ \citenamefont {Lutz}(1991)}]{muller1991single}%
  \BibitemOpen
  \bibfield  {author} {\bibinfo {author} {\bibnamefont {M{\"u}ller}, \bibfnamefont {B.}}\ and\ \bibinfo {author} {\bibnamefont {Lutz}, \bibfnamefont {H.}},\ }\bibfield  {title} {\enquote {\bibinfo {title} {Single crystal raman studies of pyrite-type rus2, ruse2, oss2, osse2, ptp2, and ptas2},}\ }\href {\doibase https://doi.org/10.1007/BF00202242} {\bibfield  {journal} {\bibinfo  {journal} {Phys. Chem. Miner.}\ }\textbf {\bibinfo {volume} {17}},\ \bibinfo {pages} {716--719} (\bibinfo {year} {1991})}\BibitemShut {NoStop}%
\bibitem [{\citenamefont {Fu}\ and\ \citenamefont {Kane}(2008)}]{FuProximity}%
  \BibitemOpen
  \bibfield  {author} {\bibinfo {author} {\bibnamefont {Fu}, \bibfnamefont {L.}}\ and\ \bibinfo {author} {\bibnamefont {Kane}, \bibfnamefont {C.~L.}},\ }\bibfield  {title} {\enquote {\bibinfo {title} {Superconducting proximity effect and majorana fermions at the surface of a topological insulator},}\ }\href {\doibase 10.1103/PhysRevLett.100.096407} {\bibfield  {journal} {\bibinfo  {journal} {Phys. Rev. Lett.}\ }\textbf {\bibinfo {volume} {100}},\ \bibinfo {pages} {096407} (\bibinfo {year} {2008})}\BibitemShut {NoStop}%
\bibitem [{\citenamefont {Stanescu}\ \emph {et~al.}(2010)\citenamefont {Stanescu}, \citenamefont {Sau}, \citenamefont {Lutchyn},\ and\ \citenamefont {Das~Sarma}}]{JQIMajorana}%
  \BibitemOpen
  \bibfield  {author} {\bibinfo {author} {\bibnamefont {Stanescu}, \bibfnamefont {T.~D.}}, \bibinfo {author} {\bibnamefont {Sau}, \bibfnamefont {J.~D.}}, \bibinfo {author} {\bibnamefont {Lutchyn}, \bibfnamefont {R.~M.}}, \ and\ \bibinfo {author} {\bibnamefont {Das~Sarma}, \bibfnamefont {S.}},\ }\bibfield  {title} {\enquote {\bibinfo {title} {Proximity effect at the superconductor--topological insulator interface},}\ }\href {\doibase 10.1103/PhysRevB.81.241310} {\bibfield  {journal} {\bibinfo  {journal} {Phys. Rev. B}\ }\textbf {\bibinfo {volume} {81}},\ \bibinfo {pages} {241310} (\bibinfo {year} {2010})}\BibitemShut {NoStop}%
\bibitem [{\citenamefont {Cook}\ and\ \citenamefont {Franz}(2011)}]{cook1}%
  \BibitemOpen
  \bibfield  {author} {\bibinfo {author} {\bibnamefont {Cook}, \bibfnamefont {A.}}\ and\ \bibinfo {author} {\bibnamefont {Franz}, \bibfnamefont {M.}},\ }\bibfield  {title} {\enquote {\bibinfo {title} {Majorana fermions in a topological-insulator nanowire proximity-coupled to an $s$-wave superconductor},}\ }\href {\doibase 10.1103/PhysRevB.84.201105} {\bibfield  {journal} {\bibinfo  {journal} {Phys. Rev. B}\ }\textbf {\bibinfo {volume} {84}},\ \bibinfo {pages} {201105} (\bibinfo {year} {2011})}\BibitemShut {NoStop}%
\bibitem [{\citenamefont {Cook}\ \emph {et~al.}(2012)\citenamefont {Cook}, \citenamefont {Vazifeh},\ and\ \citenamefont {Franz}}]{cook2}%
  \BibitemOpen
  \bibfield  {author} {\bibinfo {author} {\bibnamefont {Cook}, \bibfnamefont {A.~M.}}, \bibinfo {author} {\bibnamefont {Vazifeh}, \bibfnamefont {M.~M.}}, \ and\ \bibinfo {author} {\bibnamefont {Franz}, \bibfnamefont {M.}},\ }\bibfield  {title} {\enquote {\bibinfo {title} {Stability of majorana fermions in proximity-coupled topological insulator nanowires},}\ }\href {\doibase 10.1103/PhysRevB.86.155431} {\bibfield  {journal} {\bibinfo  {journal} {Phys. Rev. B}\ }\textbf {\bibinfo {volume} {86}},\ \bibinfo {pages} {155431} (\bibinfo {year} {2012})}\BibitemShut {NoStop}%
\bibitem [{\citenamefont {Hell}\ \emph {et~al.}(2017)\citenamefont {Hell}, \citenamefont {Leijnse},\ and\ \citenamefont {Flensberg}}]{prlProximity}%
  \BibitemOpen
  \bibfield  {author} {\bibinfo {author} {\bibnamefont {Hell}, \bibfnamefont {M.}}, \bibinfo {author} {\bibnamefont {Leijnse}, \bibfnamefont {M.}}, \ and\ \bibinfo {author} {\bibnamefont {Flensberg}, \bibfnamefont {K.}},\ }\bibfield  {title} {\enquote {\bibinfo {title} {Two-dimensional platform for networks of majorana bound states},}\ }\href {\doibase 10.1103/PhysRevLett.118.107701} {\bibfield  {journal} {\bibinfo  {journal} {Phys. Rev. Lett.}\ }\textbf {\bibinfo {volume} {118}},\ \bibinfo {pages} {107701} (\bibinfo {year} {2017})}\BibitemShut {NoStop}%
\bibitem [{\citenamefont {Reeg}\ \emph {et~al.}(2018)\citenamefont {Reeg}, \citenamefont {Loss},\ and\ \citenamefont {Klinovaja}}]{reeg2018proximity}%
  \BibitemOpen
  \bibfield  {author} {\bibinfo {author} {\bibnamefont {Reeg}, \bibfnamefont {C.}}, \bibinfo {author} {\bibnamefont {Loss}, \bibfnamefont {D.}}, \ and\ \bibinfo {author} {\bibnamefont {Klinovaja}, \bibfnamefont {J.}},\ }\bibfield  {title} {\enquote {\bibinfo {title} {Proximity effect in a two-dimensional electron gas coupled to a thin superconducting layer},}\ }\href@noop {} {\bibfield  {journal} {\bibinfo  {journal} {Beilstein J. Nanotechnol.}\ }\textbf {\bibinfo {volume} {9}},\ \bibinfo {pages} {1263--1271} (\bibinfo {year} {2018})}\BibitemShut {NoStop}%
\bibitem [{\citenamefont {Yan}\ \emph {et~al.}(2018)\citenamefont {Yan}, \citenamefont {Song},\ and\ \citenamefont {Wang}}]{MCSHTC}%
  \BibitemOpen
  \bibfield  {author} {\bibinfo {author} {\bibnamefont {Yan}, \bibfnamefont {Z.}}, \bibinfo {author} {\bibnamefont {Song}, \bibfnamefont {F.}}, \ and\ \bibinfo {author} {\bibnamefont {Wang}, \bibfnamefont {Z.}},\ }\bibfield  {title} {\enquote {\bibinfo {title} {Majorana corner modes in a high-temperature platform},}\ }\href {\doibase 10.1103/PhysRevLett.121.096803} {\bibfield  {journal} {\bibinfo  {journal} {Phys. Rev. Lett.}\ }\textbf {\bibinfo {volume} {121}},\ \bibinfo {pages} {096803} (\bibinfo {year} {2018})}\BibitemShut {NoStop}%
\bibitem [{\citenamefont {Liu}\ \emph {et~al.}(2018{\natexlab{a}})\citenamefont {Liu}, \citenamefont {He},\ and\ \citenamefont {Nori}}]{LiuMBS}%
  \BibitemOpen
  \bibfield  {author} {\bibinfo {author} {\bibnamefont {Liu}, \bibfnamefont {T.}}, \bibinfo {author} {\bibnamefont {He}, \bibfnamefont {J.~J.}}, \ and\ \bibinfo {author} {\bibnamefont {Nori}, \bibfnamefont {F.}},\ }\bibfield  {title} {\enquote {\bibinfo {title} {Majorana corner states in a two-dimensional magnetic topological insulator on a high-temperature superconductor},}\ }\href {\doibase 10.1103/PhysRevB.98.245413} {\bibfield  {journal} {\bibinfo  {journal} {Phys. Rev. B}\ }\textbf {\bibinfo {volume} {98}},\ \bibinfo {pages} {245413} (\bibinfo {year} {2018}{\natexlab{a}})}\BibitemShut {NoStop}%
\bibitem [{\citenamefont {Nayak}\ \emph {et~al.}(2008)\citenamefont {Nayak}, \citenamefont {Simon}, \citenamefont {Stern}, \citenamefont {Freedman},\ and\ \citenamefont {Das~Sarma}}]{RevModPhys.80.1083}%
  \BibitemOpen
  \bibfield  {author} {\bibinfo {author} {\bibnamefont {Nayak}, \bibfnamefont {C.}}, \bibinfo {author} {\bibnamefont {Simon}, \bibfnamefont {S.~H.}}, \bibinfo {author} {\bibnamefont {Stern}, \bibfnamefont {A.}}, \bibinfo {author} {\bibnamefont {Freedman}, \bibfnamefont {M.}}, \ and\ \bibinfo {author} {\bibnamefont {Das~Sarma}, \bibfnamefont {S.}},\ }\bibfield  {title} {\enquote {\bibinfo {title} {Non-abelian anyons and topological quantum computation},}\ }\href {\doibase 10.1103/RevModPhys.80.1083} {\bibfield  {journal} {\bibinfo  {journal} {Rev. Mod. Phys.}\ }\textbf {\bibinfo {volume} {80}},\ \bibinfo {pages} {1083--1159} (\bibinfo {year} {2008})}\BibitemShut {NoStop}%
\bibitem [{\citenamefont {Qian}\ \emph {et~al.}(2014)\citenamefont {Qian}, \citenamefont {Liu}, \citenamefont {Fu},\ and\ \citenamefont {Li}}]{qian2014quantum}%
  \BibitemOpen
  \bibfield  {author} {\bibinfo {author} {\bibnamefont {Qian}, \bibfnamefont {X.}}, \bibinfo {author} {\bibnamefont {Liu}, \bibfnamefont {J.}}, \bibinfo {author} {\bibnamefont {Fu}, \bibfnamefont {L.}}, \ and\ \bibinfo {author} {\bibnamefont {Li}, \bibfnamefont {J.}},\ }\bibfield  {title} {\enquote {\bibinfo {title} {Quantum spin hall effect in two-dimensional transition metal dichalcogenides},}\ }\href {\doibase 10.1126/science.1256815} {\bibfield  {journal} {\bibinfo  {journal} {Science}\ }\textbf {\bibinfo {volume} {346}},\ \bibinfo {pages} {1344--1347} (\bibinfo {year} {2014})}\BibitemShut {NoStop}%
\bibitem [{\citenamefont {Hsieh}\ \emph {et~al.}(2012)\citenamefont {Hsieh}, \citenamefont {Lin}, \citenamefont {Liu}, \citenamefont {Duan}, \citenamefont {Bansil},\ and\ \citenamefont {Fu}}]{hsieh2012topological}%
  \BibitemOpen
  \bibfield  {author} {\bibinfo {author} {\bibnamefont {Hsieh}, \bibfnamefont {T.~H.}}, \bibinfo {author} {\bibnamefont {Lin}, \bibfnamefont {H.}}, \bibinfo {author} {\bibnamefont {Liu}, \bibfnamefont {J.}}, \bibinfo {author} {\bibnamefont {Duan}, \bibfnamefont {W.}}, \bibinfo {author} {\bibnamefont {Bansil}, \bibfnamefont {A.}}, \ and\ \bibinfo {author} {\bibnamefont {Fu}, \bibfnamefont {L.}},\ }\bibfield  {title} {\enquote {\bibinfo {title} {Topological crystalline insulators in the snte material class},}\ }\href {\doibase https://doi.org/10.1038/ncomms1969} {\bibfield  {journal} {\bibinfo  {journal} {Nat. Commun.}\ }\textbf {\bibinfo {volume} {3}},\ \bibinfo {pages} {982} (\bibinfo {year} {2012})}\BibitemShut {NoStop}%
\bibitem [{\citenamefont {Liu}\ \emph {et~al.}(2015)\citenamefont {Liu}, \citenamefont {Qian},\ and\ \citenamefont {Fu}}]{liu2015crystal}%
  \BibitemOpen
  \bibfield  {author} {\bibinfo {author} {\bibnamefont {Liu}, \bibfnamefont {J.}}, \bibinfo {author} {\bibnamefont {Qian}, \bibfnamefont {X.}}, \ and\ \bibinfo {author} {\bibnamefont {Fu}, \bibfnamefont {L.}},\ }\bibfield  {title} {\enquote {\bibinfo {title} {Crystal field effect induced topological crystalline insulators in monolayer iv--vi semiconductors},}\ }\href {\doibase https://doi.org/10.1021/acs.nanolett.5b00308} {\bibfield  {journal} {\bibinfo  {journal} {Nano Lett.}\ }\textbf {\bibinfo {volume} {15}},\ \bibinfo {pages} {2657--2661} (\bibinfo {year} {2015})}\BibitemShut {NoStop}%
\bibitem [{\citenamefont {L{\"u}pke}\ \emph {et~al.}(2020)\citenamefont {L{\"u}pke}, \citenamefont {Waters}, \citenamefont {de~la Barrera}, \citenamefont {Widom}, \citenamefont {Mandrus}, \citenamefont {Yan}, \citenamefont {Feenstra},\ and\ \citenamefont {Hunt}}]{lupke2020proximity}%
  \BibitemOpen
  \bibfield  {author} {\bibinfo {author} {\bibnamefont {L{\"u}pke}, \bibfnamefont {F.}}, \bibinfo {author} {\bibnamefont {Waters}, \bibfnamefont {D.}}, \bibinfo {author} {\bibnamefont {de~la Barrera}, \bibfnamefont {S.~C.}}, \bibinfo {author} {\bibnamefont {Widom}, \bibfnamefont {M.}}, \bibinfo {author} {\bibnamefont {Mandrus}, \bibfnamefont {D.~G.}}, \bibinfo {author} {\bibnamefont {Yan}, \bibfnamefont {J.}}, \bibinfo {author} {\bibnamefont {Feenstra}, \bibfnamefont {R.~M.}}, \ and\ \bibinfo {author} {\bibnamefont {Hunt}, \bibfnamefont {B.~M.}},\ }\bibfield  {title} {\enquote {\bibinfo {title} {Proximity-induced superconducting gap in the quantum spin hall edge state of monolayer wte2},}\ }\href {\doibase https://doi.org/10.1038/s41567-020-0816-x} {\bibfield  {journal} {\bibinfo  {journal} {Nat. Phys.}\ }\textbf {\bibinfo {volume} {16}},\ \bibinfo {pages} {526--530} (\bibinfo {year} {2020})}\BibitemShut {NoStop}%
\bibitem [{\citenamefont {Shimamura}\ \emph {et~al.}(2018)\citenamefont {Shimamura}, \citenamefont {Sugawara}, \citenamefont {Sucharitakul}, \citenamefont {Souma}, \citenamefont {Iwaya}, \citenamefont {Nakayama}, \citenamefont {Trang}, \citenamefont {Yamauchi}, \citenamefont {Oguchi}, \citenamefont {Kudo} \emph {et~al.}}]{shimamura2018ultrathin}%
  \BibitemOpen
  \bibfield  {author} {\bibinfo {author} {\bibnamefont {Shimamura}, \bibfnamefont {N.}}, \bibinfo {author} {\bibnamefont {Sugawara}, \bibfnamefont {K.}}, \bibinfo {author} {\bibnamefont {Sucharitakul}, \bibfnamefont {S.}}, \bibinfo {author} {\bibnamefont {Souma}, \bibfnamefont {S.}}, \bibinfo {author} {\bibnamefont {Iwaya}, \bibfnamefont {K.}}, \bibinfo {author} {\bibnamefont {Nakayama}, \bibfnamefont {K.}}, \bibinfo {author} {\bibnamefont {Trang}, \bibfnamefont {C.~X.}}, \bibinfo {author} {\bibnamefont {Yamauchi}, \bibfnamefont {K.}}, \bibinfo {author} {\bibnamefont {Oguchi}, \bibfnamefont {T.}}, \bibinfo {author} {\bibnamefont {Kudo}, \bibfnamefont {K.}},  \emph {et~al.},\ }\bibfield  {title} {\enquote {\bibinfo {title} {Ultrathin bismuth film on high-temperature cuprate superconductor bi2sr2cacu2o8+ $\delta$ as a candidate of a topological superconductor},}\ }\href {\doibase https://doi.org/10.1021/acsnano.8b04869} {\bibfield  {journal} {\bibinfo  {journal} {ACS nano}\ }\textbf {\bibinfo {volume} {12}},\
  \bibinfo {pages} {10977--10983} (\bibinfo {year} {2018})}\BibitemShut {NoStop}%
\bibitem [{\citenamefont {Zhao}\ \emph {et~al.}(2018)\citenamefont {Zhao}, \citenamefont {Rachmilowitz}, \citenamefont {Ren}, \citenamefont {Han}, \citenamefont {Schneeloch}, \citenamefont {Zhong}, \citenamefont {Gu}, \citenamefont {Wang},\ and\ \citenamefont {Zeljkovic}}]{ProxZhao}%
  \BibitemOpen
  \bibfield  {author} {\bibinfo {author} {\bibnamefont {Zhao}, \bibfnamefont {H.}}, \bibinfo {author} {\bibnamefont {Rachmilowitz}, \bibfnamefont {B.}}, \bibinfo {author} {\bibnamefont {Ren}, \bibfnamefont {Z.}}, \bibinfo {author} {\bibnamefont {Han}, \bibfnamefont {R.}}, \bibinfo {author} {\bibnamefont {Schneeloch}, \bibfnamefont {J.}}, \bibinfo {author} {\bibnamefont {Zhong}, \bibfnamefont {R.}}, \bibinfo {author} {\bibnamefont {Gu}, \bibfnamefont {G.}}, \bibinfo {author} {\bibnamefont {Wang}, \bibfnamefont {Z.}}, \ and\ \bibinfo {author} {\bibnamefont {Zeljkovic}, \bibfnamefont {I.}},\ }\bibfield  {title} {\enquote {\bibinfo {title} {Superconducting proximity effect in a topological insulator using fe(te, se)},}\ }\href {\doibase 10.1103/PhysRevB.97.224504} {\bibfield  {journal} {\bibinfo  {journal} {Phys. Rev. B}\ }\textbf {\bibinfo {volume} {97}},\ \bibinfo {pages} {224504} (\bibinfo {year} {2018})}\BibitemShut {NoStop}%
\bibitem [{\citenamefont {Liu}\ \emph {et~al.}(2018{\natexlab{b}})\citenamefont {Liu}, \citenamefont {He},\ and\ \citenamefont {Nori}}]{PhysRevB.98.245413}%
  \BibitemOpen
  \bibfield  {author} {\bibinfo {author} {\bibnamefont {Liu}, \bibfnamefont {T.}}, \bibinfo {author} {\bibnamefont {He}, \bibfnamefont {J.~J.}}, \ and\ \bibinfo {author} {\bibnamefont {Nori}, \bibfnamefont {F.}},\ }\bibfield  {title} {\enquote {\bibinfo {title} {Majorana corner states in a two-dimensional magnetic topological insulator on a high-temperature superconductor},}\ }\href {\doibase 10.1103/PhysRevB.98.245413} {\bibfield  {journal} {\bibinfo  {journal} {Phys. Rev. B}\ }\textbf {\bibinfo {volume} {98}},\ \bibinfo {pages} {245413} (\bibinfo {year} {2018}{\natexlab{b}})}\BibitemShut {NoStop}%
\bibitem [{\citenamefont {Sorkun}\ \emph {et~al.}(2020)\citenamefont {Sorkun}, \citenamefont {Astruc}, \citenamefont {Koelman},\ and\ \citenamefont {Er}}]{sorkun2020artificial}%
  \BibitemOpen
  \bibfield  {author} {\bibinfo {author} {\bibnamefont {Sorkun}, \bibfnamefont {M.~C.}}, \bibinfo {author} {\bibnamefont {Astruc}, \bibfnamefont {S.}}, \bibinfo {author} {\bibnamefont {Koelman}, \bibfnamefont {J.~V.~A.}}, \ and\ \bibinfo {author} {\bibnamefont {Er}, \bibfnamefont {S.}},\ }\bibfield  {title} {\enquote {\bibinfo {title} {An artificial intelligence-aided virtual screening recipe for two-dimensional materials discovery},}\ }\href {\doibase https://doi.org/10.1038/s41524-020-00375-7} {\bibfield  {journal} {\bibinfo  {journal} {npj Comput. Mater.}\ }\textbf {\bibinfo {volume} {6}},\ \bibinfo {pages} {106} (\bibinfo {year} {2020})}\BibitemShut {NoStop}%
\bibitem [{\citenamefont {Sancho}\ \emph {et~al.}(1985)\citenamefont {Sancho}, \citenamefont {Sancho}, \citenamefont {Sancho},\ and\ \citenamefont {Rubio}}]{sancho1985highly}%
  \BibitemOpen
  \bibfield  {author} {\bibinfo {author} {\bibnamefont {Sancho}, \bibfnamefont {M.~L.}}, \bibinfo {author} {\bibnamefont {Sancho}, \bibfnamefont {J.~L.}}, \bibinfo {author} {\bibnamefont {Sancho}, \bibfnamefont {J.~L.}}, \ and\ \bibinfo {author} {\bibnamefont {Rubio}, \bibfnamefont {J.}},\ }\bibfield  {title} {\enquote {\bibinfo {title} {Highly convergent schemes for the calculation of bulk and surface green functions},}\ }\href {\doibase 10.1088/0305-4608/15/4/009} {\bibfield  {journal} {\bibinfo  {journal} {J. Phys. F: Met. Phys.}\ }\textbf {\bibinfo {volume} {15}},\ \bibinfo {pages} {851} (\bibinfo {year} {1985})}\BibitemShut {NoStop}%
\end{thebibliography}%
\appendix
\section{Minimal tight-binding models}
A minimal tight-binding model of a system supporting $|\mathcal{C}_{s}|=1$ and a trivial $\mathbb{Z}_{2}$ index is a spinful version of the celebrated Benalcazar-Bernevig-Hughes model. The Bloch Hamiltonian takes the form, 
\begin{multline}\label{eq:BBH}
    H(\mathbf{k})/t=\sin k_{x}\Gamma_{1}+\sin k_{y}\Gamma_{2}+(\cos k_{x}-\cos k_{y})\Gamma_{3}+(\Delta+ \cos k_{x}+ \cos k_{y})\Gamma_{4},  
\end{multline}
where $\Gamma_{j=1,2,3}=\sigma_{1}\otimes \tau_{j}$, and $\Gamma_{4}=\sigma_{3}\otimes \tau_{0}$, where  where $\sigma_{0,1,2,3}(\tau_{0,1,2,3})$ are the $2\times 2$ identity matrix and three Pauli matrices respectively, operating on the spin (orbital) indices.

\par
\begin{figure*}
\centering
\subfigure[]{
\includegraphics[scale=0.25]{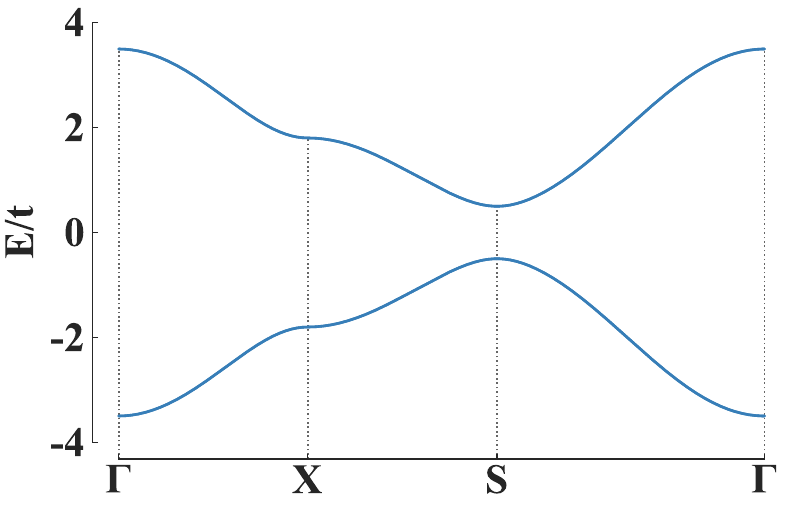}
\label{fig:BBHBands}}
\subfigure[]{
\includegraphics[scale=0.25]{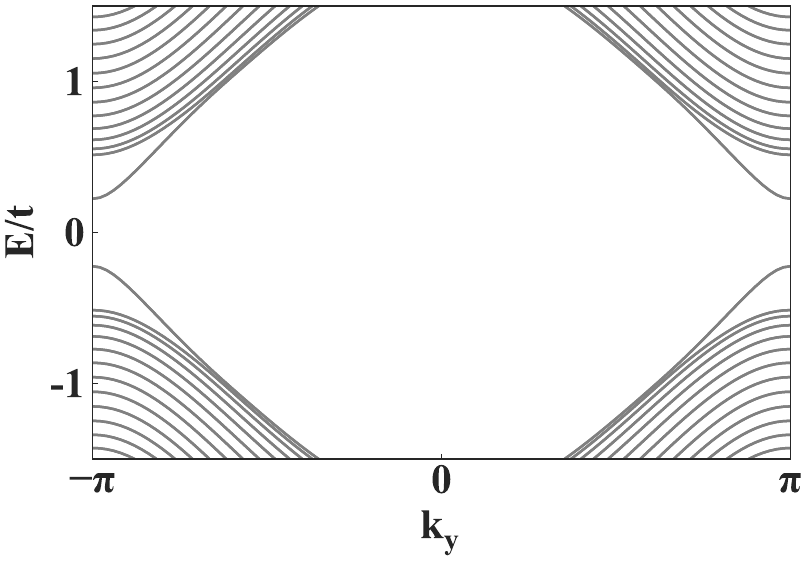}
\label{fig:BBHSS}}
\subfigure[]{
\includegraphics[scale=0.25]{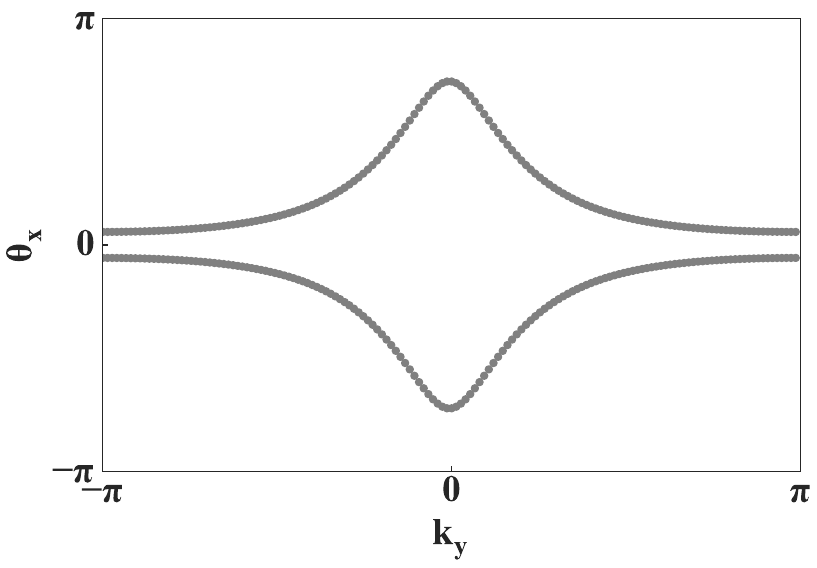}
\label{fig:BBHWCC}}
\subfigure[]{
\includegraphics[scale=0.25]{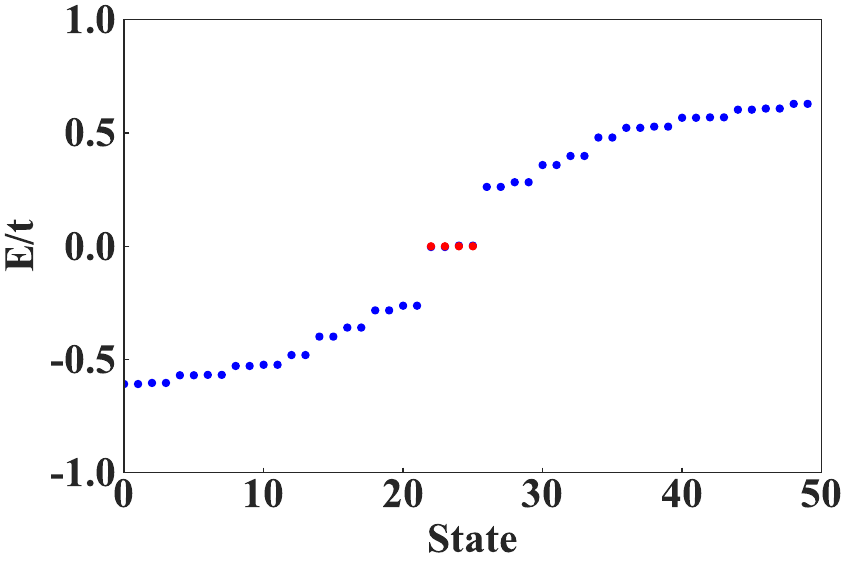}
\label{fig:BBHOBCSpectra}}
\subfigure[]{
\includegraphics[scale=0.25]{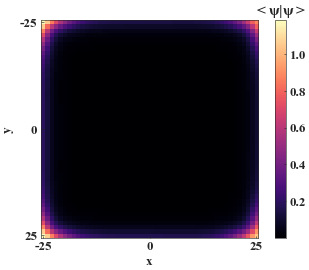}
\label{fig:BBHCLoc}}
\subfigure[]{
\includegraphics[scale=0.25]{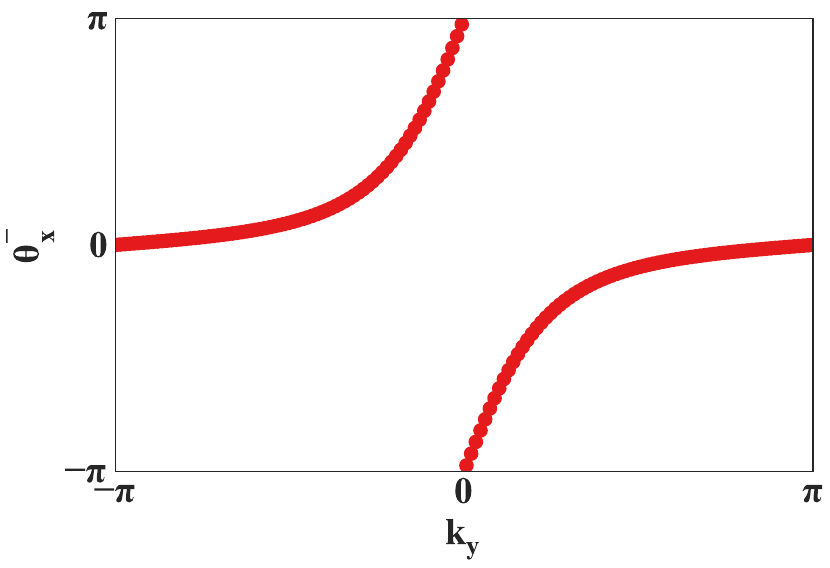}
\label{fig:BBHSpinWCC}}
\subfigure[]{
\includegraphics[scale=0.25]{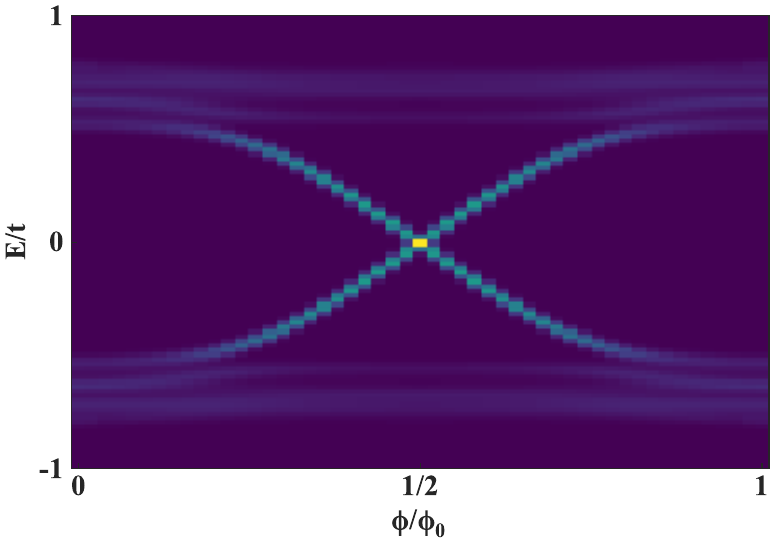}
\label{fig:BBHFlux}}
\subfigure[]{
\includegraphics[scale=0.25]{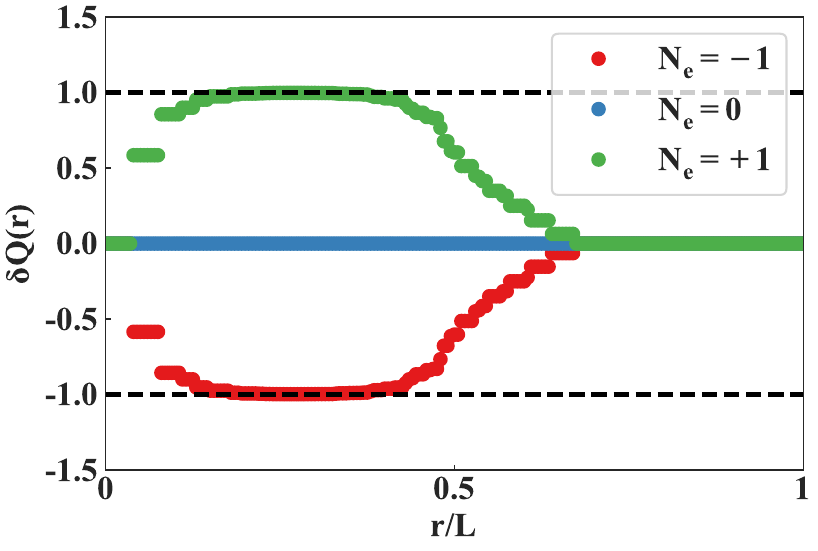}
\label{fig:BBHIC}}
\caption{(a) Band structure of the tight-binding model in eq. \eqref{eq:BBH}. The band structure considering a 20 $\times$ 1 slab with open(periodic)-boundary conditions along the $x(y)$ direction, detailing the gapped edge spectra. (c) Gapped Wannier center charge spectra disallowing non-trivial topological classification. (d) Spectra under two-dimensional open-boundary conditions for a slab of size $50 \times 50$ unit cells. Four degenerate zero modes are colored in red. Localization of the zero modes is shown in (e), demonstrating that they are corner bound. (f) Spin-resolved Wannier center charge spectra demonstrating $|\mathcal{C}_{s}|=1$. (g) Local density of states on inserted flux tube as a function of flux strength. (h) Induced charge on inserted flux tube as a function of occupied vortex-bound modes.}
\label{fig:BBH}
\end{figure*}

 
The lattice parameter is set to unity and we begin by setting $\Delta_{0}=0$ such that the model supports $C_{4}$ rotational symmetry. The time-reversal symmetry is explicitly implemented as, $\mathcal{T}=i(\sigma_{2}\otimes\tau_{0})\mathcal{K}$, where $\mathcal{K}$ indicates complex conjugation. Furthermore, both the WCC spectra and surface state spectra are gapped as seen in Fig. \eqref{fig:BBHWCC} and Fig. \eqref{fig:BBHSS} respectively. The presence of zero-energy corner states in a $C_{4}$ symmetry preserving geometry is shown in Fig. \eqref{fig:BBHOBCSpectra}. These modes are pinned at zero energy due to the presence of an additional chiral symmetry in the Hamiltonian.
\par 
Having determined that the Hamiltonian admits a trivial $\mathbb{Z}_{2}$ index, we compute the ground-state spin-Chern number following Prodan\cite{Prodan2009}. To do so we define the projected spin operator (PSO), $P(\mathbf{k})\hat{s}P(\mathbf{k})$, where $P(\mathbf{k})$ is the projector onto occupied bands and $\hat{s}=\sigma_{3}\otimes \tau_{0} $ is the preferred spin axis. As the PSO supports a spectral gap, the spin-Chern number can be computed via spin-resolved Wilson loop as detailed in Lin et. al\cite{Lin2022Spin}. The results shown in Fig. \eqref{fig:BBHSpinWCC}, demonstrate the WCC spectra when performing the spin-resolved Wilson loop along the $\hat{x}$ axis as a function of transverse momenta $k_{y}$ for the band corresponding to negative eigenvalues of the PSO, $\theta^{-}_{x}$. The conclusion is in alignment with the earlier determination that the ground states supports $|\mathcal{C}_{s}|=1$.
\par
The spin-Chern number could further have been diagnosed using the method of magnetic flux tube insertion. In order to accomplish flux insertion, we consider a slab of $60 \times 60$ unit cells and introduce the Peierls factor, $H_{ij} e^{i\phi_{ij}}$ where $\phi_{ij}=2\pi(\phi/\phi_{0})\times\Theta(x_{j}-x_{i})\delta(y_{i})$. Examining the local density of states on the flux tube as a function of $\phi/\phi_{0}$, we find the results shown in Fig. \eqref{fig:BBHFlux} detailing a pair of vortex-bound modes (VBMs) which traverse the bulk gap. The induced charge on the vortex is then measured at $\phi=\phi_{0}/2$ as a function of filling with the results shown in Fig. \eqref{fig:BBHIC}, detailing that when the VBMs are fully occupied or unoccupied, the vortex acquires a quantized induced charge in accordance with Ref. \cite{tyner2023solitons}.

\begin{figure*}[t]
\centering
\subfigure[]{
\includegraphics[scale=0.45]{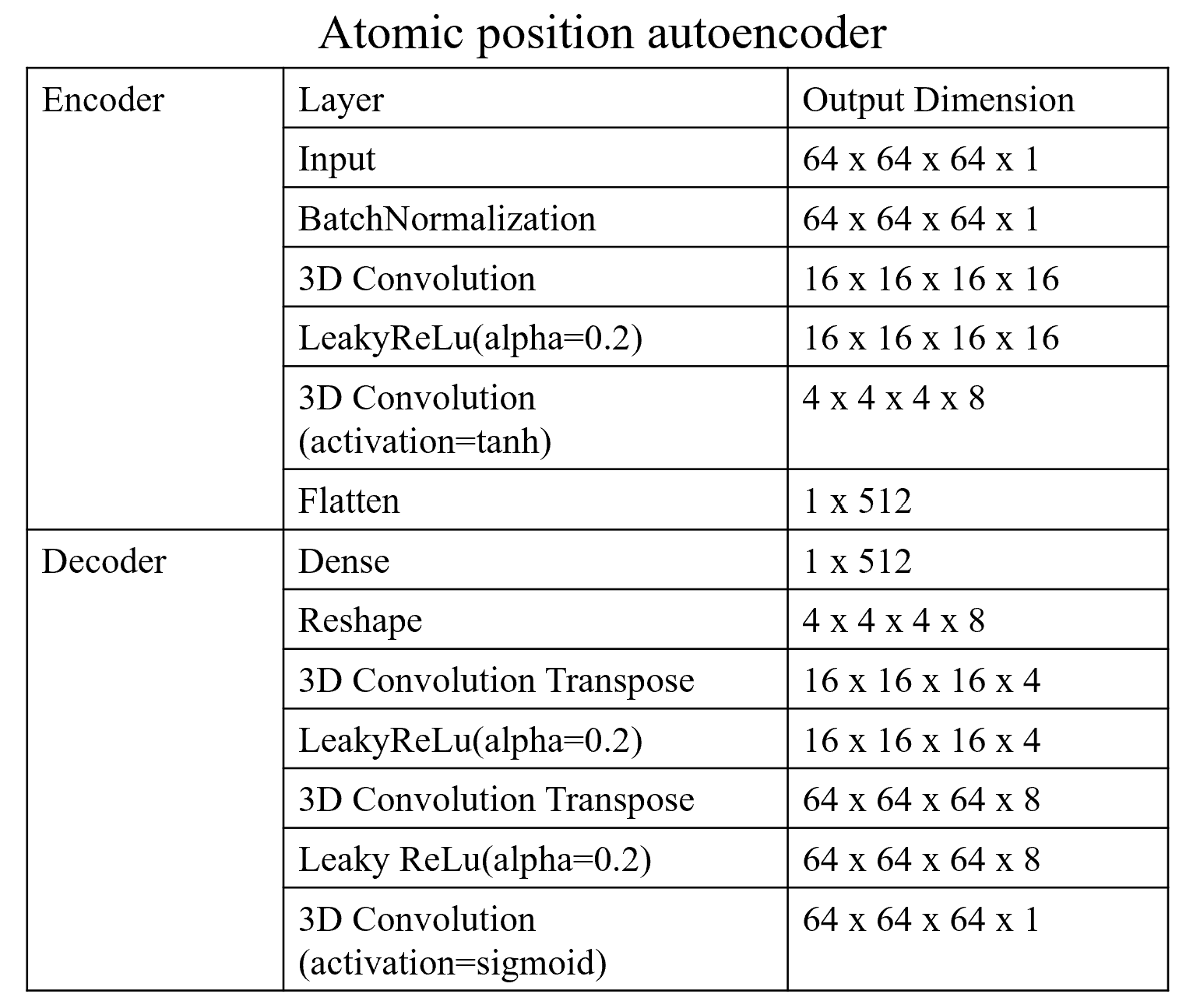}
\label{fig:AtomAuto}}
\subfigure[]{
\includegraphics[scale=0.42]{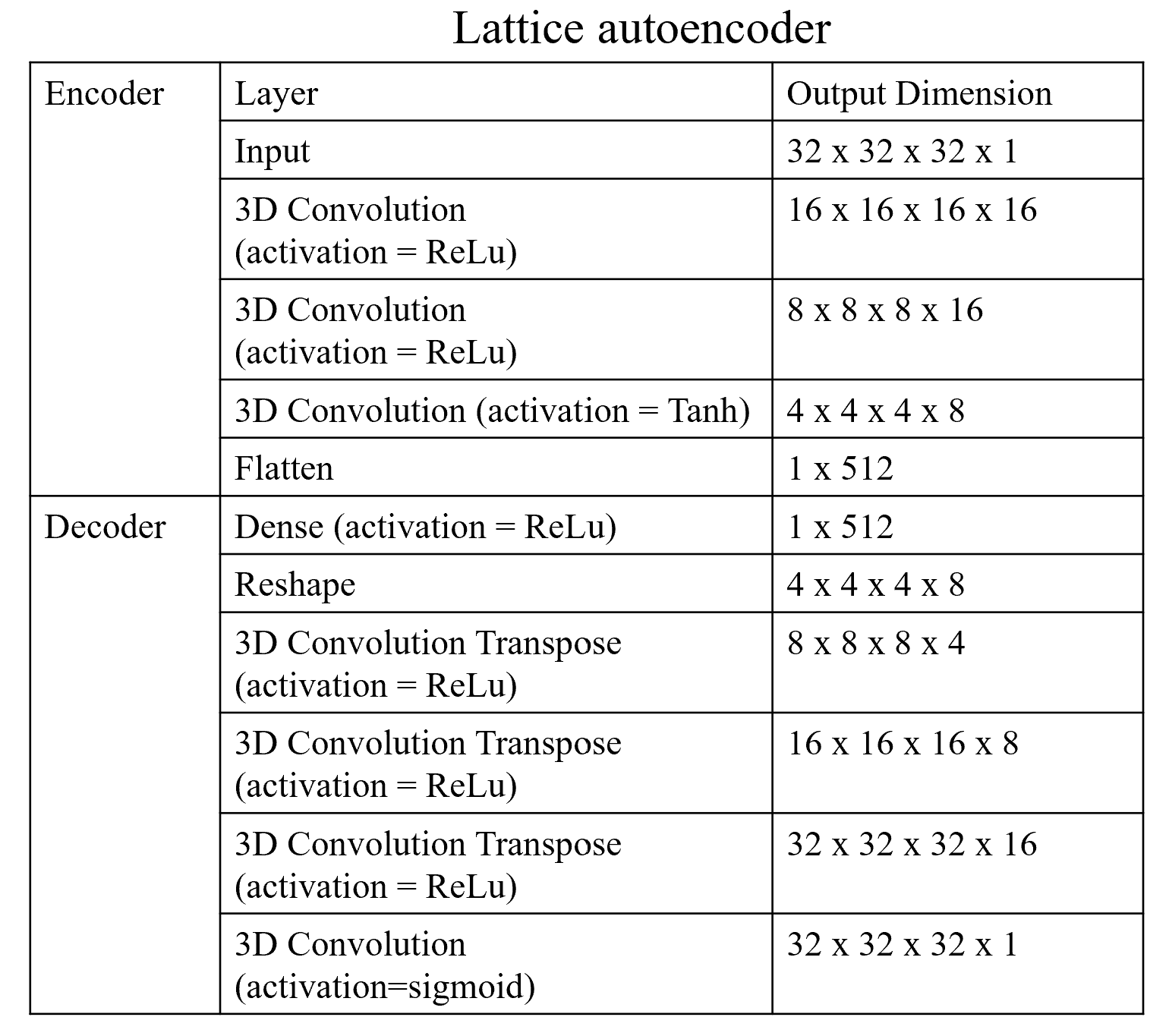}
\label{fig:LatAuto}}
\caption{(a) Architecture of encoder and decoder to recast voxel image of atomic positions to a $(1\times 512)$ vector. (b) Architecture of encoder and decoder to recast voxel image of crystal lattice to a $(1\times 512)$ vector.}
\label{fig:AutoEncoder}
\end{figure*}

\begin{figure*}[t]
\centering
\subfigure[]{
\includegraphics[scale=0.33]{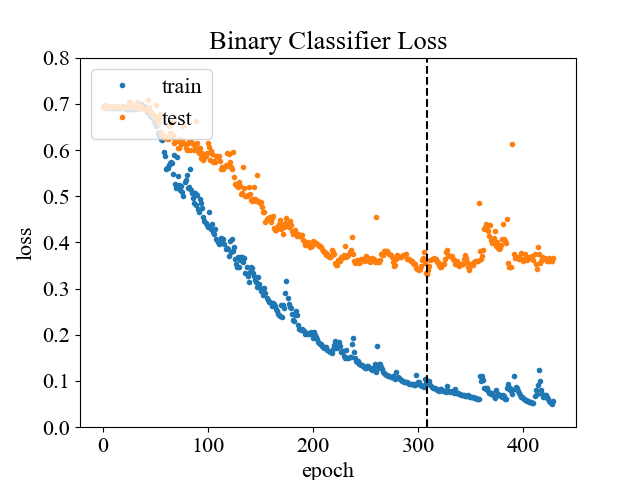}
\label{fig:BLoss}}
\subfigure[]{
\includegraphics[scale=0.33]{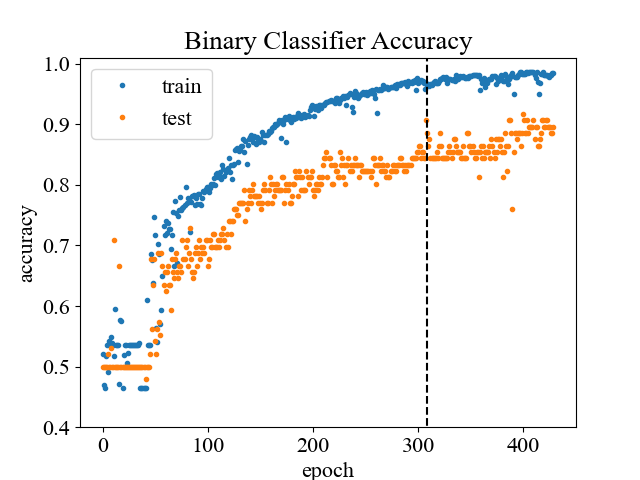}
\label{fig:BAcc}}
\subfigure[]{
\includegraphics[scale=0.33]{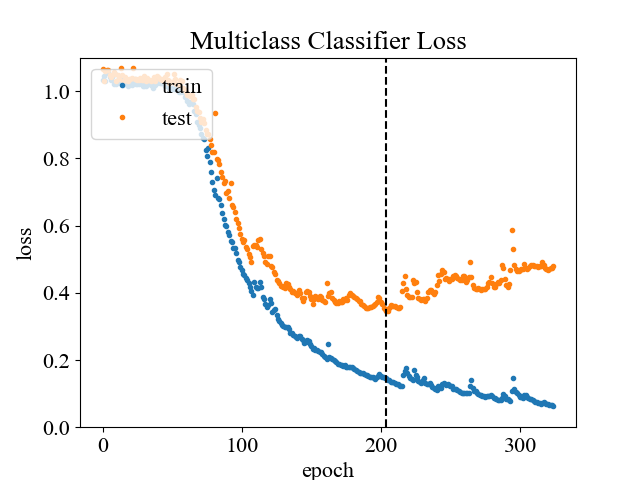}
\label{fig:MLoss}}
\subfigure[]{
\includegraphics[scale=0.33]{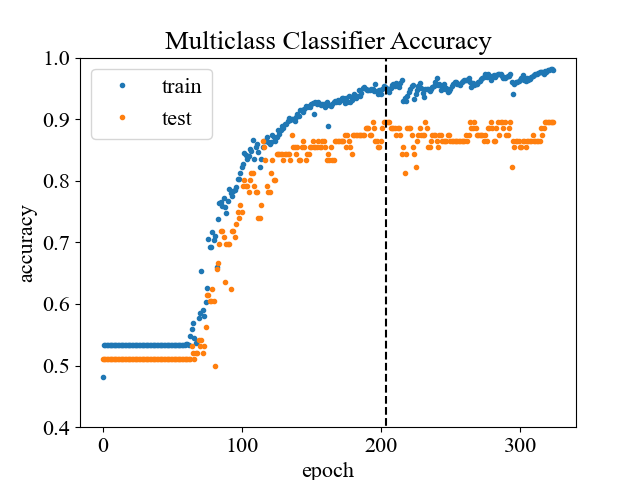}
\label{fig:MAcc}}
\caption{(a)[c] Training and validation loss as a function of epoch for binary[multiclass] classifier. Dashed lines marks location of minimum validation loss from which weights are restored for final model. (b)[d] Training and validation accuracy for binary[multiclass] classifier as a function of epoch. Dashed lines marks location of minimum validation loss from which weights are restored for final model.}
\label{fig:Training}
\end{figure*}

\section{Details of computational screening for training set construction}

In this work, all first principles calculations based on density-functional theory (DFT) are carried out using the Quantum Espresso software package \cite{QE-2009,QE-2017,QE-2020}. Exchange-correlation potentials use the Perdew-Burke-Ernzerhof (PBE) parametrization of the generalized gradient approximation (GGA) \cite{Perdew1996}. In order to facilitate automated analysis of the bulk topology, Wannier tight-binding (WTB) models, as produced via the Wannier90 software package, are constructed using the SCDM method introduced by Vitale et. al\cite{vitale2020automated}. Manipulation of Wannier tight-binding models for vortex insertion is done with a custom python program which will be made publicly available upon being developed into a stand-alone package. For all systems, a 20 x 20 x 1 Monkhorst-Pack grid of k-points is utilized as well as a plane wave cutoff of 50 Ry. Spin-orbit coupling is accounted for in all calculations. To computationally simulate the flux tube, all hopping elements $H_{ij}$, connecting lattice sites $\mathbf{r}_{i}$  and $\mathbf{r}_{j}$ are modified to $H_{ij} e^{i\phi_{ij}}$, where we define the Peierls factor,  
\begin{equation}
    \phi_{ij}=\frac{\phi}{\phi_{0}}\int_{\mathbf{r}_{i}}^{\mathbf{r}_{j}}\frac{\hat{z}\times \mathbf{r}}{\mathbf{r}^2}\cdot d\mathbf{l}. 
\end{equation}
All lattice parameters and atomic positions are used as defined in Ref. \cite{mounet2018two}. 
\par
The criteria for topological classification is based on measuring quantized induced charge on the vortex. The mechanism of spin-charge separation was shown to be insertion of a magnetic flux tube (vortex). In original works by Qi and Zhang\cite{QiSpinCharge} and Ran et. al\cite{SpinChargeVishwanath}, it was shown that a $\phi=hc/(2e)$ ($\pi$-flux) tube binds $2N$, degenerate states in a non-trivial $\mathbb{Z}_{2}$ insulator with odd integer spin Chern number $\mathcal{C}_{s,G}=N$. This concept was then extended to the situation of arbitrary spin-Chern numbers in Refs. \cite{tynerbismuthene}. For systems supporting spin-rotation symmetry, it was further demonstrated that spin charge separation can be observed by tuning filling fraction of the mid-gap vortex bound modes (VBMs). If the VBMs are half-filled, the vortex acquires induced spin but no induced charge. If we dope by $N_{e} \in [-N,+N]$ electrons away from half-filling, occupying all VBMs, the vortex acquires induced charge $\delta Q= N_{e}\times e$. If this condition is satisfied, the spin Chern number can be directly calculated by fixing $N_{e}=N$ such that $\delta Q= |\mathcal{C}_{s,G}|\times e$. While the introduction of spin-orbit coupling causes the spin bound to the vortex to become finite but non-quantized, the quantization of bound charge remains robust. As a result, \emph{quantization of bound charge is the criteria for topological classification employed in this work.} For further details please see Ref. \cite{tyner2023solitons}.
\par 
The results of the search performed in this work are given at the end of this supplementary material.

\begin{figure*}[t]
\centering
\subfigure[]{
\includegraphics[scale=0.45]{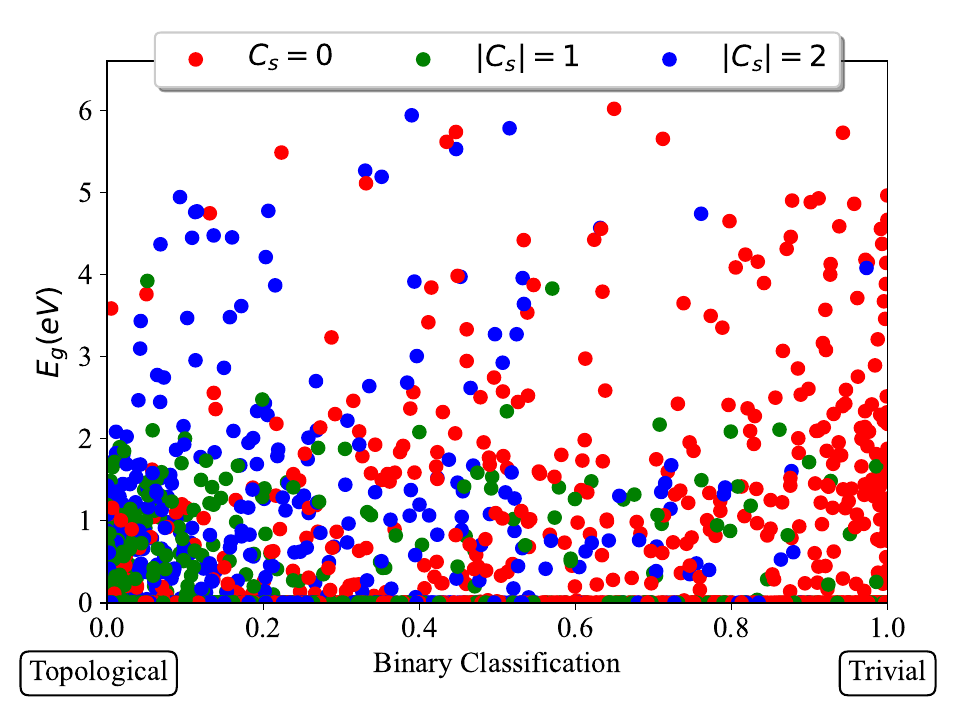}
\label{fig:GapTrend}}
\subfigure[]{
\includegraphics[scale=0.48]{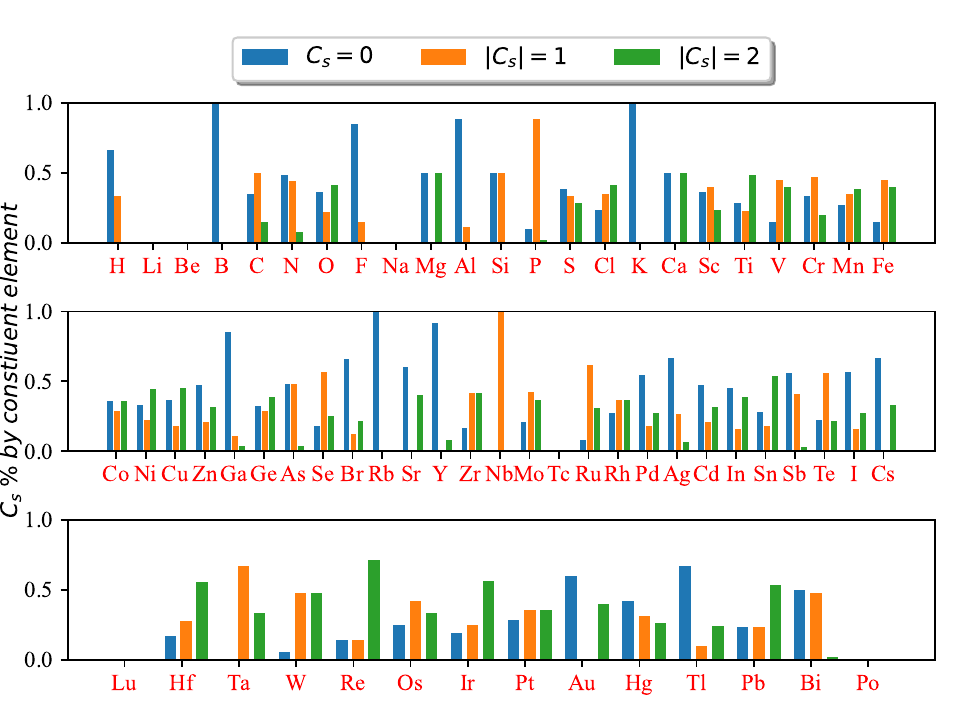}
\label{fig:AtomTrend}}
\caption{(a) Scatter plot detailing topological classification of two-dimensional insulators in the C2DB database by trained binary classifier as a function of bulk energy gap. Classification using trained multiclass network is shown via color-coding. (b) Percentage of systems containing a given element predicted to support $C_{s}=0,\pm1,\pm2$ utilizing multiclass network.}
\label{fig:Trends}
\end{figure*}

\section{Construction of 2D crystal graphs}

\par 
As discrete objects crystalline compounds are not well suited for presentation to a convolutional neural network. To overcome this obstacle and present a continuous representation of the crystal structure, voxel images of the atomic positions and the lattice are first produced following the procedure of Refs. \cite{nouira2018crystalgan,long2021constrained}. 
\par 
\emph{Voxel image of atoms:} A three-dimensional voxel image is produced for each atomic species in the compound. The atomic positions, in Cartesian coordinates are first scaled. The minimum x,y, and z value of all atoms in the compound is measured and these values are then subtracted from each atom such that the all Cartesian positions are positive and have a minimum value of zero in each direction. The voxel images are then formed by considering a three-dimensional box size 12 $\AA^{3}$, discretized into $64^{3}$ voxels. Each voxel is assigned a value as $e^{-r^2/(2\beta)}$, where $r$ is the distance from the voxel position to the nearest correct atomic species. 
\par 
Upon repeating this process for each atomic species considered in the dataset, regardless of whether they exist in the compound, the voxel images are then autoencoded into an array of size $1\times 512$. Details of the autoencoder structure are given in Fig. \eqref{fig:AtomAuto}.
\par 
\emph{Lattice autoencoder:} In order to produce a voxel image of the three-dimensional lattice in a manner that allows for the lattice to be easily reconstructed, we consider a three-dimensional box discretized into $32^{3}$ voxels. The value of each voxel is then given by $e^{-r_{ijk}/2}$, where $r_{ijk}=|a'|+|b'|+|c'|+a'\times b'+a'\times c'+b'\times c'$. We further define $a'=\mathbf{a}\cdot \{i/31,0,0\}$, $b'=\mathbf{b}\cdot \{0,j/31,0\}$, and $c'=\mathbf{c}\cdot \{0,0,k/31\}$ where $(ijk)$ index the voxels in the $32^{3}$ array. A second autoencoder is then utilized to compress this voxel image into an array of size ($1\times 512$). Details of the autoencoder are given in Fig. \eqref{fig:LatAuto}. 
\par 
The resulting one-dimensional arrays are combined to form a $80\times 512$ image that is reshaped into size $160\times 256$ before being presented to the convolutional neural network. 
\begin{figure*}[t]
\centering
\subfigure[]{
\includegraphics[scale=0.55]{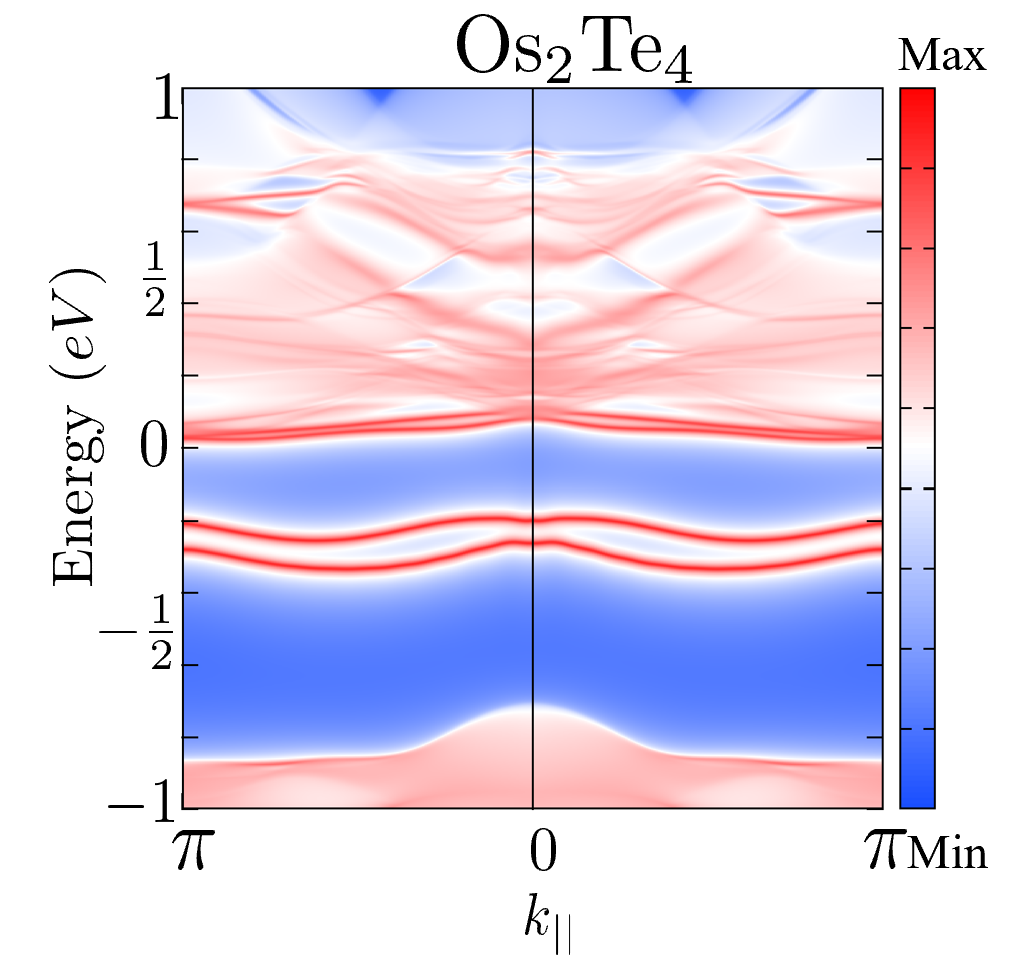}
\label{fig:oteX}}
\subfigure[]{
\includegraphics[scale=0.55]{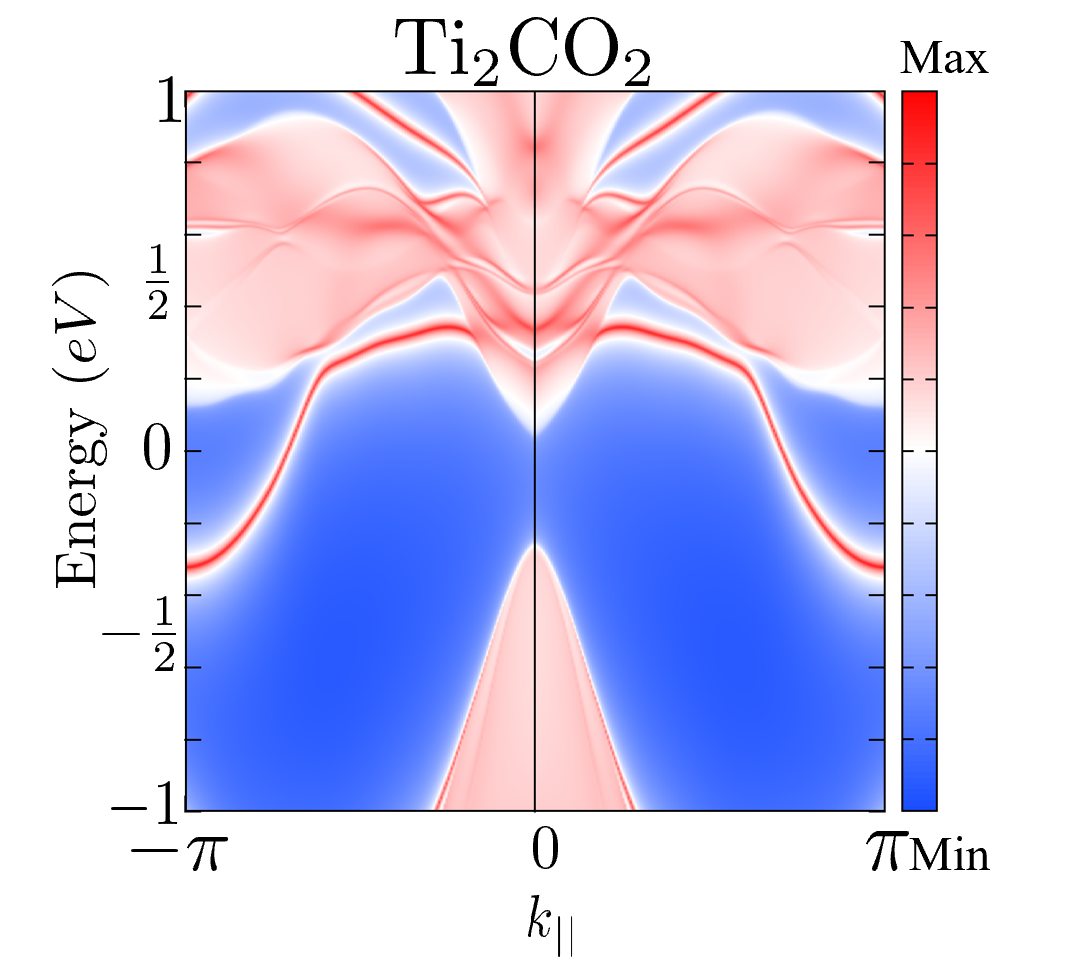}
\label{fig:TCOzz}}
\caption{ Spectral density on (a) the (10) surface of Os$_{2}$Te$_{4}$ the (b) zigzag surface of Ti$_{2}$CO$_{2}$. In both cases the surface spectra is gapped, but surface bound mid-gap states can be seen. }
\label{fig:SS}
\end{figure*}

\section{Details of network training}
\par 
In this appendix we include details of the neural network training. It is important to be aware that while the binary and multiclass networks have a similar architecture, the optimizer we use in each case are distinct. We utilize the stochastic gradient descent (SGD) optimizer as implemented in Keras for the binary classification. For the multiclass network we utilize the Adam algorithm. 
\par 
The training and validation loss and accuracy as a function of epoch for both systems is shown in Fig. \eqref{fig:Trends}. We implement early stopping based on validation loss in both cases to avoid over-fitting. This early-stopping begins after 60 epochs and is implemented with a patience of 120 epochs. The early-stopping location for which the model weights are restored after training is denoted by a vertical dashed line. We note that for epochs beyond this location the validation accuracy is generally stagnant while the training accuracy improves, a sign of the over-fitting we wish to avoid. 
\par
{\color{black} It is also important to discuss the choice of hyperparameters in the model. The initial hyperparameter choices were guided by the published data in Ref. 
\cite{long2021constrained}. These parameters were then adjusted adiabatically to optimize for the model at hand based on computational considerations such as GPU ram as well as the empirical judgement of model performance. It was found that, for the given architecture, the performance of the model was reasonable equivalent for a range of hyperparameters. The same was true when utilizing the CGCNN\cite{cgcnn,cgcnn2} architecture as will be discussed in the section VI.} 

\section{Trends in network classification}
\par
Given the limited training dataset, it is natural to inquire as to whether obvious trends appear in the behavior of the classifiers. Two such trends we wish to investigate are associated with (a) the band gap at the Fermi energy and (b) the presence of certain elements in the unit cell. 
\par 
We begin by analyzing trends associated with the band-gap at the Fermi energy. As stated in the main body, large band-gap topological insulators are rare due to the fact that the band gap is generally induced via spin-orbit coupling. We expect this correlation between small band-gaps and non-trivial topology to manifest in the behavior of the binary and multiclass predictor. To observe whether this trend is present in the classifier we supply all two-dimensional materials in the C2DB database to the binary and multiclass classifier. The results are then plotted as a function of the binary classifier classification as well as band gap at zero energy. The points are further colored by the multiclass classification. The resulting plot, Fig. \eqref{fig:GapTrend}, demonstrates a clear bias towards small band-gap insulators supporting non-trivial topology. Interestingly, this figure also demonstrates that while the multiclass and binary networks are overall consistent, there are instances in which the binary classifier labels a system as trivial and the multiclass classifier does not, and vice versa. In this way these two networks can be used as a form of cross validation. 
\par 
Next, we wish to search for clear trends in topological classification as a function of constituent elements. This is important given the limited training dataset quantity as it could be possible that if, for example, element X appears in only three entries in the training data which are all labeled trivial, that any system containing element X will be labeled trivial. To search for these trends we supply all two-dimensional materials in the C2DB database to the multiclass classifier and arrange the results based on the presence of the elements in each crystal that is classified. The results are shown in Fig. \eqref{fig:AtomTrend}. From these results it is most informative to observe those elements which display large bias. For example, all systems containing boron and rubidium have been marked trivial with $|C_{s}|=0$. In the case of these two elements, this aligns with the example scenario. All materials with boron or rubidium in the training set are labeled topologically trivial. This is a clear bias in the dataset and as such predictions made by the model on materials which include these elements should be treated cautiously. Similarly, we note all systems containing potassium and niobium are labeled as supporting $|C_{s}|=0$ and $|C_{s}|=1$ respectively. These represent more subtle cases in which the training dataset incorporates both topological and trivial systems which include these elements, however the C2DB dataset which we analyze has only four and two systems containing these elements respectively, creating large bias in the analysis.

\section{Comparison with CGCNN}

\par 

In this work a strategy of creating two-dimensional crystal graphs to present to a convolutional neural network has been utilized. This choice was primarily made as it allowed for ease in adopting the data augmentation technique presented in the main body. This is important given the limited quantity of training data available. A clear alternate choice of neural network is a crystal graph convolutional neural network (CGCNN). CGCNNs have emerged as one of the most effective methods for application of neural networks to crystalline data. As such, it is natural to inquire as to the effectiveness of CGCNNs in our work. We have trained a CGCNN using the code made publicly available by Xie\cite{cgcnn}. 
\par 
We note that a drawback of the CGCNN in our context is that the data augmentation procedure presented in the main-body is no longer effective. We thus have less data points then is recommended for a CGCNN. Nevertheless, training the model with the same train/test/validation split used in the main body and using the default setting for a classification model, we find that the CGCNN reaches a train/validation/test accuracy of 98\%/85\%/72\%. While not as high as the model presented in the main-body, it remains an impressive result and demonstrates the potential effectiveness of this powerful method if more data is made available. 

\section{Surface states of Os$_{2}$Te$_{4}$ and Ti$_{2}$CO$_{2}$ }
\par 
\emph{Surface states:} As the Wannier center spectra is gapped for both Os$_{2}$Te$_{4}$ and Ti$_{2}$CO$_{2}$, the bulk-boundary correspondence principle dictates that the surface spectra will be gapped. The surface spectra for both compounds is generated using the WannierTools software package\cite{WU2017} which utilizes the recursive Greens function algorithm of Ref. \cite{sancho1985highly}. The spectral density on the $(01)$ surface for Os$_{2}$Te$_{4}$ and Ti$_{2}$CO$_{2}$ is shown in Fig. \eqref{fig:oteX} and \eqref{fig:TCOzz} respectively. We note the presence of surface-bound mid-gap states and the absence of gapless states in both compounds. Due to the absence of gapless modes, determination of the bulk topology can not be made solely through examination of the surface spectra.

\begin{figure*}
    \includegraphics[width=16cm]{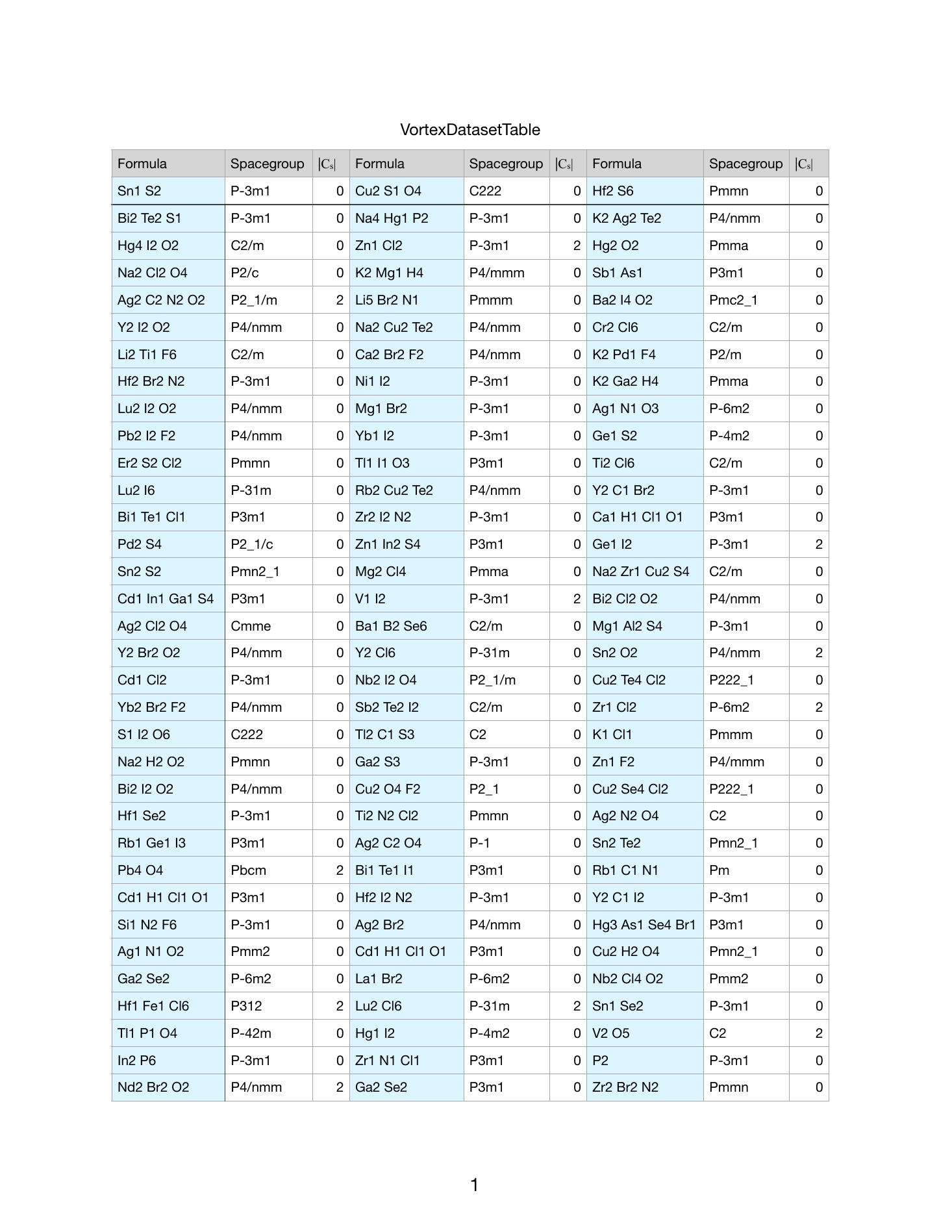}
\end{figure*}
\begin{figure*}
    \includegraphics[width=16cm]{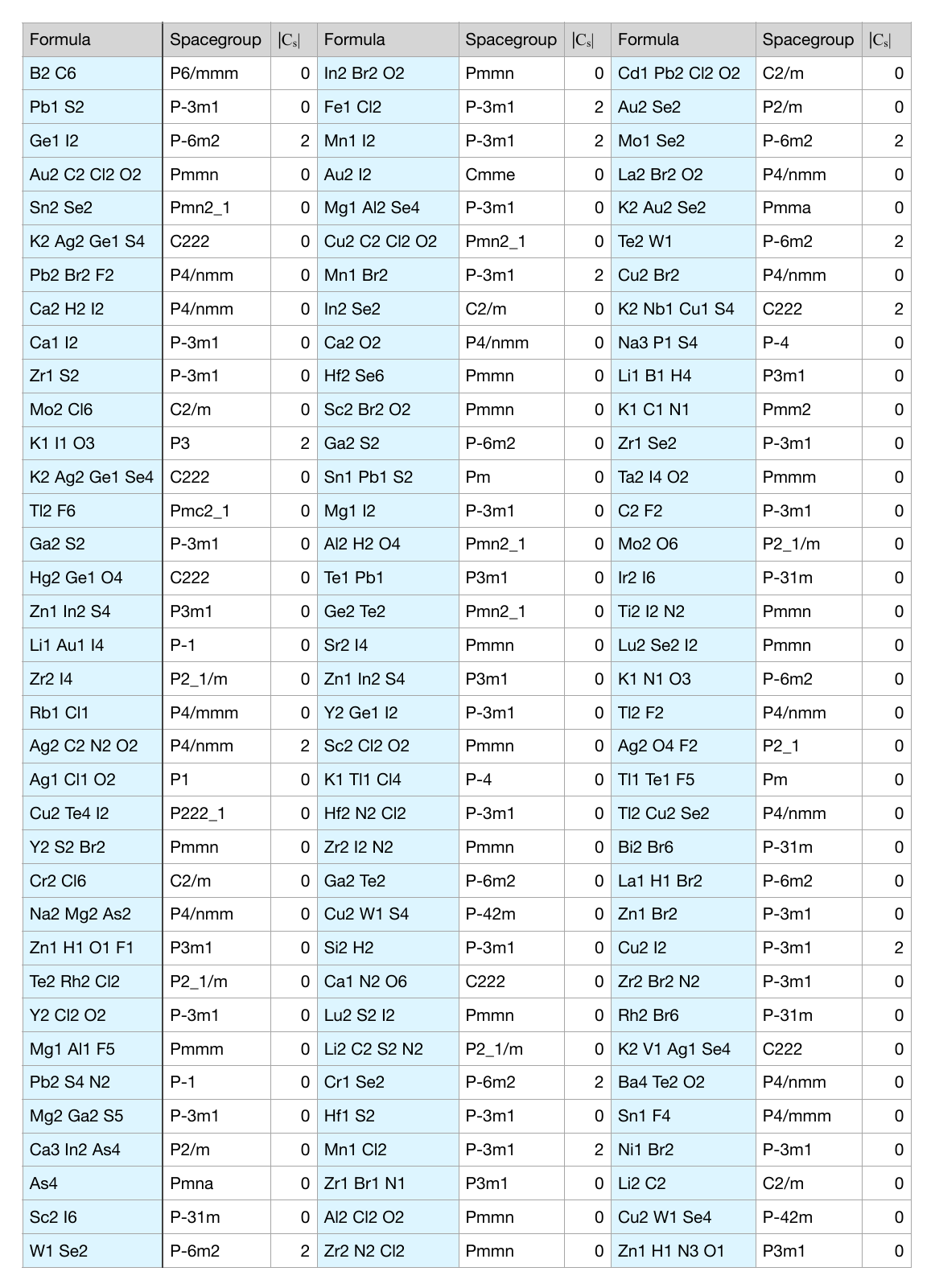}
\end{figure*}
\begin{figure*}
    \includegraphics[width=16cm]{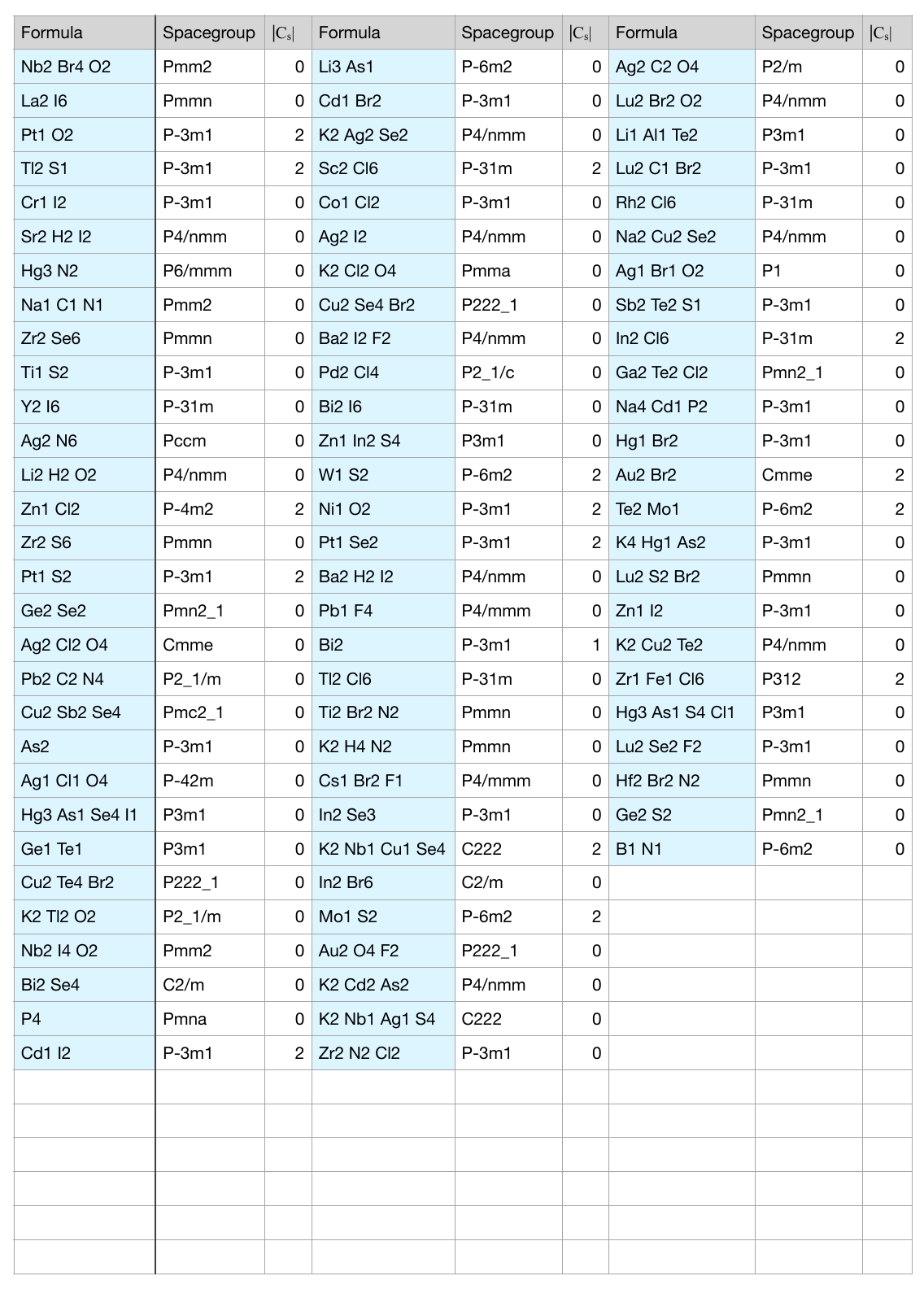}
\end{figure*}

\end{document}